\documentclass{appolb}
\usepackage[russian,english]{babel}
\usepackage{epsfig}
\usepackage{amssymb}

\newcommand{\beq}{\begin{eqnarray}}
\newcommand{\eeq}{\end{eqnarray}}
\newcommand{\nneeq}{\nonumber \end{eqnarray}}

\newcommand{\nn}{\nonumber \\}
\newcommand{\es}{& = &}
\newcommand{\rs}{\, = \,}
\newcommand{\ps}{& + &}
\newcommand{\ms}{& - &}
\newcommand{\ts}{& \times &}
\newcommand{\nt}{\nn \ts}
\newcommand{\np}{\nn \ps}

\newcommand{\cM}{ {\cal M} }
\newcommand{\cH}{ {\cal H} }

\newcommand{\cF}{ {\cal F} }

\begin{document}
\title{    Reinterpretation of gluon condensate in            \\ 
           dynamics of hadronic constituents                  }
\author{   Stanis{\l}aw D. G{\l}azek 
\address{  Institute of Theoretical Physics,
           Faculty of Physics, 
           University of Warsaw                               }}
\date{     June 30, 2011}
\maketitle
\begin{abstract}
We describe an approximate quantum mechanical
picture of hadrons in Minkowski space in the
context of a renormalization group procedure for
effective particles (RGPEP) in a light-front
Hamiltonian formulation of QCD. The picture
suggests that harmonic oscillator potentials for
constituent quarks in lightest mesons and baryons
may result from the gluon condensation inside
hadrons, rather than from an omnipresent gluon
condensate in vacuum. The resulting boost-invariant 
constituent dynamics at the renormalization group 
momentum scales comparable with $\Lambda_{QCD}$,
is identified using gauge symmetry and a crude 
mean-field approximation for gluons. Besides 
constituent quark models, the resulting picture 
also resembles models based on AdS/QCD ideas. 
However, our hypothetical picture significantly 
differs from the models by the available option 
for a systematic analysis in QCD, in which the 
new picture may be treated as a candidate for a 
first approximation. This option is outlined by 
embedding our presentation of the crude and simple 
hadron picture in the context of RGPEP and a 
brief outlook on hadron phenomenology. Several
appendices describe elements of the formalism
required for actual calculations in QCD, including
an extension of RGPEP beyond perturbation theory.
\end{abstract}

\vskip-6in
\hfill {\bf IFT/11/05}
\vskip6.5in

\section{Introduction}
\label{i}

Contrast between the simplicity of constituent
quark model (CQM) classification scheme for
hadrons in particle data tables \cite{PDG} and the
complexity of QCD, exists since quarks were
proposed to explain hadrons \cite{Zweig} and QCD
was proposed as a theory of strong interactions
\cite{ WeinbergGaugeHistory, GellMannQCD, Dubna}.
The contrast can be illustrated by the simplicity
of attempts to build relativistic quark models, in
which one introduces harmonic oscillator
potentials~\cite{Feynman}, in comparison with the
complexity of QCD sum rules \cite{SVZ}, in which
even the state without any hadron, i.e., vacuum,
is meant to contain a complex structure. The
complex structure is reflected in the concept of
condensates. 

The vacuum structure in quantum field theory (QFT)
eludes theorists \cite{DiracDeadWood} despite that
they have a complete command of perturbative
calculations. For example, in the case of QCD,
theorists can use asymptotic freedom \cite{af1,af2} 
to calculate quark and gluon scattering amplitudes
but they also have to use a parton model
\cite{partonmodel} to relate their calculations to
observables. The point is that the mechanism of
binding of partons continues to be a mystery. The
mystery is understood in principle as related to
the complex features of the theory that lie beyond
reach of perturbative approaches. Regarding these
complex features, palpable progress in calculating
hadronic masses is achieved using lattice
formulation of QCD \cite{Durr,Wilczek}. However,
so far it is not clear how to use QCD for
producing quark and gluon wave functions of
hadrons in the Minkowski space and thus provide
a picture of hadrons of comparable precision to
the picture of atoms achieved in QED.

In this situation, it is interesting to note that
there exists a formulation of QCD which uses
Dirac's light-front (LF) form of Hamiltonian
dynamics \cite{Dirac,DiracQuantum} and does not
introduce any complex vacuum structure
\cite{Wilsonetal}. How could then the LF
formulation of QCD produce the effects that in
other approaches are associated with the concept
of a complex vacuum state, including effects
associated with spontaneous symmetry breaking or
the gluon condensate? Ref. \cite{Wilsonetal}
points out that an effective Hamiltonian expected
to result from a renormalization group analysis of
the canonical LF Hamiltonian of QCD with all
counterterms that are necessary for stabilizing
the analysis, may contain new terms that provide
such effects. 

Here it is argued that the renormalization group
procedure for effective particles (RGPEP, see
Section \ref{effectiveparticles}) provides a new
way of thinking about light quarks in LF QCD. In
this new way, the harmonic oscillator potentials,
previously associated in Ref. \cite{GlazekSchaden}
with a gluon condensate in vacuum, can be
interpreted as coming from the gluon content of a
hadron rather than a vacuum. Previous reasoning
in Ref. \cite{GlazekSchaden} was developed using 
the instant form \cite{Dirac} of dynamics, i.e., the
standard Hamiltonian evolution in time. Now, the
possibility of considering harmonic oscillator
potentials for light quarks at distances
comparable with the characteristic size of
hadrons, meant here to be equivalent to the
distances on the order of $1/\Lambda_{QCD}$ in the
RGPEP scheme, is pointed out using a decomposition
of hadronic states into the LF Fock space components,
i.e., states obtained from empty vacuum using
creation operators defined in the LF quantization 
of fields rather than the standard canonical quantization
of the instant form. These Fock components contain various
numbers of virtual effective particles whose
interactions depend on the renormalization group
scale in the RGPEP scheme. The scale is denoted by
$\lambda$. One can think about $\lambda$ as having dimension 
of momentum. Low-energy features of light states in 
QCD made of light quarks are expected describable 
using an effective Hamiltonian that corresponds
to $\lambda$ comparable with $\Lambda_{QCD}$. 
Readers interested in details of RGPEP that are 
already known in the case of heavy quark dynamics, 
can consult Ref. \cite{GlazekMlynik}, where an
approximate QCD calculation of masses of heavy
quarkonia is described. Here only the case of light
quarks is discussed. 

According to Ref. \cite{BrodskyShrock}, if gluon
condensate effects are coming from the gluon
content of hadrons rather than vacuum\footnote{The
quark condensate has also been suggested to
originate in hadronic content instead of vacuum
\cite{BrodskyRobertsShrockTandy}.}, one can avoid
the problem of an excessively large vacuum energy
density in cosmology. However, the new reasoning
presented here does not deal with cosmological
issues. This article concerns only the particular
mechanism that was described in Ref.
\cite{GlazekSchaden}. In the mechanism in Ref.
\cite{GlazekSchaden}, the gluon condensate meant
to be in vacuum provided harmonic oscillator
potentials for constituent quarks in light mesons
and baryons. The potentials agreed with known
phenomenological CQMs~\cite{CQM1,CQM2,CQM3} when
the expectation value of the gluon field strength
squared was equated to the vacuum gluon condensate
value used in the QCD sum rules \cite{SVZ}. It is
argued below that it was not necessary to assume,
as it was done in Ref. \cite{GlazekSchaden}, that
the gluonic expectation value came from the empty
vacuum. The point of view developed here is that
the gluon expectation value might be coming from
the content of a hadron itself, and the relevant
content can in principle be identified in LF QCD
using RGPEP. Potentially broad implications of
such change in interpretation are not discussed
here. The hypothesis put forward in this article
requires a considerable amount of work to verify
and it would be premature to draw broad
conclusions before relevant RGPEP calculations are
completed.

The calculations described here arrive at similar 
harmonic potentials to those found in Ref. 
\cite{GlazekSchaden} but in a different way and with 
a significantly different interpretation. Namely, 
instead of starting from the non-relativistic (NR) 
Schr\"odinger equation and gluon vacuum
condensate, one starts from the LF QCD Hamiltonian
eigenvalue equation for mesons and baryons.
Canonical quark and gluon field operators are
formally transformed using RGPEP to effective
operators at the running cutoff scales that are
comparable with $\Lambda_{QCD}$. Using color gauge
invariance principle, together with a mean-field
approximation that still needs justification and is
not verified yet by a rigorous {\it ab initio}
calculation, one arrives at the eigenvalue
equations in which expectation values of effective
gluon fields are evaluated in a gluonic component
of hadronic states rather than in the vacuum. The
gluonic component is postulated to be universal.
Then, using new relative momentum variables, one
obtains LF wave functions that describe the
relative motion of hadronic effective constituents
in a boost-invariant way. In summary, the reasoning 
described here implies a relativistic picture of 
hadrons in which the same harmonic oscillator potentials 
that agree with CQMs phenomenology may result from the 
gluon content of hadrons in LF QCD rather than from the
vacuum per se.

It should be clarified that harmonic potentials
among effective constituents in lightest hadrons
apply in a situation where color charges are close
to each other. If they were separated by a
distance considerably larger than the size of a
light hadron, their invariant mass would increase
considerably due to the potential. Interactions
other than the potential term alone would be
activated, able to create additional particles and
changing the dynamics, perhaps including a string
of effective gluons. 

The article is organized in the following way.
Section \ref{effectiveparticles} describes the
concept of effective particle in the context of
RGPEP and explains why the vertex form factors
(denoted by $f_\lambda$) eliminate changes of the
number of massive virtual effective particles.
This feature is required for thinking that a
hadron can be represented by a convergent
expansion in the basis of the Fock space. The
basis that counts is not built in terms of
canonical quark and gluon operators but in terms
of the effective ones, corresponding to relatively
small scale parameter $\lambda$. The convergence
is not proven, but it is deemed not excluded given
the exponential falloff of the RGPEP form factors
$f_\lambda$ as functions of constituent invariant
masses. Section \ref{RGPEPstates} outlines the
representation of states of light hadrons in the
effective particle basis in the Fock space.
Section \ref{structureW} discusses a perturbative
expansion for the unitary transformation
$U_\lambda$ that connects current, or canonical
quarks and gluons with their CQM counterparts, and
for the transformations $W_{\lambda_1 \lambda_2}$
that connect effective particles that correspond
to different values of the scale $\lambda$. The
eigenvalue problem for light hadrons is considered
in Section \ref{eigenvalue}. This Section
describes our derivation of LF invariant mass
operators for quarks including their minimal
coupling to the gluons condensed in a hadron. Next
Section \ref{Phenomenology} shows how the RGPEP
can be applied in phenomenology of hadrons, with
emphasis on form factors, structure functions, and
connection with AdS/QCD ideas. Section \ref{c}
concludes the paper. Finally, four appendices are
provided in order to support the claim that RGPEP
can be used to verify our reinterpretation of the
gluon condensate in QCD. Appendix
\ref{RGPEPuniversalityappendix} outlines how CQMs
can be viewed as limited representatives of the
same universality class that QCD belongs to.
Appendix \ref{AppendixW} describes details
concerning perturbative calculations of RGPEP
transformations. Presentation of RGPEP beyond
perturbation theory is given in Appendix
\ref{NPRGPEP}. The last Appendix
\ref{OverlappingSwarmsModel} offers a
visualization of the RGPEP scale dependence of
hadronic structure. 

\section{ Effective particles and vertex form factors }
\label{effectiveparticles}

Reasoning described in the next sections requires
elements of RGPEP \cite{RGPEP}.\footnote{Appendix
\ref{RGPEPuniversalityappendix} describes RGPEP in
the context of asking how QCD can be transformed
into an effective theory that resembles CQMs as
approximate representatives of the same
universality class.} This section focuses on the
concept of effective particles and the role of
vertex form factors that RGPEP introduces in the
interactions of effective particles, starting from
QCD.

In QCD, RGPEP begins with the canonical LF
Hamiltonian in which the dynamically independent
quark and gluon quantum fields are expanded into
their Fourier components. These components are the
operators that create or annihilate single bare
quarks or gluons of definite momentum. For
brevity, creation and annihilation operators will
be commonly called particle operators, in order
distinguish them where necessary from quark and
gluon field operators. 

The canonical LF Hamiltonian contains singular
interaction terms. It must be regulated to avoid
infinities. This is done by introducing
regularization factors in interaction terms. These
factors are constructed to limit relative motion
of particles that participate in interactions. For
example, let the total momentum of all particles
in an interaction term have components $P^+$ and
$P^\perp$. The notation means: $P^+ = P^0 + P^z$,
$z$-axis is the direction distinguished in the
definition\footnote{LF is a hyper-plane in
space-time that is swept by a wave front of a
plane wave of light. In standard notation, the
frame of reference of the inertial observer who
develops a quantum theory is set up so that the
plane wave moves against the $z$-axis in the
chosen frame.} of the LF, $\perp$ denotes
transverse components, i.e., the components $x$
and $y$ that are transverse to the $z$-axis. Let a
selected particle have momentum components $p^+ =
xP^+$ and $p^\perp = xP^\perp + k^\perp$. Then,
the regulating factor for this particle that is
sufficient for taming logarithmic divergences in
QCD can be of the form $x^\delta
\exp{(-k^\perp\,^2 /\Delta^2)}$~\cite{gluons},
where $\delta \rightarrow 0$ and $\Delta
\rightarrow \infty$ are the small-$x$ and
ultraviolet regularization parameters,
respectively. Initially, one also introduces an
absolute infinitesimal lower bound $\epsilon^+ <
p^+ $ for all particle operators in order to
eliminate the LF zero modes. This step amounts,
however, to saying that all particles must have
positive momentum component $p^+$. This
requirement eliminates complex vacuum.
Subsequently, one demands that for every particle
in every interaction term $x > \epsilon$, with an
infinitesimal number $\epsilon$. The number
$\epsilon$ is a fraction of momentum. It is not
related to $\epsilon^+$. Then, when one works in
the limit $\epsilon \, x^{-\delta} \rightarrow 0$
point-wise in $x$ when $\epsilon$ is sent to 0,
the factor $x^\delta$ takes over the regulatory
function from $\epsilon$. As a result, the
variables $x$ range from 0 to 1 and there are
regulating factors $x^\delta$ and $(1-x)^\delta$
at the end points in all interaction vertices. The
variable $x = p^+/P^+_h$ for a particle of
momentum $p$ in a hadron of momentum $P_h$
coincides with the parton model longitudinal
momentum fraction in a hadron in the infinite
momentum frame (IMF). Variable $k^\perp$ is equal
to the transverse momentum of a parton in the IMF
when the parent hadron of the parton moves
precisely along $z$-axis. Note, however, that LF
QCD is formulated in terms of the same variables
$x$ and $k^\perp$ no matter how hadrons move,
i.e., the LF Hamiltonian theory is explicitly
invariant with respect to the 7 Poincar\'e
transformations that preserve the LF. This means
that the theory of a hadron structure in a rest
frame of the hadron has the same form as in the
IMF. 

When the regularization is being removed, $\delta
\rightarrow 0$ and $\Delta \rightarrow \infty$,
RGPEP establishes the required ultraviolet
counterterms \cite{RGPEP}. The small-$x$
regularization drops out from dynamics of
colorless states because they cannot produce
long-distance interactions along the LF.

Let the regularized LF Hamiltonian of QCD with
counterterms be denoted by $H$ and let $b$ denote
bare particle operators. RGPEP introduces effective 
particles of scale $\lambda$ through a unitary 
transformation,
\beq
\label{blambda}
b_\lambda \es U_\lambda \, b \, U_\lambda^\dagger \, . 
\eeq
The corresponding Hamiltonian operator,
\beq
\label{H}
H_\lambda(b_\lambda) \es H(b) \, ,
\eeq
is a combination of products of operators
$b_\lambda$ with coefficients $c_\lambda$ that are
different from coefficients $c$ of corresponding
products of operators $b$ in the canonical
Hamiltonian with counterterms, $H(b)$. Since
$\lambda$ is related to an upper limit on momentum
transfers in interactions, the operators $b$
correspond to $\lambda = \infty$. For the infinite
$\lambda$, we have $b_\infty = b$ and $H(b) =
H_\infty(b_\infty)$.\footnote{Where it is unlikely
to lead to a confusion, $H_\lambda(b_\lambda)$ is
abbreviated to $H_\lambda$. Later, also operators
such as $H_{\lambda_1} (b_{\lambda_2})$ occur,
meaning an operator that is a combination of
products of particle operators $b_{\lambda_2}$
with coefficients $c_{\lambda_1}$ instead of
$c_{\lambda_2}$.} 
 
RGPEP provides differential equations that produce
expressions for the coefficients $c_\lambda$ in
$H_\lambda$. The operator $U_\lambda$ is
calculable in RGPEP order-by-order in an effective
coupling constant \cite{RGPEP} in the form of
normal-ordered products of particle operators
$b_\lambda$. Appendix \ref{AppendixW} illustrates
how it is done in lowest orders. Appendix
\ref{NPRGPEP} shows how non-perturbative
calculations can be attempted. Appendices 
\ref{RGPEPuniversalityappendix}, \ref{AppendixW},
and \ref{NPRGPEP}, provide a formal background for 
the entire discussion that follows.

The key feature of $H_\lambda(b_\lambda)$ that
emerges from RGPEP is that all coefficients
$c_\lambda$ contain a form factor $f_\lambda$. The
form factor prevents changes of the total
invariant mass of effective particles undergoing
interaction by more than $\lambda$. The RGPEP
parameter $\lambda$ plays thus the role of a
momentum-space width of vertex form factors in all
interaction terms. In fact, RGPEP is designed to
work this way. When $\lambda$ is small, the form
factors suppress all interactions among massive
particles besides the terms one can call
potentials, i.e., the interaction terms that do
not change the number of massive particles (see
below). In other words, the RGPEP transformation
$U_\lambda$ is designed to identify the dynamical
relationship between nearly massless current
quarks and their field-theoretic interactions at
formally infinite $\lambda$ with massive
constituent quarks and their interactions through
potentials at $\lambda$ so small that potential
models may apply as an approximation to solutions
of the whole theory for states of smalles masses. 

If one so desires, transformation $U_\lambda$ can
be kinematically complemented with the Melosh
transformation \cite{Melosh}, in order to express
the spinors typically used for description of
constituent quarks in the constituents
frame\footnote{The CRF frame is an inertial
reference frame in which the total momentum of
constituents treated as free particles of definite
masses has only time component different from
zero. The concept of CRF in LF dynamics is
different from a similar concept in the instant
form of dynamics, because conservation of $P^+$ in
interactions makes the CRF differ from the
center-of-mass system (CMS) of a hadron in the LF
dynamics, while the CRF and CMS are the same in
the instant form of dynamics, where $P^z$ is
conserved by interactions. $P^\perp$ is preserved
in interactions in both forms of dynamics
equally.} (CRF) with the spinors one can
conveniently use in the IMF.\footnote{Ref.
\cite{GellMannQCD} is of interest here because the
last sentence in it suggests one might perhaps use
some collective coordinates as a satisfactory
method for truncating QCD, instead of ``the
brute-force lattice gauge theory approximation!''
While Ref. \cite{GellMannQCD} is quite new to the
author at the time of writing this article, it
should be noted that RGPEP can be viewed as an
attempt of this type. Namely, the effective quarks
and gluons, as constituents of size $\lambda^{-1}
\sim \Lambda_{QCD}^{-1}$, describe collective
modes in dynamics of many small quarks, anti-quarks, 
and gluons that nevertheless may be individually 
active in virtual processes characterized by large 
invariant-mass changes, allowed in a single 
interaction only when $\lambda \gg \Lambda_{QCD}$.} 

It follows from its definition that the form
factor $f_\lambda$ prevents the number of
effective particles from changing if their masses
$m_\lambda$ exceed $\lambda$. For example, suppose
that a virtual particle of mass $m_\lambda$ emits
another virtual particle with mass $m_\lambda$.
The change of invariant mass of particles in the
interaction is at least $m_\lambda$. The form
factor $f_\lambda$ is designed to quickly tend to
0 for invariant-mass changes greater than
$\lambda$. So, if $m_\lambda$ exceeds $\lambda$,
the form factor eliminates the emission. In
particular, if effective gluons are assumed to
have effective masses $m_\lambda$, the Hamiltonian
$H_\lambda$ will not include interactions that can
change the number of effective gluons when
$m_\lambda > \lambda$. A sizeable mass of
effective gluons in $H_\lambda$ is what one may
expect to happen at small values of $\lambda$ in
QCD because such effective gluon mass is a
candidate for explaining the absence of small
spacing in the spectrum of hadronic masses. There
should be small spacing in the presence of
massless gluons, as it happens in atoms described
by QED with massless photons. In contrast to atoms,
hadrons do not exhibit such small spacing.

Note that a smooth form factor $f_\lambda$ must
allow for some small, transitional range of values
of $\lambda < m_\lambda$. In the transitional
range, interactions that change the number of
effective gluons gradually disappear when
$\lambda$ is lowered below $m_\lambda$. In the
reversed RGPEP evolution in $\lambda$ from small 
to large values, gluons will gradually appear in 
the dynamics as $\lambda$ increases and $m_\lambda$
decreases so that at some point $\lambda$ becomes 
greater than $m_\lambda$.

\section{ RGPEP representation of states of light hadrons }
\label{RGPEPstates}

If QCD is to explain precisely the success of
classification of light hadrons in terms of
constituent quarks with masses on the order of 1/3
of a nucleon mass, a clearly defined procedure
must explain how the masses $m_u$ and $m_d$ of the
lightest quarks increase from their standard model
(SM) values of order 5 MeV in a local gauge theory
to their common constituent value of order
$\Lambda_{QCD}$ in a corresponding effective
theory.\footnote{ How RGPEP is able to relate a
local canonical theory to a non-local effective
theory is explained in \cite{NonlocalH}.} It is
not known yet if RGPEP leads to such increase of
effective masses of light quarks in QCD when the
parameter $\lambda$ is lowered toward
$\Lambda_{QCD}$, or even below $\Lambda_{QCD}$. We
assume here that RGPEP does lead to such
result.\footnote{Quark masses increase as momentum
scale is lowered in perturbation theory developed
in terms of the Feynman diagrams
\cite{pdgQCDmass}. Similar results are obtained
from Dyson-Schwinger equations \cite{Roberts}. LF
constituent models can incorporate running masses
\cite{GlazekNamyslowski}.} This assumption cannot
be rigorously verified yet because the domain of
$\lambda \sim \Lambda_{QCD}$ cannot be reached
using low orders of perturbation theory and
non-perturbative solutions to RGPEP equations are
still not known in QCD.

When $\lambda$ is large in comparison to
$\Lambda_{QCD}$, so that the effective coupling
$g_\lambda$ is small and perturbation theory
applies, the eigenvalue problem involves many Fock
sectors with many quarks of small Lagrangian
masses. In these circumstances, it is hard to
analyze the eigenvalue equations precisely. When
$\lambda$ is lowered, the number of necessary Fock
sectors is expected to decrease because of the
vertex form factors $f_\lambda$ but the coupling
constant $g_\lambda$ increases and it is hard to
evaluate $H_\lambda$ precisely. A way out of this
situation is to calculate $H_\lambda$ in terms of
successive approximations.\footnote{This includes
successive approximations for solutions of the
non-perturbative RGPEP equations in Appendix
\ref{NPRGPEP}.}

A workable RGPEP method for building successive
approximations has been outlined including results
for hadron masses and wave functions in the case 
of heavy quarkonia \cite{GlazekMlynik}, where the 
number of heavy quarks is nearly constant, equal 2. 
In that case, one can take advantage of the NR 
approximation for relative motion of quarks, based 
on smallness of the effective coupling constant 
$g_\lambda$ when $\lambda$ is comparable with heavy 
quark masses that are much larger than $\Lambda_{QCD}$. 

However, no counterpart for such approximation
scheme is available yet for light quarks in LF QCD
using RGPEP. The reason, stated already above, is
that, in distinction from the case of heavy
quarks, the light quark masses are so much smaller
than $\Lambda_{QCD}$ that many light quarks can a
priori be created or annihilated in the
interactions that limit virtual energy changes by
about $\Lambda_{QCD}$. Therefore, one needs to
approximate reasonably well a great deal of
relativistic dynamics of virtual particles that
are constantly created and annihilated by strong
interactions. It is suggested below that a
reinterpretation of the gluon condensate as a part
of a hadron provides a guideline for moving in that 
direction.
 
It is a conceptual jump that requires a connection
between constituent quarks and structure functions
(see Section \ref{structurefunctions}), but one
can observe here that parton distribution
functions at small $Q^2$ are typically very stable
as functions of $Q^2$ and quarks always carry only
about half of the nucleon momentum
\cite{pdMSR1994, pdMSR1995, pdGRV1995, pdGRV1998,
DurhamPDF}. The gluon component of hadrons, which
plays the role of a gluon condensate in the
picture described below, may only be responsible
for a part of the other half of hadron momentum.
Namely, the constituent quarks also contain
gluons. The issue of how heavy is the part of a
hadron that is made of effective quarks at $
\lambda \lesssim \Lambda_{QCD} $ and how heavy is
the corresponding gluon condensate component, is
discussed in Section \ref{Phenomenology} and
Appendix \ref{OverlappingSwarmsModel}. Firm
conclusions require RGPEP calculations that have
not been done yet.

In the picture described below, the masses of
effective quarks and gluons at some scale $\lambda
\lesssim \Lambda_{QCD}$ are assumed greater than 
$\lambda$ itself. All interactions that change the 
number of the effective particles are absent and 
the only interactions left in $H_\lambda$ with 
$\lambda$ near and below $\Lambda_{QCD}$ are 
potentials. These potentials are expected to match 
potentials used in the CQMs. Our reinterpretation 
of the gluon condensate is associated with a definite
result for the potentials that act among the
constituent quarks. In summary, the LF Hamiltonian 
of QCD with some $\lambda = \lambda_c \lesssim
\Lambda_{QCD}$ is assumed to describe quarks of
mass $m_c \gtrsim \lambda_c$. These quarks interact
only through potentials, i.e., interactions that do 
not change the number of constituents. The subscript
$c$ refers to the word constituent. 

An eigenstate $|\psi \rangle$ of the effective 
Hamiltonian $H_{\lambda_c} (b_{\lambda_c})$ that 
represents a hadron of momentum $P_h$, is a 
superposition of effective-particle basis states 
($n=2$ for mesons and $n=3$ for baryons)
\beq
| 1...n \rangle_{\lambda_c} \es \prod_{i=1}^n 
b^\dagger_{i \lambda_c} | 0 \rangle \, , \\
|\psi \rangle \es \sum_{1...n}
\psi_{\lambda_c}(1,...,n)| 1...n \rangle_{\lambda_c} \, .
\label{psi}
\eeq
The arguments 1 to $n$ of the wave function
provide a shorthand notation for momenta, spins,
flavors, and colors of the corresponding
particles. For example, 2 as an argument of a wave
function stands for the three-momentum, spin
projection on $z$-axis, flavor, and color of the
particle number 2. Summation over numbers 1 to $n$
is a shorthand notation for summation over the
quantum numbers of the corresponding particles
including integration over their momenta. In this
abbreviated notation, the eigenvalue equation for
a mass and a wave function of a hadron with momentum
$P_h^+$ and $P_h^\perp$ built from $n$ constituents 
of scale $\lambda_c$, takes the form
\beq
\label{eqpsi}
_{\lambda_c}\langle 1...n | (P_h^+H_{\lambda_c} - P_h^\perp\,^2) |
\psi \rangle
\es 
M^2 \,\, 
_{\lambda_c}\langle 1...n | \psi \rangle \, .
\eeq
This eigenvalue equation with $n=2$ or $n=3$ is 
meant to correspond to the CQM picture of hadrons 
as built from 2 or 3 constituent quarks.

Note that the vacuum state $|0\rangle$ is not 
changed when $\lambda$ changes. This is a unique
feature of LF dynamics. In the standard, instant 
form of dynamics, a change in particle operators 
must be accompanied with a change of the 
corresponding vacuum state.

One can write the same eigenvalue equation for the
same states using other values of $\lambda$ than
$\lambda_c$. The eigenstate $|\psi\rangle$ is the
same for all values of $\lambda$ but the basis
states and corresponding wave functions depend on
$\lambda$. For $\lambda \gg \Lambda_{QCD}$, the
state $|\psi \rangle$ may contain a giant number of
multi-particle Fock components built by acting with
creation operators $b^\dagger_\lambda$ on the
vacuum, while when $\lambda = \lambda_c$, the entire 
state contains only 2 or 3 constituent quarks. 

Using RGPEP, one can express constituent quarks at 
scale $\lambda_c$ in terms of effective quarks and 
gluons corresponding to $\lambda > \lambda_c$, 
\beq
\label{blq}
b_{\lambda_c} \es W \, b_\lambda \, W^\dagger \, , \\
\label{W}
W \es U_{\lambda_c}U_\lambda^\dagger \, .
\eeq
Momenta, spins, and izospins of the effective particles 
are the same, irrespective of the change in parameter 
$\lambda$. Since $W|0\rangle = W^\dagger |0\rangle = |0\rangle$,
one has
\beq
| 1...n \rangle_{\lambda_c} \es W \prod_{i=1}^n 
b^\dagger_{i \lambda} | 0 \rangle \, ,
\eeq
and the same eigenstate can be written as 
\beq
\label{psilambda}
|\psi \rangle \es 
\sum_{1...n}
\psi_{\lambda_c}(1,...,n) \, W \, | 1...n \rangle_\lambda \, . 
\eeq
The result of action of $W$ on the basis state
with $n$ constituents corresponding to scale
$\lambda$, with $\lambda > \lambda_c$, is a
coherent slew of Fock sectors with various
particle numbers, with the number of quarks 
grater or equal to the minimal constituent 
number $n$ for a hadron. The momentum
space wave functions of the resulting Fock
components are determined by the bound-state
eigenvalue condition of $H_\lambda$. 

Although a precise structure of $H_\lambda$ is beyond 
insight of perturbation theory for $\lambda \sim
\Lambda_{QCD}$ and $W$ at such scales also contains 
non-perturbative dynamics, some generic features
of the Fock wave functions in Eq. (\ref{psilambda}) 
can be assessed on the basis of generic features of 
$W$ implied by properties of operators $U_{\lambda_c}$ 
and $U_\lambda$ that are visible already in 
perturbation theory \cite{GlazekMaslowski}.
Operator $W$ conserves momentum. $W$ can replace
a single effective particle with a bunch of other
particles which carry together the same momentum 
as the replaced particle. $W$ can also annihilate 
a whole set of particles and create a new set with 
the same momentum. 

Hence, the colorless meson (M) and baryon (B) states 
$|\psi\rangle$ that correspond to the CQM picture 
(color factors are written explicitly),
\beq
\label{Mlambdaq}
|\psi \rangle_M 
\es 
\sum_{12} \psi_{\lambda_c}(1,2) 
\, {\delta^{ab}\over \sqrt{3}} \,
b_{1 \lambda_c}^{a \dagger}
d_{2 \lambda_c}^{b \dagger} 
|0\rangle \, , \\
\label{Blambdaq}
|\psi \rangle_B 
\es \sum_{123} \psi_{\lambda_c}(1,2,3) 
\, {\epsilon^{abc}\over \sqrt{6}} \,
b_{1 \lambda_c}^{a \dagger}
b_{2 \lambda_c}^{b \dagger} 
b_{3 \lambda_c}^{c \dagger} 
|0\rangle \, ,
\eeq
are equal to, respectively, 
\beq
\label{Mlambda}
|\psi \rangle_M 
\es 
\sum_{12} \psi_{\lambda_c}(1,2) 
\, {\delta^{ab}\over \sqrt{3}} \,
\, W \, 
b_{1 \lambda}^{a \dagger}
d_{2 \lambda}^{b \dagger} 
|0\rangle \, , \\
\label{Blambda}
|\psi \rangle_B 
\es 
\sum_{123} \psi_{\lambda_c}(1,2,3) 
\, {\epsilon^{abc}\over \sqrt{6}} \,
\, W \, 
b_{1 \lambda}^{a \dagger}
b_{2 \lambda}^{b \dagger} 
b_{3 \lambda}^{c \dagger} 
|0\rangle \, .
\eeq
The representation of meson and baryon states in
Eqs. (\ref{Mlambdaq}) and (\ref{Blambdaq}) in
terms of only 2 or 3 effective constituents at
scale $\lambda_c$, is thus transformed into the
representation of the same states in Eqs.
(\ref{Mlambda}) and (\ref{Blambda}) that are built
from effective particles of scale $\lambda$. The
resulting Fock space wave functions for effective
particles of size $\lambda^{-1}$ involve functions
$\psi_{\lambda_c}(1,2)$ and
$\psi_{\lambda_c}(1,2,3)$ through convolutions
that emerge from the sums in Eqs. (\ref{Mlambda})
and (\ref{Blambda}) and the structure of $W$. This
is how the complex Fock-space structure of QCD
hadrons is supposed to contain information about a
simple CQM picture.

\section{ Structure of $W$ }
\label{structureW}

In order to provide an illustration of how $W$ may act
on quark and gluon states, this Section explains 
the structure of $W$ obtained just in first-order 
perturbation theory in the effective coupling
constant $g_\lambda$. The perturbative expression
is obtained starting from Eqs. (\ref{blambda}) and
(\ref{differentialeq}), which produce together
\beq
{d \over d\lambda} {\cal H}_\lambda 
\es 
\left[ {\cal H}_\lambda , {\cal T}\right] \, , 
\eeq
where ${\cal T} = {\cal U}_\lambda^\dagger {d \over d\lambda} 
{\cal U}_\lambda$ and ${\cal U}_\lambda = U_\lambda(b_\infty)$. 
This means that the functional $\cF_\lambda$ in Eq.
(\ref{differentialeq}) is set to have the form 
\beq 
F_\lambda \left[ {\cal H}_\lambda \right] 
\es
\left[ {\cal H}_\lambda , {\cal T}\right]
\, ,
\eeq
and the key to RGPEP is the dependence of $\cal T$
on ${\cal H}_\lambda$ \cite{RGPEP}. This
dependence is designed keeping in mind that the
resulting Hamiltonian $H_\lambda(b_\lambda)$
should contain the form factor $f_\lambda$ in each and
every interaction term. Suitable notation for this
condition is provided by writing $H_\lambda =
f_\lambda G_\lambda$, where $H_\lambda$ contains
the form factor $f_\lambda$ in interaction
vertices while $G_\lambda$ does not. If an
operator $G_\lambda$ contains a product of
creation and annihilation operators $b_\lambda$
with a coefficient $c_\lambda$, the operator
$H_\lambda = f_\lambda G_\lambda$ contains exactly
the same product with coefficient $f_\lambda
c_\lambda$, where $f_\lambda$ depends on the
difference between the invariant mass squared of
particles annihilated by annihilation operators in
the product and the invariant mass squared of
particles created by creation operators in the
product. Both invariant masses are calculated 
using the particle kinematical momentum variables 
and eigenvalues of part $H_{0 \lambda} = G_{0
\lambda}$ of the Hamiltonian $H_\lambda$. 

Thus, using ${\cal H}_\lambda = H_\lambda(b)$
and omitting $\lambda$, one writes 
\beq 
{\cal H} \es f {\cal G} \rs {\cal G}_0 + f {\cal
G}_I \, .
\eeq
Differentiating $\cal H$ with respect to $\lambda$ 
one arrives at the equation that defines $\cal T$ 
in such a way that it must vanish when interactions 
vanish (this guarantees that $\cal T$ is expandable 
in powers of the coupling constant) and that its 
matrix elements between eigenstates of ${\cal G}_
{0\lambda}$ vanish when the differences between 
the corresponding eigenvalues vanish \cite{RGPEP},
\beq
\label{Tau}
{[} {\cal T}, {\cal G}_0 {]}  
\es 
\left[ (1-f){\cal G}_I \right]' \, .
\eeq
Prime denotes differentiation with respect
to $\lambda$. The solution is denoted by
\beq
{\cal T}
\es
\left\{ \left[ (1-f){\cal G}_I \right]'
\right\}_{{\cal G}_0} \, ,
\eeq
where the curly braces with subscript ${\cal G}_0$
indicate the energy difference (difference of 
eigenvalues of ${\cal G}_0$) in denominator that 
results from the commutator of $\cal T$ with 
${\cal G}_0$ on the left-hand side in Eq. (\ref{Tau}). 
Then $\cal U$ satisfies
\beq
\label{U}
{\cal U}'
\es
{\cal U}
\left\{ \left[ (1-f){\cal G}_I \right]'
\right\}_{{\cal G}_0} \, ,
\eeq
with condition ${\cal U}_\infty = 1$. Eq. (\ref{U})
can be expanded in powers of the coupling constant.

\subsection{ Perturbative expansion for $U_\lambda$ }
\label{PTforU}

Let us assume here that ${\cal G}_0$ does not
depend on the coupling constant to all orders of
perturbation theory. This means that perturbative
self-interaction terms are included in ${\cal
G}_I$ and ${\cal G}_0$ is actually independent of 
$\lambda$.\footnote{ \label{ConformalWindow} This way
of proceeding is not necessary, but it considerably 
simplifies the discussion that follows because there 
is no need to expand the form factor $f_\lambda$ and 
the energy denominators in powers of the coupling 
constant.} While this condition may seem quite 
restrictive, note that it allows for appearance of 
mass terms that are proportional to positive powers 
of $\Lambda_{QCD}$, since $\Lambda_{QCD}$ vanishes to 
all orders of perturbation theory. Writing expansions 
in the bare coupling constant,
\beq
{\cal U}_\lambda \es 1 + g u_{\lambda 1} 
+ g^2 u_{\lambda 2} + ... \, , \\
{\cal G}_{I\lambda} \es g {\cal G}_{I\lambda 1} +
g^2 {\cal G}_{I\lambda 2} + ...  \, , 
\eeq
one obtains equations (for convenience of notation, 
the prime is now put outside the curly braces,
which is justified because eigenvalues of ${\cal G}_0$ 
are independent of $\lambda$) 
\beq
\label{u1'}
u'_{\lambda 1}
\es
\left\{ (1-f) {\cal G}_{I\lambda 1} 
\right\}_{{\cal G}_0}' \, , \\
u'_{\lambda 2}
\es
u_{\lambda 1} \left\{ (1-f) {\cal G}_{I\lambda 1}
\right\}_{{\cal G}_0}'
+
\left\{ (1-f) {\cal G}_{I\lambda 2}
\right\}_{{\cal G}_0}' \, , 
\eeq
etc., with solutions of the form
\cite{GlazekMaslowski}
\beq
\label{u1}
u_{\lambda 1}
\es
\left\{ (1-f) {\cal G}_{I\lambda 1} 
\right\}_{{\cal G}_0} \, , \\
\label{u2}
u_{\lambda 2}
\es
{1 \over 2} 
u_{\lambda 1}^2  
+ 
{1 \over 2}
\int_\infty^\lambda ds \, [u_{s1}, u_{s1}']
+
\left\{ (1-f) {\cal G}_{I\lambda 2}
\right\}_{{\cal G}_0} \, ,
\eeq
etc. This expansion can be rewritten in terms of
the effective coupling $g_\lambda$ for the purpose
of eliminating ultraviolet divergences involved in
the definition of the bare coupling constant and
the corresponding counterterm. However, for $g =
g_\lambda + O(g_\lambda^3)$, one can simply
replace $g$ by $g_\lambda$ in the terms explicitly 
listed in Eqs. (\ref{u1}) and (\ref{u2}). In addition,
$U_\lambda$ is unitary by construction. Therefore,
$ {\cal U}_\lambda = U_\lambda (b_\infty) =
U_\lambda(b_\lambda)$ and one can freely replace
$b_\infty$ by $b_\lambda$ in these equations,
obtaining thus first two terms in expansion of
$U_\lambda$ in powers of $g_\lambda$.

\subsection{ Perturbative expansion for $W$ }
\label{PTforW}

Now consider $\lambda_c$ and $\lambda \ge \lambda_c$. 
Eq. (\ref{W}) implies $W(b_\infty) =
U_{\lambda_c}(b_\infty)
U^\dagger_\lambda(b_\infty)$, while in Eqs.
(\ref{Mlambda}) and (\ref{Blambda}) the operator
$W$ acts on states created from the vacuum by
operators $b_\lambda$. It is therefore convenient 
to use an equivalent expression 
\beq
W(b_\infty)
\es
 U^\dagger_\lambda(b_\lambda) \, 
U_{\lambda_c}(b_\lambda) \\
\es 
\left[ 1 + g_\lambda u^\dagger_{\lambda 1} + g_\lambda ^2
u^\dagger_{\lambda 2} + ... \right] 
\left[ 1 + g_\lambda u_{\lambda_c 1} + g_\lambda^2 u_{\lambda_c 2}
+ ... \right] \, , 
\eeq
where $b_\infty$ in every $u$ on the right-hand side is replaced 
by $b_\lambda$. Thus,  
\beq
\label{W12}
W
\es
1 + g_\lambda W_1 + g_\lambda^2 W_2 + ... \, , \\
\label{W1}
W_1 
\es 
\left\{ (f_\lambda - f_{\lambda_c} ) 
{\cal G}_{I\infty 1} \right\}_{{\cal G}_0} \, , \\
W_2 
\es
{1 \over 2} 
\left( u_{\lambda 1} - u_{\lambda_c 1} \right)^2  
+
{1 \over 2} \left[ u_{\lambda_c 1},  u_{\lambda 1}
\right]
+
{1 \over 2}
\int_{\lambda_c}^\lambda ds \, [u_{s1}, u_{s1}']
\np
\label{W2}
\left\{ (1-f_{\lambda_c}) {\cal G}_{I\lambda_c 2}
\right\}_{{\cal G}_0} 
-
\left\{ (1-f_\lambda) {\cal G}_{I\lambda 2}
\right\}_{{\cal G}_0} \, ,
\eeq
where $b_\infty$ is replaced by $b_\lambda$ 
everywhere on the right-hand sides. 

The perturbative expansion can be similarly
carried out to higher orders. The term of
focus here is $W_1$, as an illustration of
how $W$ acts on quark and gluon states. 
This illustration is valid when the coupling 
constant $g_\lambda$ is very small, which
means that $\lambda \gg \Lambda_{QCD}$ in the
RGPEP scheme. 

To some extent, $W_1$ is also useful as an
indicator of how $W$ acts on states when $\lambda$
approaches $\lambda_c \sim \Lambda_{QCD}$. In this
case, $g_\lambda$ is relatively large. Although
the size of contribution of terms containing $W_k$
with $k > 1$ in comparison to the size of
contribution of the term with $W_1$ for such
small $\lambda$ is not known, it is certain that
all these terms together contribute 0 when
$\lambda = \lambda_c$. They vanish no matter how
large is $g_\lambda$ because integrals in them
effectively range from $\lambda_c$ to $\lambda$
and thus vanish when $\lambda$ tends to
$\lambda_c$. On the other hand, the higher power
of $g_\lambda$ the larger number of integrals 
involved. All these integrals must effectively
range from $\lambda_c$ to $\lambda$, so that they
vanish when $\Delta \lambda = \lambda - \lambda_c$
tends to 0 as the appropriate power of $\Delta
\lambda$. The first term in the expansion in
$\Delta \lambda$ must be linear in $\Delta
\lambda$ and $W_1$ is such. 

Regarding the size of $g_\lambda$, the range of
$\lambda$ right above $\lambda_c$ is a region
where $g_\lambda$ in the RGPEP scheme in QCD may
be limited in size, instead of having large values
that one may expect on the basis of a
straightforward extrapolation of low-order
expressions obtained in the region of large
$\lambda$ \cite{alphac}. Such limitation of
$g_\lambda$ from above is certainly observed in
simple mathematical models that include asymptotic
freedom and bound states in a renormalization
group procedure for Hamiltonians \cite{AFandBS}.
Similar limitation of $g_\lambda$ is also
suggested in phenomenology of non-perturbative
running couplings based on AdS/QCD ideas and LF
holography \cite{gAdS}. New theoretical
information about the size of $g_\lambda$ at small
$\lambda$ is expected to follow from a
non-perturbative RGPEP equations described in
Appendix \ref{NPRGPEP}.

\subsection{ $W_1$ for quarks and gluons }
\label{W1inQCD}

According to Eq. (\ref{W1}), the first-order 
term in $W = 1 + g_\lambda W_1 + O(g_\lambda^2)$,
is given by 
\beq
\label{W1example}
W_1 
\es 
\left\{ (f_\lambda - f_{\lambda_c} ) 
{\cal H}_{I\infty 1} \right\}_{{\cal H}_0} \, , 
\eeq
where ${\cal H}_0$ is the term independent of the
bare coupling constant $g$ and ${\cal H}_{I\infty 1}$ 
is a regulated interaction term proportional to the 
first power of $g$ in the canonical LF QCD Hamiltonian, 
${\cal H}_\infty = {\cal H}_{0\infty} + g {\cal H}_
{I\infty 1} + O(g^2)$, derived using gauge $A^+=0$ from 
the Lagrangian for QCD,
$ {\cal L} = \bar \psi(i\hspace{-4pt}\not\!\!D - m)
\psi - {1\over 2}Tr F^{\mu\nu}F_{\mu\nu}$. Thus,
\beq
\label{calHinfty}
{\cal H}_\infty \es \int dx^- d^2 x^\perp 
\left[ h_{0\infty} + g h_{I\infty 1} +
O(g^2)\right] \, ,
\eeq
where the Hamiltonian density terms independent of $g$ are 
\beq
\label{hinfty0}
h_{0\infty}
\es  
{1\over 2} \bar \psi \gamma^+ {-\partial^{\perp \, 2} + 
  m^2 \over i\partial^+} \psi
\, - \,  
Tr A^\perp (\partial^\perp)^2 A^\perp \, ,
\eeq
and the interaction terms order $g$ are 
\beq
\label{hinfty1}
h_{I\infty 1} 
\es 
\bar \psi \hspace{-4pt}\not\!\!A \psi
+ 
2i Tr \partial_\alpha A_\beta [A^\alpha,A^\beta] \, ,
\eeq
with $SU(3)$ color notation $ A = A^a t^a$,
$[t^a,t^b] = i f^{abc} t^c$, $Tr(t^a t^b) =
{1\over 2} \delta^{ab}$. The calculation of $W_1$
starts with the above expressions and proceeds as
described in Appendix \ref{AppendixW}. 

Appendix \ref{AppendixW} also argues that $W$ for
$\lambda \gtrsim \lambda_c \sim \Lambda_{QCD}$
contains all the terms in ${\cal H}_\lambda$ that
change the number of particles. This means that
$W$ can create a coherent slew of the effective
Fock components at scale $\lambda$ in Eqs.
(\ref{Mlambda}) and (\ref{Blambda}) that are
implied by the 2 or 3 constituent quark wave
functions at $\lambda_c$ in Eqs. (\ref{Mlambdaq})
and (\ref{Blambdaq}). In turn, this means that the
result of action of $W$ is fully encoded through
the Hamiltonian in the wave functions
$\psi_{\lambda_c}(1,2)$ and
$\psi_{\lambda_c}(1,2,3)$ that correspond to the
CQM. These wave functions are solutions to the
eigenvalue problem for $H_{\lambda_c}$, which does
not change the number of constituents. But the
larger $\lambda$ the closer $H_\lambda$ to the
canonical LF Hamiltonian of QCD and the more
complex the states generated by $W$ in Eqs.
(\ref{Mlambda}) and (\ref{Blambda}).

\section{ Eigenvalue problem for light hadrons }
\label{eigenvalue}

According to previous sections, the same
eigenvalue problem for light hadrons can be
written in different ways when one uses different
Hamiltonian expressions that correspond to
different values of $\lambda$. Three values are
distinguished in further discussion: $\lambda =
\infty$, $\lambda \gtrsim \lambda_c$, and $\lambda
= \lambda_c$. The corresponding eigenvalue
problems are equivalent if RGPEP equations are
solved exactly. However, since exact solutions are
not available, one is forced to guess plausible
candidates for the Hamiltonian expressions at
different values of $\lambda$. 

At large $\lambda$, insight comes from the
perturbative expansion of RGPEP, using regulated
canonical LF QCD Hamiltonian with counterterms to
start with at $\lambda = \infty$ and taking
advantage of asymptotic freedom. At small
$\lambda$, one has to guess a first approximation.
Phenomenology suggests that some form of a
Hamiltonian for constituent quarks bound in a
potential well is a good candidate for a first
approximation to the QCD Hamiltonian with $\lambda
= \lambda_c$ for hadron states of smallest masses.

These two extreme regions of $\lambda$ need to be
connected to each other within the RGPEP framework
in terms of some interpolation.\footnote{A similar
reasoning applies also in the case of Yukawa theory,
which is not asymptotically free, when one
considers a limited range of scales and effective
coupling constant is sufficiently small from
theoretical point of view to use perturbative
RGPEP \cite{GlazekWieckowski}.} The size of
calculable corrections to any first approximation
constructed this way will eventually tell us how
far from a true RGPEP solution such first
approximation can be. The candidate for the first
approximation that is developed in this article
suggests that the concept of a gluon condensate
may be interpreted in a new way. 

\subsection{ Three ways of writing the eigenvalue problem }
\label{3ways}

The first of the three ways, which uses the
regulated canonical Hamiltonian for LF QCD with
all due counterterms, $H_\infty$, amounts to
formal writing of the eigenvalue problem in terms
of Fock states created by products of operators
$b^\dagger_\infty$, $d^\dagger_\infty$, and
$a^\dagger_\infty$ from the vacuum state
$|0\rangle$ in the limit of regularization being
removed,
\beq
\label{hadroninfty}
H_\infty |\psi \rangle 
\es 
P_h^- |\psi \rangle \, .
\eeq
In this equation, a reasonably accurate
description of the eigenstate $|\psi \rangle$
presumably requires a very large number of wave
functions for many significant bare-particle Fock
components. The number of such components is
expected to grow when the regularization is
lifted. Although the full structure of $|\psi
\rangle$ is not known precisely, one can assume
some model for one part of it, such as a Fock
sector with a smallest possible number of bare
particles in a meson or a baryon, and try to
determine another part, such as a component with
one additional gluon, using perturbation theory.
Such approach is useful in description of
exclusive processes that involve large momentum
transfers \cite{LepageBrodsky}.\footnote{A
different way from RGPEP to attack the eigenvalue
problem of a LF Hamiltonian for QCD head on is to
use discretized light-cone quantization (DLCQ) in
which one introduces a finite minimal unit of
$p^+$ momentum, $\epsilon^+$, which is the inverse
of the size of a periodicity box for fields as
functions of the position variable $x^-$ on the
LF. Since the total $P^+$ of an eigenstate of
$H_\infty$ is conserved, the unit $\epsilon^+$
naturally limits the number of bare particles in
an eigenstate from above by the ratio
$P^+/\epsilon^+$ \cite{PauliBrodsky1,
PauliBrodsky2, BrodskyPauliPinsky}. Additional
cutoffs need to be imposed in DLCQ in order to
limit transverse momenta of constituents that may
appear in an eigenvalue problem with some
simultaneously fixed eigenvalues of the total
momentum components $P^+$ and $P^\perp$.}

The second of the three ways is designed for the
opposite end of the scale for $\lambda$ in RGPEP,
i.e., when $\lambda = \lambda_c$. The same
Hamiltonian is expected to be expressible there 
in terms of operators for constituent quarks. In 
this case, one has the eigenvalue equation of the form 
\beq
\label{hadronlambdac}
H_{\lambda_c} |\psi \rangle 
\es 
P^-_h |\psi \rangle \, ,
\eeq
in which the state $| \psi \rangle $ of a single
light hadron is represented by Eq.
(\ref{Mlambdaq}) for mesons and Eq.
(\ref{Blambdaq}) for baryons. This means that the
light hadron eigenstates are described by only one
wave function, $\psi_c(1...n)$, with $n=2$ or
$n=3$. These wave functions appear in the Fock states
built using creation operators $b^\dagger_{\lambda_c}$ 
and $d^\dagger_{\lambda_c}$. There are no components
with effective gluons or additional quark-anti-quark 
pairs. In other words, the complexity of structure of 
light hadrons in QCD is encapsulated in the structure 
of constituent quarks. This means that the operators
$b^\dagger_{\lambda_c}$ and $d^\dagger_{\lambda_c}$ 
in Eqs. (\ref{Mlambdaq}) and (\ref{Blambdaq}) are 
complex combinations of products of operators 
$b^\dagger_\infty$, $d^\dagger_\infty$, $a^\dagger_\infty$ 
and their conjugates. 

Using Eqs. (\ref{eqpsi}), (\ref{Mlambdaq}) and
(\ref{Blambdaq}), the eigenvalue problem of Eq.
(\ref{hadronlambdac}) for the $n=2$ or $n=3$
constituents is obtained in the form 
\beq
\label{Vcn}
\left( \sum_{i = 1}^n p_i^- + V_{cn}/P_h^+ \right)
\,
{_{\lambda_c}\langle 1...n | \psi \rangle } \es P_h^- \, 
{_{\lambda_c}\langle 1...n | \psi \rangle } \, .
\eeq
The unknown element in this form is the
interaction potential term $V_{cn}$. In order to
derive the structure of $V_{cn}$ in mesons and
baryons, we shall employ the third way of writing
the same eigenvalue problem using RGPEP. The third
way suggests the reinterpretation of gluon
condensate that is the subject of this article.
When all gluons considered in the third way are
incorporated in the constituent quarks at
$\lambda_c$, one is left only with the potential
$V_{cn}$, see Appendix
\ref{OverlappingSwarmsModel}.

The third way of writing the same eigenvalue
problem uses $H_\lambda$ with $\lambda \gtrsim
\lambda_c$. The size of $\lambda$ is assumed to be
such that gluons are already somewhat active as
participants in the dynamics while the effective
quarks are still heavy enough for a NR analysis to
apply to their relative motion in a light hadron.
The NR approximation is thinkable for a slowly
moving hadron according to previous Sections,
since interactions in the effective Hamiltonian at
scale $\lambda \gtrsim \lambda_c$ contain form
factors. The effective particles of sizable masses
cannot change their momenta through interactions
by large amounts. Thus, an eigenvector
corresponding to a slowly moving light hadron can
be expected to be described by the wave functions
that have significant values only for small
momenta of the constituents.

The third picture at some scale $\lambda \gtrsim
\lambda_c$ is considered a candidate for a first
approximation to hadrons in RGPEP. The
corresponding Hamiltonian will be proposed below
using ideas motivated by gauge symmetry in its NR
form. Difference between this approximate NR form
and the interactions that can be systematically
calculated using RGPEP in LF QCD, is to be treated
as a perturbation.\footnote{This strategy is
similar to the one described in Ref.
\cite{Wilsonetal} with an artificial linear
potential. RGPEP introduces new elements: the
dynamical transformation from bare to effective
particles available at different scales,
construction of corresponding effective fields,
possibility of using NR approximation at $\lambda
\gtrsim \lambda_c$ for identifying interactions
through gauge symmetry as indicated later in the
text (instead of introducing an artificial
potential), extension beyond perturbation theory
(see Appendix \ref{NPRGPEP}), and invariance with
respect to 7 kinematical LF symmetries for the
resulting interaction terms and wave functions
\cite{NonlocalH}. As a result, the first
approximation potential is proposed to be not a
linear but a quadratic function of a relative
distance between color charges (see next
Sections). The quadratic potential is expected to
cause creation of additional particles when a
distance between two colored particles increases
and the net linear increase of energy with
distance must ultimately result from the energy of
particles created along a line that connects the
two widely separated color charges.} The
approximate picture at $\lambda \gtrsim \lambda_c$
is described in more detail in the next Sections.
Here we only suggest that RGPEP provides a scheme
that may be used in future calculations of
$V_{cn}$ in Eq. (\ref{Vcn}) at $\lambda_c$ by
evolving the third-way picture at $\lambda \gtrsim
\lambda_c$ down to $\lambda_c$ and taking
advantage of the RGPEP reinterpretation of the gluon
condensate that is described in the next Sections.

\subsection{ Light hadron states at $\lambda \gtrsim \lambda_c$ }
\label{states}

When $\lambda \gtrsim \lambda_c$, the meson and
baryon states are described by Eqs. (\ref{Mlambda}) 
and (\ref{Blambda}), respectively. In these equations, 
the operator $W$ acts on the Fock components with 
2 or 3 quarks of scale $\lambda$ and creates additional 
components. The meson state is simpler than the
baryon state and it will be discussed first. 

The meson state is
\beq
\label{mesonWork}
|\psi_M\rangle \es 
\sum_{12} \psi_{\lambda_c}(1,2) \,
\, W \, 
b_{1 \lambda}^\dagger
d_{2 \lambda}^\dagger 
|0\rangle \, .
\eeq
From Eqs. (\ref{W1exact}), (\ref{WY1ra}) and 
(\ref{WY1a}), it is clear that $W$ shares 
properties of the interaction Hamiltonian. 
For example, if $H_{\lambda I}$ changes a 
group of particles to another group, a similar 
change with additional factors results from 
action of $W$. So, out of 2 effective quarks 
in a meson (3 in a nucleon) more quarks and 
gluons are created. In order to use local gauge 
symmetry to propose the approximate form of 
the state that results from action of $W$, 
it is useful to represent the state of 
Eq. (\ref{mesonWork}) in position space. 
This is done using quantum fields built from 
effective particle operators at scale $\lambda$.

The effective quantum fields are constructed
using Eq. (\ref{quantumfieldpsi}) for quarks and
Eq. (\ref{quantumfieldA}) for gluons and replacing
operators $b$, $d$ and $a$ with $b_\lambda$,
$d_\lambda$ and $a_\lambda$, respectively. Thus,
the operator $\psi_\lambda(x)$ on the LF is built
in the same way from $b_\lambda$ and $d^\dagger_
\lambda$ as the canonical operator $\psi(x)$ is 
built in Eq. (\ref{quantumfieldpsi}) from the 
operators $b$ and $d^\dagger$ that are equal 
$b_\infty$ and $d^\dagger_\infty$, respectively; 
\beq
\label{quantumfieldpsilambda}
\psi_\lambda(x) 
\es 
\sum_{\sigma c f} \int [k] 
\left[ \chi_c u_{\lambda fk\sigma} b_{k\sigma c f \lambda } e^{-ikx} + 
\chi_c v_{\lambda fk\sigma} d^\dagger_{k\sigma c f \lambda } e^{ikx}
\right] \, .
\eeq
The spinors $u_{\lambda fk\sigma}$ and $v_{\lambda
fk\sigma}$ include corresponding vectors in flavor
space. Spinors might depend on $\lambda$ if the
effective quark masses in the field expansion are
allowed to depend on $\lambda$.\footnote{Note that
the independent field components, $\psi_+ = \Lambda^+
\psi$, $\Lambda^+ = (1/2)\gamma^0 \gamma^+$, are 
independent of the quark masses, and inclusion of 
effective masses in spinors is merely a way of 
useful notation for some effects of interactions.}
Similarly, operator $A^\mu_\lambda(x)$ on the LF
is built from $a_\lambda$ and $a^\dagger_\lambda$
as the canonical operator $A^\mu(x)$ is built in
Eq. (\ref{quantumfieldA}) from the operators $a$
and $a^\dagger$ that are equal $a_\infty$ and
$a^\dagger_\infty$, respectively. Gluon
polarization vectors do not depend on $\lambda$.
\beq
\label{quantumfieldAlambda}
A_\lambda^\mu (x)
\es 
\sum_{\sigma c} \int [k] \left[ t^c \varepsilon^\mu_{k\sigma} 
  a_{k\sigma c \lambda } e^{-ikx} + t^c \varepsilon^{\mu *}_{k\sigma} 
  a^\dagger_{k\sigma c \lambda } e^{ikx}\right] \, . 
\eeq
As a consequence, the dynamically independent
components, $\psi_\lambda = \Lambda^+
\psi_\lambda$ and $A^\perp_\lambda$ in $A^+=0$ gauge 
on the LF $x^+=0$ have the same commutation relations 
as in a canonical theory. 

With particle operators at $\lambda = \lambda_c$, an 
effective constituent quark field operator is
constructed in the same way. When this field is
used to describe a slowly moving meson, it is useful 
to write the field as
\beq
\psi_{\lambda_c}
\es
\left[ \begin{array}{c} U_{\lambda_c}(\vec x\,) \\ 
                        V_{\lambda_c}(\vec x\,) 
\end{array} \right] \, ,
\eeq
where the upper two-component field $U_{\lambda_c}$
annihilates quarks and the lower two-component 
$V_{\lambda_c}$ creates anti-quarks.\footnote{The
three-vector notation $\vec x$ or $\vec k$ refers 
here to the LF co-ordinates $(x^-, x^\perp)$ in position 
space. A similar notation is sometimes also used for
momentum variables $(k^+,k^\perp)$. However, when we 
proceed later to the Schr\"odinger eigenvalue problem 
for effective Hamiltonians for light hadrons, the same 
notation will be adopted also for three-vectors built 
from $\perp$ and $+$ or $-$ components in such a way
that the standard three-dimensional notation respecting 
rotational symmetry will be natural.}

Thus, by inverting the Fourier transforms through 
integration over the LF hyperplane and using 
conventions explained in Appendix \ref{AppendixW}, 
the state in Eq. (\ref{Mlambdaq}) for a meson of 
momentum $P_h$ can be written as
\beq
\label{mesonpositionc}
|\psi \rangle_M
\es
{1 \over \sqrt{3}}
\int { d^3x_1 \, d^3x_2 \over 4m^2}
\int[12] \,
16\pi^3 P_h^+ \delta^3(P_h-k_1-k_2) \,
\nt
 e^{-i (k_1 x_1 + k_2 x_2)} \,
U^\dagger_{\lambda_c}(x_1) \, 
\psi_{2\times 2}(\vec k_{12}) \, 
V_{\lambda_c}(x_2)
|0\rangle  \, ,
\eeq
where $\psi_{2\times 2}(\vec k_{12})$ denotes the
$2\times 2$ matrix wave function of relative
motion of the quarks. The three-vector $\vec
k_{12}$ can be defined as a relative momentum of the
quarks in the CRF. Details of the definition of 
$\vec k_{12}$ are not essential at this point
but later discussion will include relevant details.

In order to obtain analogous position representation 
of the meson state in Eqs. (\ref{Mlambda}) or 
(\ref{mesonWork}), one needs to apply $W$ to the 
expression in Eq. (\ref{mesonpositionc}). The result is
\beq
\label{meson71}
|\psi \rangle_M
\es
{1 \over \sqrt{3}}
\int { d^3x_1 \, d^3x_2 \over 4m^2}
\int[12] \,
16\pi^3 P_h^+ \delta^3(P_h-k_1-k_2) \,
\nt 
e^{-i (k_1 x_1 + k_2 x_2)} \,
W \,  U^\dagger_\lambda(x_1) \, 
\psi_{2\times 2}(\vec k_{12}) \, 
V_\lambda(x_2)
|0\rangle  \, .
\eeq
A similar expression is generated in terms of
three quark fields and $W$ for baryons.
Integration over momenta in these expressions
generates position wave functions as coefficients
in the expansion of meson or baryon states into
basis states that are created from the LF vacuum 
by action of a product of two or three quark 
fields at scale $\lambda$ and $W$. 

Continuing with the meson states, the central 
hypothesis about action of $W$ is that gauge 
symmetry forces the result of action of $W$ 
on a state of two quarks at scale $\lambda$ to 
have the form 
\beq
&& W \, \int d^3x_1 d^3x_2 \, U^\dagger_\lambda(x_1) \, 
\psi(x_1,x_2) \, 
V_\lambda(x_2)
|0\rangle  \\
\es
\int d^3x_1 d^3x_2 d^3 x_0 \,
U^\dagger_\lambda(x_1) \, W(x_1, x_2, x_0) \,
 V_\lambda(x_2) |0\rangle \, , 
\nn
\eeq
where the operator $W(x_1, x_2, x_0)$ is responsible 
for creating components generated by $W$. These
components can be of two kinds. One kind is formed 
by operators that carry the color of initial two
quarks. For example, if $W$ generates a gluon from
a quark, the generated gluon and the emerging quark 
together carry the color of the initial quark. 
The other kind is formed by colorless operators.
For example, $W$ may cause emission of two gluons
by a quark and the gluons may form a color
singlet. 

These two kinds of contributions are encapsulated 
in $W(x_1, x_2, x_0)$ by writing
\beq
\label{mesonW}
W(x_1, x_2, x_0)
\es
\psi(x_1,x_2,x_0) \, T(x_1, x_2) \, G^\dagger(x_0) \, ,
\eeq
where $\psi(x_1,x_2,x_0)$ denotes a new wave
function at scale $\lambda \gtrsim \lambda_c$,
$G^\dagger(x_0)$ denotes the colorless component
of the state, and 
\beq
\label{mesonT}
T(x_1, x_2) \es 
P \exp \left[-ig \int_{x_2}^{x_1} dx^\mu A_\mu(x)
\right]\, ,
\eeq
is the color-transport factor along a straight
line between quarks that maintains local gauge 
symmetry by bringing in required gluon 
fields.\footnote{In the LF gauge $A^+=0$, one 
may expect only transverse separation between
quarks to count. However, the dependent (constrained)
components of fields, $\psi_-$ and $A^-$, contribute
to the effective interactions in a non-trivial
way and one has to keep in mind that the complete
effective theory at small $\lambda$ should have 
full rotational symmetry restored. Therefore, a 
complete RGPEP expression for $W$ must account 
for the dynamics along $x_1^- -x_2^-$ as well as 
along $x_1^\perp - x_2^\perp$. This issue will be 
addressed later by introducing effective interactions 
that respect rotational symmetry inside slowly 
moving hadrons through a new definition of
three-dimensional relative momenta of constituents
to which the minimal gauge coupling rule can
be applied in a rotationally symmetric way.} 
This factor is constructed in analogy with Ref.
\cite{GlazekSchaden}.

The operator that generates the colorless 
component of a hadron state, $G^\dagger(x_0)$,
will provide the contribution in dynamics of
quarks that is associated with gluon condensation
in hadrons, rather than in a vacuum. Namely,
instead of the vacuum expectation values of
operators considered in Ref. \cite{GlazekSchaden},
such as $\langle \Omega | A^i A^j | \Omega
\rangle$ with $i,j = 1,2$, the Schr\"odinger 
equation for the wave function $\psi(x_1,x_2,x_0)$ 
will involve expectation values 
\beq
\label{expectation}
\langle A^i A^j \rangle_G 
\es
{ \langle G | A^i A^j | G \rangle \over 
  \langle G | G\rangle } \, , \\
|G\rangle \es G^\dagger(x_0) |0\rangle \, .
\eeq

A similar reasoning is followed regarding baryons. 
In the case of baryons, one has to deal with three
color-transport factors that are constructed in 
analogy to Ref. \cite{GlazekSchaden}. The operator
$G^\dagger$ in baryons generates the state 
$G^\dagger|0\rangle$ that plays the same role that 
the vacuum state $|\Omega \rangle$ played in Ref.
\cite{GlazekSchaden}. It is assumed that the 
colorless components of mesons and baryons are 
approximately the same. The issue of universality 
of expectation values such as in Eq. (\ref{expectation})
for mesons will be further discussed below when
we come to the construction of the effective
Hamiltonian.

In summary, the claim of gauge symmetry regarding 
the colorless basis states of quarks and gluons 
at scale $\lambda$ from which mesons and baryons 
are made, is that they are of the form (all fields 
are effective at scale $\lambda \gtrsim \lambda_c$) 
\beq
\label{Mnewbasisx0}
|\vec x_1, \vec x_2, \vec x_G \rangle
\es 
\sum_{ab} {\delta^{ab} \over \sqrt{3}} \,  
(u_{1\lambda}^\dagger T_1)^a \, 
(T_2^\dagger v_{2\lambda})^b \,
G^\dagger |0\rangle \, , \\
\label{Bnewbasisx0}
|\vec x_1, \vec x_2, \vec x_3, \vec x_G \rangle
\es 
\sum_{abc} {\epsilon^{abc} \over \sqrt{6} } \,
(u_1^\dagger T_1)^a\, 
(u_2^\dagger T_2)^b\, 
(u_3^\dagger T_3)^c\, 
G^\dagger |0\rangle \, ,
\eeq
where
\beq
\label{generalT}
T_i 
\es
e^{-ig \int_{\underline{x}}^{x_i} dx_\mu A^\mu } \, , 
\eeq
and $\underline{x} = (\vec x_1 + \vec x_2)/2$ in
a meson and $\underline{x} = (\vec x_1 + \vec x_2
+ \vec x_3)/3$ in a baryon. The factor $T$ is 
defined here along a straight path in such a way 
that it reproduces the color-transport factor in 
mesons in Eq. (\ref{mesonT}). The path dependence 
is ignored as an unnecessary complication at the
level of mean-field approximation. The operator 
$G$ with argument $\vec x_G$ represents the white
energy density background called glue. It is 
assumed to be a scalar boson field with
corresponding commutation relations. 

There is a trouble on the LF with non-locality of
the boson commutation relations in the direction
of $x^-$, which requires a solution. However, the
effective dynamics that will be constructed in
next Sections will explicitly circumvent this
difficulty by using an operator $G$ that is a
function of a three-dimensional position space
variables associated with slowly moving hadronic
constituents in a slowly moving hadron. Thus, the
problem ultimately requiring a solution is not so
much the non-locality of a scalar field but how LF
QCD can generate a rotationally symmetric
effective theory for light hadrons. The
construction offered in next Sections proposes to
treat $G$ as a field depending on a suitable
three-dimensional position variables in which it
can be local in a way that respects rotational
symmetry.

The quanta of $G$ are meant to represent
excitations of the states of gluons condensed
inside hadrons. The quanta of $G$ are not
point-like. Their size is characterized by
$1/\lambda$ and can be considered roughly on the
order of $1/\Lambda_{QCD}$, i.e., the quantum
extends over the volume of an entire
hadron.\footnote{ The effective glue degree of
freedom represents contributions of all white Fock
sectors of effective particles that share the
hadron momentum to a varying degree as $\lambda$
varies.} There exists a possibility to associate
vector or even higher spin nature with $G$,
contributing to the hadron spin, but there is no
need to do so here.

The basis states are orthonormal in the sense that, for 
all quarks being different, 
\beq
\langle 
\vec x_1, \vec x_2, \vec x_G | 
\vec x_{1'}, \vec x_{2'}, \vec x_{G'} 
\rangle
\es
\delta_{11'} \delta_{22'} \delta_{GG'} \, , \\
\langle 
\vec x_1, \vec x_2, \vec x_3, \vec x_G | 
\vec x_{1'}, \vec x_{2'}, \vec x_{3'}, \vec x_{G'} 
\rangle
\es
\delta_{11'} \delta_{22'} \delta_{33'} \delta_{GG'} \, ,
\eeq
and $\delta_{kk'}$ includes $\delta^3(\vec x_k -
\vec x_{k'})$.\footnote{With the qualification
that locality in the $x^-$ direction requires
proper definition of $z$-components in a
rotationally invariant effective theory to be
discussed in next Sections.} If quark quantum
numbers besides color are not different (this 
comment concerns only the baryon case), one 
has to adjust the normalization of basis states 
by including all permutations that contribute 
to the scalar products. 

In the abbreviated notation used in the remaining 
part of the article,
\beq
|\vec x_1, \vec x_2, \vec x_G \rangle
\es 
|12G\rangle \, , \\
\label{Bnewbasisx}
|\vec x_1, \vec x_2, \vec x_3, \vec x_G \rangle
\es 
|123G\rangle \, .
\eeq
In these abbreviated notation, the 
meson and baryon states read:
\beq 
\label{mesonab}
|\psi \rangle_M
\es
\sum_{12G} \psi(12G) \, 
|12G\rangle \, , \\
\label{baryonab}
|\psi \rangle_B
\es
\sum_{123G} \psi(123G) \, 
|123G\rangle
 \, .
\eeq

\subsection{ Preliminaries concerning Hamiltonian 
             density at $\lambda \gtrsim \lambda_c$ }
\label{dynamicsl}

Using the concept of effective quark and gluon 
fields, one can propose that the LF Hamiltonian 
at $\lambda \gtrsim \lambda_c $ be expressible,
by analogy to the canonical theory, in terms of 
an integral over the LF of a density that is a 
function of the fields. Thus, 
\beq
\label{Hlambda}
H_\lambda \es \int_{LF} \, \left( h_{0\lambda} + 
 \, h_{I\lambda} \right)
\, ,
\eeq
where the density $h_{0\lambda}$ is bilinear in 
the fields and renders single-particle operators 
while the density $h_{I\lambda}$ describes 
interactions in terms of products of at least 
three fields (the interactions involve 
at least three effective particles). 

Using power-counting \cite{Wilsonetal}, the 
bilinear density at scale $\lambda$ can
be assumed in the form (all fields at $x^+=0$)
\beq
\label{hlambda0}
h_{0\lambda}
\es  
{1\over 2} \bar \psi_\lambda \gamma^+ 
{-\partial^{\perp \, 2} + m^2 
\over i\partial^+} \psi_\lambda \, + \,  
Tr A^\perp_\lambda (- \partial^\perp \,^2 
+ m_g^2) A^\perp_\lambda + CT_{I\lambda}\, ,
\eeq
where the masses correspond to $\lambda$. The
symbol $CT_{I\lambda}$ denotes mass-like terms
needed to make the mass parameters $m$ and $m_g$
to count as masses in the eigenvalue equations
for light mesons and baryons. In other words,
$CT_{I\lambda}$ are by construction cancelled by
self-interactions in light colorless states.\footnote{
There is no need to specify these terms further
here. Such terms can be identified explicitly in
perturbation theory, e.g., proceeding in a similar 
way to how it is done in the eigenvalue problem 
for heavy quarkonia in Ref. \cite{GlazekMlynik}.}

The interaction density $h_{I\lambda}$ in Eq.
(\ref{Hlambda}) must be non-local in the sense
that it involves products of fields at different
points.\footnote{The non-locality discussed here
is due to the RGPEP vertex form factors of width
$\lambda$ in momentum space. The RGPEP-induced
non-locality must not be confused with the
canonical non-locality of LF Hamiltonians in which
the dependent parts of fields, such as $\psi_- =
\Lambda_- \psi$ and $A^-$, involve inverse powers
of $i\partial^+$, which is a non-local integral
operator.} In particular, the minimal coupling 
between quarks and gluons, which in the
canonical gauge theory is of the form 
\beq
H_{MC\infty} 
\es 
\int d^3x \, h_{MC\infty}(x) \, , 
\eeq
with
\beq
\label{hiinfty}
h_{MC\infty}(x) 
\es
g \, \bar \psi(x) \not\!\!A(x) \, \psi(x) \, ,
\eeq
in the effective theory must take a non-local form 
that in a lowest-order approximation is of the
type \cite{NonlocalH}
\beq
\label{HpsiApsi}
H_{MC\lambda} 
\es 
\int d^3x_1 \, d^3 x_2 \, d^3 x_3 \,
h_{MC\lambda}(x_1, x_2, x_3)
\, ,
\eeq
with
\beq
\label{hpsiApsi}
h_{MC\lambda} 
(x_1, x_2, x_3)
\es g_\lambda f_\lambda (x_2-x_1, x_3-x_1) \,
\bar \psi_\lambda(x_1) \, \hspace{-4pt}\not\!\!A_\lambda(x_2) \,
\psi_\lambda(x_3)  \, .
\eeq
This non-local interaction term appears with 
other non-local terms in the Hamiltonian 
obtained from RGPEP at scale $\lambda$. 
The dependent and independent components of 
the fields are grouped in Eq. (\ref{hpsiApsi})
according to a free theory. The dependent 
components involve the inverse of $i\partial^+$, 
which is a non-local operator. This non-locality 
is of the same type as in a canonical theory and 
the RGPEP non-locality appears here on top of 
the canonical one. 

The non-local interaction terms with small $\lambda$ 
can be simplified considerably if the domain of 
action of the Hamiltonian is restricted to slowly 
moving effective quarks. Namely, consider 
again the case of Eqs. (\ref{HpsiApsi}) and (\ref{hpsiApsi}) 
and introduce a gradient expansion of the form 
\beq
h_{MC\lambda} 
(x_1, x_2, x_3)
\es 
g_\lambda f_\lambda (x_2-x_1, x_3-x_1) \, \bar \psi_\lambda(x_1) \, 
\nt
\left[ \not\!\!A_\lambda(x_1) + 
\partial \hspace{-1pt}\not\!\!A_\lambda(x_1)(x_2-x_1) + ...\right]
\nt
\left[ \psi_\lambda(x_1) + \partial\psi_\lambda(x_1) (x_3-x_1) +
... \right] \, .
\eeq
The three dots indicate terms with higher
derivatives. If all effective particles move 
slowly and the derivative terms are small, 
the Hamiltonian can be approximated by the 
first term in the expansion, 
\beq
H_{I\lambda} 
\es 
\int d^3x_1 \, d^3 x_2 \, d^3 x_3 \, h_{I\lambda} \\
& \sim &
g_\lambda \int d^3x \,
\bar \psi_\lambda(x) \not\!\!A_\lambda(x) \, \psi_\lambda(x) \, 
\int d^3y \, d^3 z \, f_\lambda (y, z) + ...\, .
\eeq
This means that the non-local effective interaction 
in a slowly moving, NR system still looks like a local 
one except that its strength is determined not solely 
by the coupling constant $g_\lambda$ but also by the 
integral of the non-local form factor on the LF hyperplane,
\beq
\tilde g_\lambda 
\es 
g_\lambda 
\int d^3y \, d^3 z \, f_\lambda (y, z) \, .
\eeq 

The point is that the effective Hamiltonian
density at $\lambda \gtrsim \lambda_c$ may partly
resemble a Hamiltonian of local gauge theory in
its terms that couple quarks with gluons even though
the actual effective interactions are non-local,
provided that one limits the domain of the
Hamiltonian to slowly moving hadrons. In this
case, construction of approximate candidates for
the LF Hamiltonian density of coupling between
quarks and gluons, before they are corrected using
RGPEP, may proceed in analogy to QED
\cite{BjorkenKogutSoper}. This means that one uses
the gauge $A^+=0$. The derivative
$i\partial^\perp$ in the quark kinetic energy in
Eq. (\ref{hlambda0}) is supplied with the minimal
coupling addition of $g A^\perp$ (the product 
of derivatives is separated by the inverse of 
$i\partial^+$, which is not altered). In addition, 
one includes terms dictated by constraints that imply
the result of Eq. (\ref{hiinfty}) for the
quark-gluon coupling. 

Much less is understood about the gluon part of
the effective Hamiltonian density for light
hadrons at $\lambda \gtrsim \lambda_c$. The lack
of understanding of the Hamiltonian is reflected
in the lack of understanding of the gluon
components in its eigenstates. One would have to 
calculate the operator $H_\lambda$ precisely in order to 
uncover the information it contains about how to 
build a model approximating light hadrons in LF QCD. 
While RGPEP provides tools for such calculations, 
the remaining part of the paper is only devoted to 
deriving the model that can be treated as a first 
approximation. 

A simple candidate for an approximate Hamiltonian
for light hadrons in LF QCD is constructed in the
next Section assuming that: (1) the result of
action of $W$ in RGPEP can be represented by
inclusion of the color-transport factors $T$ that
maintain local gauge symmetry of hadronic states,
(2) the operator $G$ represents the condensation
of gluons inside hadrons, (3) a mean field
approximation can be applied to the gluon field
operator $A$ in the minimal coupling of effective
quarks to the gluons condensed in a hadron, and
(4) the effects of quark and gluon binding that 
are not treatable in the mean field approximation 
can be included by an ad hoc Gaussian approximation
to the wave function of relative motion of
constituent quarks with respect to the condensed
gluons.

\subsection{ Mass squared with minimal coupling
             at $\lambda \gtrsim \lambda_c$ }
\label{dynamics2}

The leading idea is that the approximate LF Hamiltonian 
for 2 or 3 quarks at scale $\lambda \gtrsim \lambda_c$
in light hadrons should have a form compatible with 
minimal coupling between quarks and gluons and it 
should respect Poincar\'e symmetry. Realization of 
the idea employs the four assumptions listed at
the end of the previous Section in the following
way.

The LF Hamiltonian eigenvalue problem at $\lambda 
\gtrsim \lambda_c$ for a light hadron built from 
2 or 3 quarks and condensed gluons,
\beq
H_\lambda |\psi\rangle \es {M^2 + P_h^{\perp \, 2} \over
P^+_h} |\psi\rangle \, ,
\eeq
can be written in terms of the eigenvalue equation 
for the effective invariant mass operator of the 
hadron, $\cM^2_\lambda$, in the form
\beq
\cM^2_\lambda |\psi\rangle \es M^2 |\psi\rangle \, .
\eeq
One can think about the operator on the left-hand side 
of this equation in terms of the invariant mass squared 
of free particles plus  interaction terms. 

For a free quark of mass $m_1$ and a free anti-quark of 
mass $m_2$, one has
\beq
\cM^2_{free} \es 
\left( \sqrt{m_1^2 + \vec p_1^{\,2}}
     + \sqrt{m_2^2 + \vec p_2^{\,2}} \right)^2
- (\vec p_1 + \vec p_2)^2 \, .
\eeq
For slowly moving particles, neglecting terms smaller than
the ones that are quadratic in momenta, the NR approximation 
renders
\beq
\cM^2_{free} 
\es 
( m_1+ m_2)^2 + {m_1+ m_2 \over m_1} \, \vec p_1^{\,2}
             + {m_1+ m_2 \over m_2} \, \vec p_2^{\,2}   
- \vec P^{\,2}_{q \bar q} 
\, ,
\eeq
where $\vec P$ denotes the total momentum of the
two quarks. Using 
\beq
\beta_1 \es {m_1 \over m_1 + m_2} \, , \\
\beta_2 \es {m_2 \over m_1 + m_2} \, , 
\eeq
one has
\beq
\cM^2_{free} 
\es 
( m_1+ m_2)^2 + {1 \over \beta_1} \, \vec p_1^{\,2}
              + {1 \over \beta_2} \, \vec p_2^{\,2} - \vec P^{\,2}_{q \bar q}  \, .
\eeq
According to the gauge rule of minimal coupling,
the interaction of quarks with the condensed
gluons should be described by 
\beq
\label{minimalM}
\cM^2_{minimal} 
\es 
( m_1+ m_2)^2 + { \left( \vec p_1 - g\vec A_1 \right)^2  \over \beta_1} \,
              + { \left( \vec p_2 - g\vec A_2 \right)^2  \over \beta_2} \,
- \vec P^{\,2}_{q \bar q}  \, ,
\eeq
where $\vec A_i$ is an abbreviation for $\vec
A(x_i)$. Analogous reasoning in the case of 
baryons yields
\beq
\label{minimalB}
\cM_{minimal}^2 
\es 
\left( \sum_{i=1}^3 m_i \right)^2 
+
\sum_{j=1}^3 { \left( \vec p_j - g \vec A_j\right)^2 \over \beta_i} 
- \vec P^{\,2}_{3q} \, ,
\eeq
where $\beta_i = m_i/(m_1 + m_2 + m_3)$.

For all quarks having the same mass $m$,
$\beta_M = \beta_1 = \beta_2 = 1/2$ in mesons and 
$\beta_B = \beta_1 = \beta_2 = \beta_3 = 1/3$ in baryons.
Assuming these simplifications, the minimal coupling 
terms in Eqs. (\ref{minimalM}) and (\ref{minimalB}) 
can be interpreted following Ref. \cite{GlazekSchaden}
in the context of standard Hamiltonian dynamics as 
resulting from the Hamiltonian operator with a proper 
$\beta$, i.e., $\beta_M$ in mesons and
$\beta_B$ in baryons,
\beq
\label{Hmin}
H_{min} \es {1 \over \beta} 
\int d^3 x \, : \bar \psi(\vec x \,)
[ -i \vec \nabla_x - g \vec A(\vec x\,) ]^2 \psi(\vec x\,) :
\, .
\eeq

\subsection{ Reinterpretation of the gluon condensate }
\label{dynamics3}

Using Eqs. (\ref{mesonab}), (\ref{baryonab}),
and (\ref{Hmin}), one can consider the eigenvalue 
problems
\beq 
H_{min} \sum_{12G}  \psi(12G)  \, |12G\rangle 
\es
M^2_{q \bar q}  \, \sum_{12G}  \psi(12G)  \, |12G\rangle  \, , \\
H_{min} \sum_{123G} \psi(123G) \, |123G\rangle
\es
M^2_{3q}  \, \sum_{123G} \psi(123G) \, |123G\rangle \, ,
\eeq
for quark subsystems in mesons and baryons and
project them on the corresponding basis states. 
The calculation proceeds in a similar way to the 
one in Ref. \cite{GlazekSchaden}, with three exceptions. 

The first difference is that the eigenvalues
$M^2_{q \bar q}$ and $M^2_{3q}$ refer only to the 
$q\bar q$ subsystem in mesons, instead of the entire 
meson mass squared, and to the $3q$ subsystem in a 
baryon, instead of the entire baryon mass squared.
Note also that the eigenvalue equations considered 
in Ref. \cite{GlazekSchaden} were for usual energies 
while we now consider the invariant mass squared 
operators that include the quark free invariant mass
squared and minimal coupling interactions. 
The second difference is that the vacuum state 
$|\Omega \rangle$ is replaced by the state of 
gluons condensed in a hadron, 
\beq
|G\rangle \es G^\dagger(\vec
x_G)|0\rangle \, . 
\eeq
The vacuum state $|0\rangle$ does not develop any condensate.
The third difference is that in the mean-field
approximation for the gluon field,\footnote{This 
approximation replaces the Schwinger gauge formula
for the background gluon field that reproduces 
QCD sum rules for heavy quarkonia in the LF 
Hamiltonian formulation of the theory 
\cite{SDGcondensates}.}
\beq
\label{Ameanfield}
\vec A (\vec x) \es {1 \over 2} \, \vec B(\vec x_G) 
\times (\vec x - \vec x_G) \, , 
\eeq
one introduces the color magnetic field operator 
at the center of the gluon component $G$ of
a hadron, $\vec B(\vec x_G)$, instead of the
vacuum operator $\vec B$ at an arbitrary point 
$\vec x_0$. However, these differences do not 
change the formal calculation of the matrix 
elements. The field $\vec B(\vec x_G)$ is assumed
factorized into a product of vectors, one with space 
and the other with color components as in Ref.
\cite{GlazekSchaden}. This assumption is meant 
to reflect the assumed lack of correlation between
position space and color space directions in 
the mean-field approximation and it renders a 
simple, Abelian form of the approximate model 
for the effective Hamiltonian. Matrix elements 
of terms linear in $A$ are set to zero because 
they change under gauge transformations whereas 
the white component of a hadron does not.

The matrix elements one obtains for a 
$q\bar q$ pair in a meson and $3q$ in 
a baryon read\footnote{Eq. (14) in Ref.
\cite{GlazekSchaden} misses the factor 1/3 
in front of the condensate term that is
correctly printed here in Eq. (\ref{mM}).}
\beq
\label{mM}
\langle 12G | H_{min} |\psi \rangle_M 
\es
{1 \over \beta_M} 
\sum_{i=1}^2
\left\{ -\Delta_i 
+{g^2 \over 3}
\,
Tr
{ \langle G| \left( A_1 - A_2 \right)^2 |G'\rangle 
  \over \langle G | G' \rangle }
\right\} 
\langle 12 G| \psi\rangle \, , 
\nn
&& \\
\label{mB}
\langle 123G | H_{min} |\psi \rangle_B  
\es
{1 \over \beta_B} \, 
\sum_{i=1}^3( -\Delta_i) \langle 123 G| \psi\rangle_B 
+
{1 \over \beta_B} \, 
\sum_{i=1}^3
{g^2 \over 3}
\nt
Tr
{ 
\langle G|
\left( A_i - A_{j+k\over 2} \right)^2 
+ {1 \over 12} (A_j-A_k)^2 |G'\rangle 
\over \langle G | G' \rangle  } \,
\langle 123 G| \psi\rangle_B  \, . \nn
\eeq
All gluon field matrix elements contain squares of 
differences of the effective gluon field operator
at different points. Therefore, the position of the 
gluon body, $\vec x_G$, drops out. In the mean-field 
approximation, one has
\beq
\vec A(\vec x\,) - \vec A(\vec y\,) 
\es
{1 \over 2} \, \vec B \times (\vec x - \vec y\,)
\, ,
\eeq
where $\vec B = \vec B(\vec x_G)$. Consequently,
\beq
Tr \langle G| {g^2 \over 4\pi^2} 
(\vec A_x - \vec A_y)^2 |G'\rangle
\es
2 \, (\vec x-\vec y\,)^2 \, C_{glue} \, 
\langle G | G' \rangle  \, ,
\eeq
with
\beq
\langle G|G'\rangle \es \delta_{GG'} \, .
\eeq
The expectation value $C_{glue}$ plays here 
the same role that the vacuum gluon condensate 
value $C_{vacuum} = \varphi^2_{vacuum}/96$ with
\beq
\varphi_{vacuum}^2 \es \langle \Omega|
(\alpha/\pi) G^{\mu \nu c} G_{\mu \nu}^c |\Omega
\rangle \, ,
\eeq
plays in Ref. \cite{GlazekSchaden}. Replacement 
of the constant $C_{vacuum}$ by the constant 
$C_{glue} = \varphi^2_{glue}/96$, implies our
reinterpretation of the phenomenologically 
useful quantity $\varphi_{vacuum}$ on the order of 
$\Lambda_{QCD}$ as coming from the quantity
$\varphi_{glue}$ that originates according to 
the RGPEP model in the gluon condensation only 
inside a hadron instead of the entire space. 
Hence,
\beq
\label{mMf}
\langle 12G | H_{min} |\psi \rangle_M 
\es
{1 \over \beta_M} 
(-\Delta_1 - \Delta_2)\, \langle 12 G| \psi\rangle_M 
\np
{1 \over \beta_M} \,
\left( {\pi \varphi_{glue} \over 3}
\right)^2
 { r_{12}^2 \over 2} \,
\langle 12 G| \psi\rangle_M \, , \\
\label{mBf}
\langle 123G | H_{min} |\psi \rangle_B  
\es
{1 \over \beta_B} \,
( -\Delta_1 -\Delta_2 -\Delta_3) \langle 123 G| \psi\rangle_B 
\np
{1 \over \beta_B} \, {5 \over 8} \, 
\left( {\pi \varphi_{glue} \over 3 } \right)^2 
\left( {r_{12}^2 \over 2} +  { 2 r_3^2 \over 3} \right) \, 
\langle 123G |\psi \rangle_B \, , 
\eeq
where 
\beq
r_{12} \es x_1 - x_2 \, , \\
r_3    \es x_3 - (x_1+x_2)/2  \, .
\eeq
The results for $H_{min}$ with new interpretation of 
the gluon condensate inside hadrons are used in the 
next Section to construct candidates for effective 
LF Hamiltonians of light mesons and baryons as 
first approximations to solutions of RGPEP in QCD. 

\subsection{ Approximate LF Hamiltonian for quarks 
             at $\lambda \gtrsim \lambda_c$ in 
             light hadrons }
\label{dynamics4}

Candidates for approximate Hamiltonians for light
hadrons can be obtained starting from inclusion of
results in Eqs. (\ref{mM}) and (\ref{mB}), or
(\ref{mMf}) and (\ref{mBf}), in Eqs.
(\ref{minimalM}) and (\ref{minimalB}),
correspondingly. The meson case is simpler than
the baryon case and explains the leading idea of
constructing the effective Hamiltonians using
the minimal coupling dictated by gauge symmetry. 
We have
\beq
\cM^2_{q \bar q} 
\es 
( m_1+ m_2)^2 - (\vec p_1 + \vec p_2)^2 
\np
  {1 \over \beta_1} \, 
  \left[ \vec p_1^{\,2} + \left< {g^2 \over 3} Tr (\vec A_1 - \vec A_2)^2 \right>_G \right]  
\np 
  {1 \over \beta_2} \, 
  \left[ \vec p_2^{\,2} + \left< {g^2 \over 3} Tr (\vec A_1 - \vec A_2)^2 \right>_G \right]  \, .
\eeq
Using the variables
\beq
\vec P_{12} \es \vec p_1 + \vec p_2 \, , \\
\vec k \es \beta_2 \vec p_1 - \beta_1 \vec p_2 \, , \\
\vec p_1 \es \beta_1 \vec P_{12} + \vec k \, , \\
\vec p_2 \es \beta_2 \vec P_{12} - \vec k \, , 
\eeq
one arrives at 
\beq
\label{mqq}
\cM^2_{q \bar q} 
\es 
( m_1+ m_2)^2 
+
{1 \over \beta_1 \beta_2} \, 
\left[ \vec k^{\,2} + 
\left< {g^2 \over 3} Tr (\vec A_1 - \vec A_2)^2 \right>_G \right] \, . 
\eeq
The LF counterpart of this result is obtained by 
considering three-vectors with $+$ and $\perp$ 
components instead of $z$ and $\perp$. So, one
writes
\beq
P_{12}^{+,\perp} \es p_1^{+,\perp} + p_2^{+,\perp}  \, , \\
x_1 \es p_1^+/P^+_{12} \, , \quad x_2 \rs p_2^+/P^+_{12} \, , \\
\kappa^\perp \es x_2 p_1^\perp - x_1 p_2^\perp \, , \\
p_1^\perp \es x_1 P^\perp_{12} + \kappa^\perp  \, , \\
p_2^\perp \es x_2 P^\perp_{12} - \kappa^\perp  \, .
\eeq
These are standard expressions in LF 
description of relative motion of two 
constituents carrying the total momentum 
$P$. Using these variables for two free 
particles named 1 and 2, one obtains their 
free invariant mass squared in the form
\beq
\cM^2_{12}
\es
{\kappa^{\perp \, 2} + m_1^2 \over x_1}
+
{\kappa^{\perp \, 2} + m_2^2 \over x_2} \\
\es
(m_1 + m_2)^2 
+
{1 \over x_1 x_2}
\left[
\kappa^{\perp \,2} + (m_2 x_1 - m_1 x_2)^2
\right] \, .
\eeq
Comparison with Eq. (\ref{mqq}) identifies
the relative momentum $\vec k$,
\beq
\label{kperp}
k^\perp
\es 
\sqrt{ \beta_1 \beta_2  \over x_1 x_2} \, \kappa^\perp  \, , \\
\label{kz}
k^z
\es 
\sqrt{ \beta_1 \beta_2  \over x_1 x_2} \, [ m_2
x_1 - m_1 x_2 ] \, .
\eeq
This is a new way of parameterizing relative motion
of two constituents in LF approach to quantum
mechanics and field theory. The new variables 
match non-relativistic relative momenta that 
include effects of particle masses. On the other 
hand, it is known that in the relativistic relative 
motion of two constituents the mass parameters 
appear not significant. For example, the standard 
relative momentum, in which $k^\perp = \kappa^\perp$, 
has length
\beq
\label{standardk}
\vec k_{standard}^{\,2} 
& = & 
{1 \over 4 \cM^2_{12} }
\left[ \cM^2_{12} - (m_1+m_2)^2 \right]
\left[ \cM^2_{12} - (m_1-m_2)^2 \right] \, ,
\eeq
and a relatively complicated expression is
obtained for $k^z$ if one insists on the
identification $k^\perp = \kappa^\perp$. But when
$\cM_{12}$ is large, one has $\vec k_{standard}^2
= \cM_{12}^2/4$ independently of the quark masses.
At the same time, the variable $k^z_{standard}$
depends on $\kappa^\perp$. The length of the
standard relative three-momentum in the
constituent rest frame differs from the length of
the new one,
\beq
\vec k^{\,2}_{standard} 
\es 
{ \vec k^{\,2} \over 4 \beta_1 \beta_2}
\left[ 1 - {(m_1 - m_2)^2 \over \cM^2_{12}}  \right] \, ,
\eeq 
and the angular orientations of the three-vectors 
$\vec k_{standard}$ and $\vec k$ also differ. 
This means that the new choice of LF three-momentum 
variables in relativistic cases involves rotation 
by some polar angle (the azimuthal angles are the 
same). Analysis of rotational symmetry is influenced
in the sense that in order to recover a simple
NR quantum mechanical picture of hadrons from  
quantum field theory one is motivated by the effective 
picture in RGPEP to use the new variable $\vec k$
rather than $\vec k_{standard}$ as arguments of 
potentials for constituent quarks. One should also 
remember that when a change of variables is made 
in the expressions involving effective quark wave 
functions, the new variables produce Jacobian 
factors in phase-space integration. 

Using Eq. (\ref{mMf}), one obtains
\beq
\left< {g^2 \over 3} Tr (\vec A_1 - \vec A_2)^2 \right>_G 
\es
{1 \over 2} \left( {\pi \varphi_{glue} \over 3}
\right)^2
{ r_{12}^2 \over 2} \, ,
\eeq
where $\vec r_{12}$ should be the relative position
variable that is canonically conjugated to
$\vec k$. This means in quantum mechanics that
\beq
\vec r_{12} \es i \, {\partial \over \partial \vec k} \, .
\eeq
Thus, the LF mass squared for a constituent quark-anti-quark
pair at scale $\lambda \gtrsim \lambda_c$ interacting in 
a gauge minimal way with gluons condensed inside a meson and
treated in a mean-field approximation, has the form 
\beq
\label{mqqfinal}
\cM^2_{q \bar q} 
\es 
( m_1+ m_2)^2 
+
{1 \over \beta_1 \beta_2} \, 
\left[ \vec k^{\,2} + 
{1 \over 2}\left( {\pi \varphi_{glue} \over 3}
\right)^2
{1 \over 2} \left( i \, {\partial \over \partial
\vec k} \right)^2
\right] \, ,
\eeq
where the vector $\vec k$ is defined by Eqs. 
(\ref{kperp}) and (\ref{kz}). 

In particular, the effective quarks $u$ and $d$ 
are expected to have practically the same mass
$m$ order $\Lambda_{QCD}$ in the RGPEP scheme. 
For them, $\beta_1 = \beta_2 = 1/2$ and the
associated Jacobi variables are 
\beq
\vec \rho   \es \vec r_{12} / \sqrt{2} \, , \\
\vec p_\rho \es \vec k \sqrt{2} \, .
\eeq
In terms of these variables,
\beq
\label{mqqJacobi}
\cM^2_{q \bar q} 
\es 
4m^2 
+
2 \, p_\rho^2 
+ 
2 \left( {\pi \varphi_{glue} \over 3} \right)^2 \,
\rho^2 \, .
\eeq
For small relative momenta and corresponding 
distances, this result approximately matches
the NR oscillator dynamics,
\beq
\label{mqqJacobiNR}
\cM_{q \bar q} 
\es 
2m 
+
{ p_\rho^2  \over 2 m }
+ 
{1 \over 2} \, m \, \left( {\pi \varphi_{glue} \over 3 m} \right)^2 \,
\rho^2 \, ,
\eeq
with frequency $\omega_M = \pi
\varphi_{glue}/(3m)$. 

On the one hand, this result relates 
the RGPEP reasoning to the CQM phenomenology
since the reinterpreted value of the gluon 
condensate produces physically reasonable 
frequency $\omega_M$ \cite{GlazekSchaden}.
On the other hand, the variables $\vec k$
and $\vec r$ identified here include the 
characteristic factor $\sqrt{x(1-x)}$ that
is required in AdS/QCD holographic variables 
\cite{FAdS1,FAdS2,FAdS3}. Thus, it becomes
plausible that also the RGPEP reasoning
regarding the $\lambda$-dependence of
effective interactions may provide insight
into the significance of soft wall (SW) models 
\cite{SW} of hadron spectrum. 

As a result of analogous reasoning in the baryon 
case, we obtain
\beq
{\cal M}_{123}^2 - (m_1 + m_2 + m_3)^2 
\es
{1 \over \beta_3 (1-\beta_3)} \, \vec Q^{\,2}
+ {1-\beta_3 \over \beta_1 \beta_2} \, \vec
K^{\,2} \, , 
\eeq
where
\beq
\label{xi}
x_i        \es p_i^+/P^+_{123} \, , \\
\beta_i    \es {m_i \over m_1 + m_2 + m_3} \, , \\
\label{qperp}
p_3^\perp  \es x_3 P^\perp_{123} + q^\perp \, , \\
\label{kperp2}
p_2^\perp  \es x_2 P^\perp_{123} - {x_2 \over 1-x_3} \, q^\perp - k^\perp \, , \\
\label{kperp1}
p_1^\perp  \es x_1 P^\perp_{123} - {x_1 \over 1-x_3} \, q^\perp + k^\perp \, , \\
\label{Qperp}
Q^\perp    \es \sqrt{ \beta_3 (1-\beta_3) \over x_3 (1-x_3)} \,\, q^\perp \, , \\
\label{Qz}
Q^z        \es \sqrt{ \beta_3 (1-\beta_3) \over x_3 (1-x_3)} \,\,
               \left[ (m_1 + m_2) x_3 - m_3(x_1 + x_2)\right] \, , \\
\label{Kperp}
K^\perp    \es \sqrt{  {\beta_1 \beta_2 \over x_1 x_2} {1-x_3 \over 1-\beta_3} } \,\, k^\perp \, , \\
\label{Kz}
K^z        \es \sqrt{  {\beta_1 \beta_2 \over x_1 x_2} {1-x_3 \over 1-\beta_3} } \,\, 
               \left( m_2 \, {x_1 \over 1-x_3} - m_1 \, {x_2 \over 1-x_3} \right) \, .
\eeq  
For slow relative motion of constituents, the new
momentum variables $\vec K$ and $\vec Q$ match the
non-relativistic three-momenta used in quark models,
but in fact they apply in the entire range of relativistic
kinematics of motion of constituents inside
baryons, and for arbitrary motion of baryons as a
whole, thanks to the kinematical symmetries of LF
formulation of the theory. However, one needs to
remember that the lengths and angular orientations
of the new momentum variables differ from the
standard LF variables among which, in particular, 
$q_{12}^\perp$ and $q_3^\perp$ (see below) could 
be thought directly useable for obtaining some 
effective constituent picture for baryons from 
LF QCD.

The LF mass squared for 3 constituent quarks of 
one and the same mass $m$ at scale $\lambda \gtrsim 
\lambda_c$ interacting in a gauge minimal way with 
gluons condensed inside a baryon in a mean-field
approximation, has the form 
\beq
\label{m3qfinal}
{\cal M}_{3q}^2 \es 9m^2 + 6 \, \vec K^{\,2} 
                         + {9 \over 2} \, \vec Q^{\,2}  
- 3m^2 \left( {\pi \varphi \over 3 m }
\right)^2 {5 \over 8} (\Delta_K^2/2 + 2 \Delta_Q/3 )
\, .
\eeq
We identify relations
\beq
\vec K 
\es 
\vec q_{12} \rs \vec p_\rho/\sqrt{2} \, , \\
-i {\partial \over \partial \vec K} 
\es 
\vec r_{12} \rs \sqrt{2} \, \vec \rho \, ,  \\
\vec Q
\es
\vec q_3  \rs - \sqrt{2/3} \, \vec p_\lambda \, , \\
-i {\partial \over \partial \vec Q} 
\es
\vec r_3  \rs - \sqrt{3/2} \, \vec \lambda \, ,
\eeq
in terms of the Jacobi co-ordinates for 3 quarks
with equal masses, $\vec \rho$, $\vec p_\rho$,
$\vec \lambda$, and $\vec p_\lambda$.
One has
\beq
P_{123} \es p_1 + p_2 + p_3 \, , \quad
q_{12} = (p_1 - p_2)/2 \, , \\
P_{12} \es p_1 + p_2 \, , \quad
q_3 = (2 p_3 - P_{12})/3 \, , \\
P_{12} \es 2P_{123}/3 - q_3 \, , \quad p_3 \rs P_{123}/3 + q_3 \, , \\
p_1 \es P_{12}/2 + q_{12} \rs  P_{123}/3 + q_{12} - q_3/2 \, , \\
p_2 \es P_{12}/2 - q_{12} \rs  P_{123}/3 - q_{12} - q_3/2 \, , \\
p_3 \es P/3 + q_3 \, . 
\eeq
\beq
p_1^2 + p_2^2 + p_3^2 
\es
P_{123}^2/3 + 2 q_{12}^2 + 3q_3^2/2 \, .
\eeq
\beq
R_{123} \es (x_1 + x_2 + x_3)/3 \, , \quad 
r_{12} \rs x_1 - x_2 \, , \quad
r_3    \rs x_3 - R_{12} \, , \\
x_3 \es R_{123} + 2r_3/3\, , \quad 
R_{12} \rs (x_1+x_2)/2  \rs R_{123} -  r_3/3 \, , \\
x_1 \es R_{12} + r_{12}/2 \rs R_{123} -  r_3/3  + r_{12}/2 \, , \\
x_2 \es R_{12} - r_{12}/2 \rs R_{123} -  r_3/3  - r_{12}/2 \, .
\eeq
\beq
\sum_{i=1}^3 p_ix_i 
\es P_{123}R_{12} + q_{12} r_{12} + q_3 r_3 \, . 
\eeq
In the case of the same masses, the relativistic
LF result for three quarks in a baryon reads, 
\beq
\label{b3qJacobi}
{\cal M}_{3q}^2 \es 9m^2 + 3 \, \vec p_\rho^{\,2} 
                        + 3 \, \vec p_\lambda^{\,2}  
+ 3m^2 \left( {\pi \varphi \over 3 m }
\right)^2 {5 \over 8} (\vec \rho^{\,2} + \vec \lambda^{\,2} )
\, .
\eeq
In the NR limit,
\beq
\label{b3qJacobiNR}
{\cal M}_{3q} 
\es 
3m + {\vec p_\rho^{\,2} \over 2m} 
   + {\vec p_\lambda^{\,2} \over 2m}  
   + {1 \over 2} \, m \, {5 \over 8} \, \left( {\pi \varphi \over 3 m }
\right)^2  (\vec \rho^{\,2} + \vec \lambda^{\,2} )  \, ,
\eeq
which matches precisely the result of Ref. 
\cite{GlazekSchaden}, and
\beq
\omega_B^2 \es {5 \over 8} \, \omega_M^2 \, .
\eeq
This numerical relation is in a reasonable 
agreement with phenomenology of constituent 
quark models \cite{CQM1,CQM2,CQM3}, as noticed
already in Ref. \cite{GlazekSchaden} (see also 
Section \ref{AdS}).  

\subsection{ Dynamics of quarks and glue in hadrons 
             at $\lambda \gtrsim \lambda_c$ }
\label{dynamics5}

The constituent models do not, however, include
the gluon component $G$. Model potentials are also
not assigned energy or momentum, while gluons do
carry energy and momentum. Bag models do include a
bag in terms of boundary conditions on quark wave
functions on the bag walls and a constant
contribution of the bag to the total hadron
energy, but the bag is not placed yet in the 
context of QCD \cite{bag}. The author does not 
know how to interpret results of lattice formulation 
of QCD regarding the constituent model and the gluon 
component in terms of hadronic wave functions. 
Summarizing, it is not clear how the addition of $G$ 
will affect phenomenology familiar through the 
CQMs without $G$, although some foreseeable 
changes appear welcome (see next Section). 

Regarding the contribution of mass of $G$ to the
mass squared of the whole hadron, one cannot say
anything but hypothesize that the mass of $G$ may
depend on $\lambda$. It can be small, due to
interactions that hold gluons among themselves in
the form of $G$. The mass of $G$ may even decrease
when $\lambda$ drops down to $\lambda_c$. A
graphical picture that makes this idea plausible
as a result of overlapping of quarks of size
$1/\lambda$ (a quark and an anti-quark in a meson
or three quarks in a baryon) and $G$ in position
space, is described in Appendix
\ref{OverlappingSwarmsModel}. However, a simple
concept of a constituent $G$ may be insufficient
in the case of $\pi$-mesons, considering that
$\pi$-mesons may differ from other hadrons as a
consequence of their relationship to chiral
symmetry and its breaking. The symmetry breaking
can be addressed in LF QCD \cite{Wilsonetal} but
it is not addressed in this article. Such
discussion requires reinterpretation of the quark
condensate and $G$ built from gluons alone is not
sufficient. 

Regarding the motion of the glue component $G$
inside a hadron, one may observe that if the state
$G^\dagger|0\rangle$ in RGPEP is to correspond to
the vacuum in the instant dynamics way of thinking
and $G$ is associated through RGPEP with a hadron
in an effective theory of scale $\lambda \gtrsim
\lambda_c$, the effective quarks at this scale
should not be free to move farther away from $x_G$
than about $1/\lambda$ (see also Appendix
\ref{OverlappingSwarmsModel}). The mean field
approximation misses the fact that the gluon
component has a size comparable with a hadron. The
Schwinger gauge expression for the gluon field
potential $A_\lambda(x)$ with only one term linear
in $\vec x - \vec x_G$ and $\vec B(\vec x_G)$, Eq.
(\ref{Ameanfield}), must be missing important effects
at distances comparable with the diameter of a
hadron, cf. \cite{SDGcondensates}. This means that
the gauge transport factors $T(x,{\underline x})$
cannot be approximated well in the entire volume
of a hadron using the Schwinger gauge with a
linear term alone. Thus, the mean-field
calculation does not report well on what happens
at the boundary of the glue or the boundary of a
hadron. While one cannot determine in advance what
will eventually result from RGPEP calculations of
effective Hamiltonians and eigenvalue problems for
these Hamiltonians, one should certainly expect
that there will be forces that keep the center of
the quarks and the glue together. 

There is no simpler choice for the needed
interaction term between the quarks and $G$ than a
quadratic function of the distance between
$\underline x$ and $x_G$. Such distance is easy
to write in NR quantum mechanics. However, one has
to carefully define its analog in the hadron mass 
squared operator on the LF. Solution of this 
technical issue described here, is found by looking 
at hadrons as built only from two constituents. 
One constituent is the effective quarks treated as 
one particle of mass resulting from eigenvalues 
of $\cM^2_{q \bar q}$ in mesons or $\cM^2_{3q}$ in 
baryons. The other constituent is $G$ treated as a 
particle. The mass of $G$ is unknown, may depend on 
$\lambda$, and is free to adjust in a process of 
defining a first approximation to hadrons at 
$\lambda \gtrsim \lambda_c$. With this setup, one 
applies the same method for constructing the LF mass 
squared operator that we used for a quark and an 
anti-quark treated as two constituents in a 
harmonically bound state. There is no minimal 
gauge coupling to use for the colorless state of quarks
interacting with a white $G$, and a large degree 
of ambiguity must be confronted without clear 
guidance from QCD.\footnote{It is precisely this type
of ambiguity at small energies that RGPEP is meant
to eventually help in resolving.} But there is no
problem now with introducing the appropriate
oscillator potential operator using the distance
that is quantum-mechanically conjugated to a 
relative momentum.

The first-approximation to LF Hamiltonian for light 
hadrons at $\lambda \gtrsim \lambda_c$ is suggested 
to have the form
\beq
\label{masshadron}
H_{\lambda \gtrsim \lambda_c} 
\es
{ \cM^2_{quarks} + P_{quarks}^{\perp \, 2} \over P_{quarks}^+ }
+ 
{ \cM^2_G + P_G^{\perp \, 2} \over P_G^+ }
+
{\cM^2_{qG} \over P^+_{hadron}} \, .
\eeq
In order to define a suitable candidate 
for the last term, one can introduce 
\beq
\label{momentahadron}
P_{hadron}^{+ , \perp} \es P^{+, \perp}_{quarks} + P^{+,
\perp}_G \, ,  \\
x_q \es P^+_{quarks}/P^+_{hadron} \, , \\
x_G \es P_G^+/P^+_{hadron} \, , \\
\kappa_q^\perp \es x_G P_{quarks}^\perp - x_q P_G^\perp \, , \\
P_{quarks}^\perp \es x_q P^\perp_{hadron} + \kappa_q^\perp  \, , \\
P_G^\perp \es x_G P_{hadron}^\perp - \kappa_q^\perp  \, .
\eeq
In terms of these variables, the hadron mass squared
reads
\beq
\label{mhfinal}
\cM^2_{hadron}
\es
{ \cM^2_{quarks} + \kappa_q^{\perp \, 2} \over x_q }
+ 
{ \cM^2_G + \kappa_q^{\perp \, 2} \over x_G }
+
\cM^2_{qG} \, ,
\eeq
where $\cM^2_{quarks}$ is now to be identified with the 
eigenvalue $M_q^2$ for the mass squared of the quark 
subsystem in a hadron and $\cM_G$ is set to the corresponding
eigenvalue $M_G^2$, which could be considered to have a ground
state value or an excited value, if the condensed gluons were 
in an excited state. The constant $M_q^2$ is either $4m^2$ or 
$9m^2$ plus an appropriate number times $m\omega$ with 
$\omega = \omega_M$ or $\omega = \omega_B$, respectively. 
Having introduced
\beq
\beta_q \es { M_q \over M_q + M_G} \, , \\
\beta_G \es { M_G \over M_q + M_G} \, , 
\eeq
one can define
\beq
\label{khperp}
k_h^\perp
\es 
\sqrt{ \beta_q \beta_G  \over x_q x_G} \, \kappa_q^\perp  \, , \\
\label{kzh}
k_h^z
\es 
\sqrt{ \beta_q \beta_G  \over x_q x_G} \, [ M_G
x_q - M_q x_G ] \, .
\eeq
Proceeding as in the case of two quarks in a meson,
Eq. (\ref{mqqfinal}), one establishes
\beq
\label{hMint}
\cM^2_{qG}
\es
{1 \over \beta_q \beta_G} \, 
{1 \over 2}\left( {\pi \varphi_h \over 3}
\right)^2
{1 \over 2} \left( i \, {\partial \over \partial
\vec k_h} \right)^2 - M_\varphi^2 \, , \\
\cM^2_{hadron}
\es
(M_q + M_G)^2 - M_\varphi^2
\np
\label{hadronmassfinal}
{1 \over \beta_q \beta_G} \,
\left[ \vec k_h^{\,2} + 
{1 \over 2}\left( {\pi \varphi_h \over 3}
\right)^2
{1 \over 2} \left( i \, {\partial \over \partial
\vec k_h} \right)^2
\right] \, ,
\eeq
where two new parameters, $\varphi_h$ and
$M_\varphi$, are introduced.

The parameter $\varphi_h$ is used in the similar
way to how $\varphi_{glue}$ is used in the
invariant mass of quarks in mesons. However,
$\varphi_h$ serves merely the purpose of notation
for the unknown quantity of binding between quarks
and $G$ whose value must be adjusted in the process 
of correcting the candidate for the first approximation. 
In RGPEP, $\varphi_h$ can be expected to depend on 
$\lambda$. Intuitive arguments are offered in 
Appendix \ref{OverlappingSwarmsModel}. This candidate 
for a first approximation will only be satisfying 
if it turns out in future calculations in RGPEP that
some optimal value of $\varphi_h$ can be adjusted 
as a function of $\lambda/\Lambda_{QCD}$ and the 
corresponding coupling constant, assuming that the 
light quark mass parameters do not matter at small 
$\lambda$ where the effective quark masses are on 
the order of $\Lambda_{QCD}$ anyway and do not vary 
with $\lambda$ so rapidly that no average constituent 
picture can correspond to QCD. Assuming that $\varphi_h$ 
does not vary from hadron to hadron, one can introduce
\beq
\label{omegah}
\omega_h
\es
{\pi \varphi_h \over 3 \mu_h} \, , \\
\label{muh}
\mu_h 
\es
{M_q M_G \over M_q + M_G} \, .
\eeq
The parameter $\mu_h$ is the reduced
mass for the two-body system quarks-$G$.
It depends on a hadron as far as $M_q$ depends
on a hadron, assuming $M_G$ can be considered
universal in the first approximation. Thus,
the mass-squared eigenvalues for the whole hadron
takes the form 
\beq
\label{hadronspectrum}
M^2_h
\es
(M_q + M_G)^2 - M_\varphi^2
+
{\mu_h \over \beta_q \beta_G} \,
(n_h + 3/2) \omega_h \, ,
\eeq
where $n_h$ denotes the excitation quantum number
in relative motion of quarks with respect to
the gluon condensate in a hadron, being zero
in a ground state. Thus the momentum width of 
relative motion o quarks with respect to $G$ is
order $\varphi_h$.

The mass parameter $M_\varphi$ is introduced by
fiat and brings in another considerable degree of
ambiguity. Physical motivation for $M_\varphi$ is 
that for the gluon condensation to occur in hadrons 
it must be favorable energetically. If one just added
the free energy of $G$ and a harmonic quarks-glue
potential energy to the energy of quarks, the
condensation of gluons would only add energy to
the system of quarks. The effect of condensation
should rather be opposite: the concept of
condensation of gluons inside hadrons implies
thinking that the mass of a hadron state is
lowered by condensation of gluons in comparison
with a state in which condensation is absent.
Since the kinematical minimum of energy of quarks
and glue is $M_q + M_G$, which corresponds to
$\kappa_q^\perp = 0$ and $x_q = \beta_q$, the
interaction mass parameter $M_\varphi$ may be
estimated by inspecting the condition
\beq
M_q^2  & > & (M_q + M_G)^2 + {3 \over 2}
(M_q + M_G)\omega_h - M_\varphi^2 \, .
\eeq
Assuming that the resulting lower bound on 
$M_\varphi$ provides an estimate of its
likely magnitude in QCD, we obtain
\beq
\label{hadronestimate}
M_\varphi^2  & \sim & 
M_q M_G \left ( 2 + {M_G \over M_q} + { 3 \omega_h
\over 2 \mu_h} \right) \, , 
\eeq
where 3 signifies the number of spatial
dimensions. Assuming that $M_q$ equals nucleon
mass $m_N$, and $M_G \sim \omega_h \sim m_N/3$,
one obtains $M_\varphi \sim 1.2 \, m_N$. If $M_G 
\sim M_q \sim m_N$ and $\omega_h \sim m_N$, we 
have $M_\varphi \sim 2.4 \, m_N$. This means that 
$M_\varphi$ can be adjusted to obtain a hadron 
mass as coming mainly from the quark mass 
eigenvalue $M_q$ provided that $M_\varphi$ is 
quite sizable. Such sizable energy benefit from 
gluon condensation eliminates a large contribution 
of $G$ to a hadron mass and sustains the possibility 
that a simple oscillator quark model without any 
glue can reproduce masses of light hadrons assuming 
that the effective quark masses are on the order of 
$\Lambda_{QCD}$.\footnote{If the condensation mass
advantage, $-M_\varphi^2$, were associated also 
with condensation of quark-anti-quark pairs, one
might expect a pronounced reduction in masses of
the hadron states in which a bilinear effective
quark field expectation value plays a significant
role in the dynamics, with $\pi$ mesons being the
primary candidates.} 

We wish to stress that Eq. (\ref{hMint}) does not
imply one and the same interaction between quarks
and glue $G$ in all hadrons even if the parameters
concerning $G$ are assumed the same for all
hadrons. Different hadrons are actually having
different Hamiltonians characterized by different
quantum numbers of their quark component, such as
radial and orbital excitations (elementary spin
effects are discussed in the next Section) and the
mass eigenvalue for the quarks, $M_q$, depends on
these quantum numbers. This mass enters the
definition of $\vec k_h$ and thus also the
definition of harmonic potential between the
quarks and $G$ that is defined in terms of
$\partial / \partial \vec k_h$. As a result,
potentials between quarks and $G$ vary from hadron
to hadron in a well-defined pattern: the higher
excitation of quarks, the greater their mass and
the more momentum of a hadron carried by
quarks.\footnote{Provided that the
first-approximation quantity $M_\varphi$ is kept
constant.} For example, quarks in excited nucleons
are expected to carry a larger fraction of the
resonance momentum than quarks carry in a ground-state 
nucleon. And vice versa, an excitation of the gluons
condensed in a hadron increases $M_G$ and the
fraction of a hadron momentum they thus carry.
Nevertheless, a CQM for light
hadrons should be viewed as corresponding to 
$n_h=0$ when $\lambda \gtrsim \lambda_c$ is 
actually lowered to $\lambda_c$. As suggested in 
Appendix \ref{OverlappingSwarmsModel}, when $\lambda$ 
is lowered to $\lambda_c$, the role of $G$ may be
imagined taken over entirely by the content 
of extended constituent quarks that overlap 
each other heavily, so that no separate glue 
component can exist in hadron besides what is 
already contained in the extended effective 
quarks. Instead of the speculation, however, 
the proper problem is what will result from 
attempts to solve the RGPEP equation for the 
effective Hamiltonian at $\lambda = \lambda_c$.
Strictly speaking, nothing is known currently 
about the solution.

\section{ RGPEP in QCD and phenomenology }
\label{Phenomenology}

The discussion that follows is limited to the
case of all effective quarks having the same 
mass $m$. The common mass is expected to be a 
reasonable approximation for quarks $u$ and $d$ 
at $\lambda \sim \lambda_c$. Calculational 
complications due to differences in mass 
between these quarks are not discussed.
 
Assuming that $\varphi_{glue}$ is on the order of
$\Lambda_{QCD}^2$, which means that it has a
similar value to the values obtained in QCD sum
rules for $\varphi_{vacuum}$, one obtains
phenomenologically attractive values of $\omega_M$
and $\omega_B$ \cite{GlazekSchaden}. Eqs.
(\ref{mqqfinal}), (\ref{m3qfinal}), (\ref{mhfinal}), 
(\ref{hMint}), and (\ref{hadronmassfinal}), imply 
together eigenvalues and wave functions for light
hadrons in the form of harmonic oscillator solutions 
that include the glue component $G$. The generic
oscillator eigenvalue problem for two particles has 
the form
\beq
\left( {\vec p^{\,2} \over 2 m} + {1 \over 2} m
\omega^2 \vec r\,^2 \right) \psi \es  E \, \psi \, ,
\eeq
with eigenvalues $E_n = (n+3/2)\omega$ and the
ground-state wave function $\psi_0 = N \exp{[-\vec
p^{\,2}/(2 m \omega)]}$ for $n=0$, where $\vec p$
and $\vec r$ are canonically conjugated variables. 

For two effective constituent quarks in  mesons, 
using Eq. (\ref{mqqJacobiNR}) that explains what 
happens in the LF eigenvalue equation for 
$\cM^2_{q \bar q}$ in terms of a NR approximation 
to $\cM_{q\bar q}$, Eq. (\ref{mqqJacobi}) implies 
the first LF approximation of the form
\beq
M^2_{q \bar q \, n} 
\es 
4m^2 
+ 4 (n+3/2) m \omega_M \, , \\
\psi_{q \bar q \, 0} 
\es N_{q \bar q \, 0} \exp{[-\vec k^{\,2}/(m\omega_M)]} \\
\es N_{q \bar q \, 0} \exp{\left\{ - 
{ 1 \over 4m\omega_M } 
\left[{ \kappa^{\perp \, 2} + m^2 \over x(1-x) } - 4m^2 \right] \right\} } \, ,
\eeq
with frequency $\omega_M = \pi \varphi_{glue}/(3m)$
and wave functions of excited states, $\psi_{q \bar q \, n}$, 
generated by building harmonic oscillator 
excitation operators from the vector $\vec k$ and 
gradient $\partial /\partial \vec k$ in a standard way 
and applying them to the ground state wave function 
$\psi_{q \bar q \, 0}$.

For three effective constituent quarks in baryons, 
using Eq. (\ref{b3qJacobiNR}) that explains what 
happens in the LF eigenvalue equation for 
$\cM^2_{3q}$ in terms of a NR approximation 
to $\cM_{3q}$, Eq. (\ref{b3qJacobi}) implies 
the first LF approximation of the form
\beq
M^3_{3q n_1 n_2} 
\es 
9m^2 
+ 6(n_1+n_2+3)m \omega_B \, , \\
\psi_{3 q \, 0} 
\es N_{3q \, 0} \exp{   \left\{ -{1 \over 6 m\omega_B} 
\left[ 
{9 \over 2} \, \vec Q^{\,2}
+ 6  \, \vec
K^{\,2} \,
\right]
\right\} } \\
\es N_{3q \, 0} 
\exp 
\left\{ 
- {1 \over 6 m\omega_B} 
\left[ { (1 - x_3) \, k^{\perp \, 2} \over x_1 x_2 } 
     + { q^{\perp \, 2} \over x_3 (1-x_3) } 
\right.
\right.
\np
\left.
\left.
m^2\left(  {1 \over x_1} + {1 \over x_2} + {1 \over x_3} - 9 \right) \right] \right\}  \, ,
\eeq
with baryon oscillator frequency $\omega_B = \sqrt{5/8} \,\, \omega_M$
and wave functions of excited states, $\psi_{3q\, n_1 n_2}$, 
generated in a standard way by building harmonic oscillator 
excitation operators from the vectors $\vec K$ and
$\vec Q$ and gradients $\partial /\partial \vec K$ and 
$\partial /\partial \vec Q$ and applying them to the 
ground state wave function $\psi_{3q \, 0}$. The LF
momentum variables are defined in Eqs. (\ref{xi}),
(\ref{qperp}), (\ref{kperp2}) and (\ref{kperp1}). 

In summary, the factor in a hadron 
ground-state wave function that depends on the 
relative motion of $n$ quarks has the form 
\beq
\label{finalGaussian}
\psi_{n q \, 0}
\es
N_{n q \, 0} \, 
\exp{ \left\{ - {1 \over 2 n m
\omega_n} \left[ \left( \sum_{i=1}^n p_i \right)^2
- (nm)^2 \right] \right\} }\, , 
\eeq
with $n=2$ for mesons, $n=3$ for baryons, $\omega_2 = 
\omega_M$, $\omega_3 = \omega_B$, and $p_i$ the 
on-maas-shell four-momentum for quark number $i$.

Exponentials of an invariant mass squared of
quarks are popular as wave functions in
phenomenological studies. Here such exponentials
are related to gauge symmetry and reinterpretation 
of the gluon condensate as a part of a hadron. 

The spectrum of light hadron masses in Eq.
(\ref{hadronspectrum}), i.e., the spectrum
corresponding to the ground state of relative
motion of quarks with respect to the glue $G$,
becomes equal to the spectrum of masses of the
quark component alone when the estimate
(\ref{hadronestimate}) is adopted for the benefit
of the gluon condensation in a hadron. This result
reproduces success of the constituent quark
models.\footnote{The oscillator functions of
relative quark motion are independent of quark
spins. A Coulomb part of the wave function must
depend on spins. For small $\lambda$, effective
quark spin wave function can be treated as a
separate factor in the oscillator part of the wave
function. Spin factors can be constructed in a
boost-invariant way using LF spinors and following
the example of heavy quarkonia \cite{GlazekMlynik}
in the case of mesons or Ioffe currents
\cite{GlazekNamyslowski} in the case of baryons.
However, in contrast to models not related to QCD,
RGPEP provides a scheme for systematic study of
spin-dependent corrections to the first oscillator
approximation.} 

At the same time, when the relative motion of 
quarks with respect to the glue part is described 
by Eq. (\ref{hadronmassfinal}), the hadron mass is 
given by Eq. (\ref{hadronspectrum})
and the corresponding wave function is generated 
from the ground-state wave function
\beq
\label{psihadron}
\psi_h
\es
\psi_{n q\,0} \, \psi_{qG \, 0} \, , \\
\label{psiqG}
\psi_{qG \, 0}
\es
N_{qG\,0}
\exp{ 
\left[ - { (P_q+P_G)^2 - (M_q+ M_G)^2  
\over 2(M_q + M_G) \omega_h } \right]} \, ,
\eeq
in a standard way for harmonic oscillators,
first for the quark component and then, using
the quark component mass eigenvalue, for the
whole hadron if the quarks are excited in 
their motion with respect to $G$. Thus, the
invariant mass squared of quarks and $G$
together, $(P_q+P_G)^2$ is evaluated in a standard
way using their LF momenta and their minus
components calculated using quark mass eigenvalue
$M_q$ and the glue mass $M_G$, respectively. 
Normalization factors are fixed by normalizing 
probability to 1, or fixing the charge to the 
appropriate value (if the charge is not zero 
these normalization conditions are the same). 

\subsection{ Form factors }
\label{ff}

\begin{figure}
\begin{center}
\includegraphics[scale=.6]{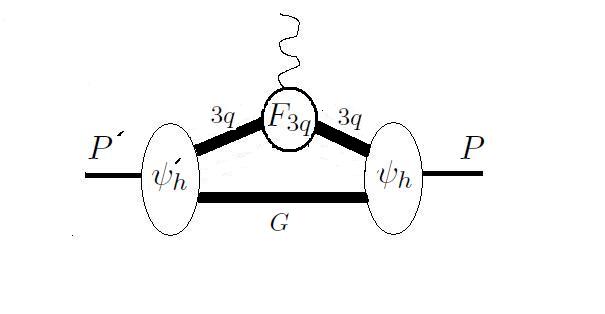}
\end{center}
\caption{The RGPEP calculation of a baryon 
form factor at $\lambda \sim \lambda_c$.}
\label{fig:FormFactor}
\end{figure}

The relative motion of quarks with respect to the
glue $G$ smears the quark observables that follow
from the quark factor in the hadron wave function
alone. Consider the ground state of quarks' motion
with respect to $G$. Calculation of the baryon
form factor at a small momentum transfer $q$, $q^2
= - Q^2$, $ 0 \le Q^2 \le \Lambda^2_{QCD}$, is 
represented graphically in Fig. \ref{fig:FormFactor}. 
Meson form factor calculation involves 2 instead of 
3 quarks but otherwise proceeds in the same way.

Using the Breit frame with $q^+=0$, which makes 
$q^-$ that depends on masses irrelevant, one has 
$Q^2 = q^{\perp \, 2}$ and
\beq
&& F_h(Q^2) \rs  \int {dx_q d^2 \kappa^\perp_q \over 16\pi^3 x_q(1-x_q)} 
\nt
\psi^*_{qG \, 0}[x_q, \kappa_q + (1-x_q)q^\perp]\,
F_{3q}(Q^2)\,
\psi_{qG \, 0}(x_q, \kappa_q) \, ,
\label{ff1}
\eeq
with normalization condition
\beq
1 \es  \int {dx_q d^2 \kappa^\perp_q \over 16\pi^3
x_q(1-x_q)} \,
|\psi_{qG \, 0}(x_q, \kappa_q)|^2 \, .
\eeq
Using the wave function in Eq. (\ref{psihadron})
and changing variables to $x = x_q$ and $\kappa^\perp
= \kappa_q^\perp + (1-x)q^\perp/2$, one obtains
\beq
\label{ff2}
F_h(Q^2) \es  
F_{3q}(Q^2)\, f(Q) \, , \\
\label{fQ}
f(Q) \es \int {dx \, d^2 \kappa^\perp \over 16\pi^3 x(1-x)} 
|\psi^*_{qG \, 0}(x, \kappa^\perp)|^2\, e^{- {1-x
\over x} {Q^2 \over 4(M_q + M_G) \omega_h} } \, .
\eeq
After integration over $\kappa^\perp$, one is 
left with 
\beq
f(Q) \es
{ \int_0^1 dx \,
e^{ - a \left({ M^2_q \over x} + { M_G^2 \over 1-x } 
           + {1-x \over x}\,{Q^2 \over 4} \right) } 
\over
\int_0^1 dx \,
e^{ - a \left({ M^2_q \over x} + { M_G^2 \over 1-x } \right)} 
} \, ,
\eeq
where $a^{-1} = (M_q + M_G)\omega_h$. It is now
visible that $f(Q^2)$ is hardly different from
1 for small $Q^2$: the inclusion of the 
glue component $G$ does not significantly alter 
the result for a form factor calculated as if the 
glue component was absent. Moreover, in the picture 
discussed in Appendix \ref{OverlappingSwarmsModel}, 
the glue component may disappear in favor of 
overlapping constituent quarks when $\lambda$ is 
lowered below $\lambda_c$. The momentum fraction 
carried by the quarks, $x=x_q$, becomes very close 
to 1 and the factor $f(Q^2)$ becomes 1.\footnote{Even 
if the component $G$ contained quark-anti-quark 
pairs, and were used to account for the quark 
condensate inside hadrons \cite{BrodskyRobertsShrockTandy}, 
its neutrality would imply a small contribution to 
form factors of charged hadrons, such as proton. It 
might, however, contribute a detectable piece for 
chargeless hadrons, such as neutron.}

At large $Q$, much greater than a hadron mass, the
hadron state needs to be represented in terms of
the Fock components created by quark and gluon
operators at $\lambda$ comparable with $Q$ itself,
in order to obtain a simple picture based on the
smallness of an asymptotically small coupling
constant that determines the strength of
interactions which are responsible for
distributing the large momentum transfer $q$ to a
minimal set of constituents required to build a
hadron. The minimal quark component may carry practically 
the whole momentum of a hadron, $x_q \rightarrow 1$. 
When this configuration dominates the transition
amplitude, the whole hadron is turned from the total 
momentum $P$ to $P' = P+q$ in a combination of two 
mechanisms. One is $x_q \rightarrow 1$, and the other
is the distribution of $q$ in the quark component. 
The former contributes to soft effects, and the 
latter is responsible for hard exclusive processes, 
cf. \cite{LepageBrodsky}.

\subsection{ Structure functions }
\label{structurefunctions}

\begin{figure}
\begin{center}
\hspace{-1cm}
\includegraphics[scale=.6]{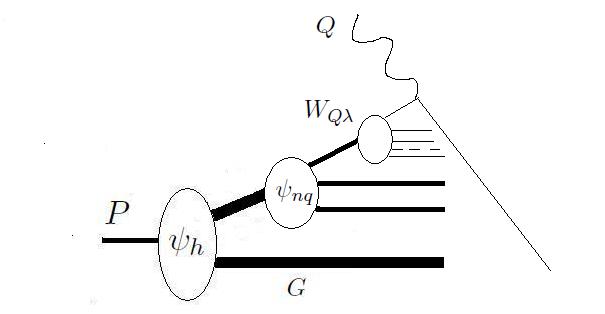}
\end{center}
\caption{The RGPEP calculation of a baryon 
structure function.}
\label{fig:StructureFunction}
\end{figure}

The transition amplitude for deep inelastic
lepton-hadron scattering is illustrated in Fig.
\ref{fig:StructureFunction}.\footnote{The lepton
is not shown. Also, Fig.
\ref{fig:StructureFunction} ignores the
possibility that the impinging boson is absorbed
by a constituent inside the component $G$, which
may be thought here to be only made of gluons. If
$G$ included quark-anti-quark pairs, it would
contribute to the sea parton distributions as well
as the quark condensate in hadrons, cf.
\cite{BrodskyRobertsShrockTandy}.} A hard photon
or other boson is suddenly absorbed by a
constituent characterized by the scale $Q$. The
new element RGPEP introduces is the possibility of
using the transformation $W_{Q\lambda}$ of Section
\ref{structureW} to calculate the probability
amplitude for finding such constituent in a
hadron. The scale $\lambda$ refers to the particle
operators in terms of which the hadron wave
function is obtained from the eigenvalue problem.
The scale $Q$ refers to the hard boson. The final
state in Fig. \ref{fig:StructureFunction} is made
of many constituents produced by $W_{Q\lambda}$.
Fig. \ref{fig:StructureFunction} does not show
that $W_{Q\lambda}$ actually acts on all
constituents.

The final state in Fig.
\ref{fig:StructureFunction} is in fact a virtual
state whose evolution factor into observable
particles is assumed to amount to unity. The
energy, or $P^-$ of the constituents that appear
in the final state in Fig.
\ref{fig:StructureFunction}, is dominated by the
constituent at scale $Q$ that absorbs the boson.
This situation corresponds to the leading operator
expansion terms, i.e., hand-bag diagrams that are
more important than cat-ears diagrams. One can use
the Hamiltonian $H_Q(b_Q)$ to account for $P^-$ of
the virtual state. This Hamiltonian is actually
the same as $H_\lambda(b_\lambda)$. Therefore,
$H_Q(Q)$ may count the energy of particles in the
final state including the interactions that are
responsible for grouping constituents at scale $Q$
into spectators at scale $\lambda$. Thus, the
inclusive sum over final states may be replaced by
summing over a relatively small number of
spectators at scale $\lambda$ and a potentially
large number of constituents at scales between
$\lambda$ and $Q$. The operator $W_{Q\lambda}$ in
RGPEP is hence expected to describe the evolution
of structure functions with $Q^2$. 

The above qualitative reasoning must be verified
by new type of calculations using RGPEP evolution
equations in place of other evolution equations in 
$Q^2$ \cite{GribovLipatov,Dokshitzer,AltarelliParisi}.
Since it is known that RGPEP incorporates the well-known 
splitting functions in QCD \cite{gluons}, there is no 
obvious reason for expecting that RGEPEP evolution will 
significantly differ from known results where they apply. 
On the other hand, RGPEP provides the framework for 
combining the perturbative evolution with a non-perturbative
hadron wave functions obtained by solving eigenvalue 
problem for $H_\lambda(b_\lambda)$. The eigenvalue problem
describes saturation at small $\lambda$. But the evolution 
parameter $\lambda$ determines dependence of wave functions 
on invariant masses of constituent states. The invariant 
masses depend on transverse momenta and fractions $x$ 
of longitudinal momentum in a specific way. One can thus 
expect that the evolution in $x$ \cite{BFKL1, BFKL2} 
is uniquely related in RGPEP with the evolution in 
$Q^2$.

\subsection{ Other processes }
\label{OtherProcesses}

The approximate picture of a hadron at scale $\lambda
\gtrsim \lambda_c$ as built from the effective quark 
component and from $G$ treated as a scalar boson 
suggests thinking that the dynamics of $G$ can be 
further approximated by suitable effective Hamiltonian 
interaction terms. These terms would result in   
processes in which quanta of type $G$ participate 
in strong interactions, softened by RGPEP form factors 
in vertices. Nothing can be said at this point about 
physical relevance of the approximate concept of 
quanta of the type $G$ being exchanged between hadrons. 
In particular, it is not excluded that an exchange of 
such quanta may contribute to diffractive processes 
in hadron-hadron collisions \cite{pomeron}, or even 
in photon-hadron scattering \cite{GolecBiernat} if a 
photon were allowed to contain its own glue component 
coupled to quarks.

\subsection{ Connection with AdS/QCD }
\label{AdS}

The NR form of minimal coupling in LF Hamiltonians
at small $\lambda$ apparently leads to variables
and harmonic oscillator potentials that are
similar to the ones that Brodsky and Teramond
\cite{FAdS1, FAdS2,FAdS3} identified as providing
a correspondence between formulae for hadronic
form factors in LF QCD and in field theory in
5-dimensional AdS space
\cite{WittenHolography,PolczynskiStrassler}. The
new variables $\vec k$ for mesons, Eqs.
(\ref{kperp}) and (\ref{kz}), and $\vec K$ and
$\vec Q$ for baryons, Eqs. (\ref{Qperp}),
(\ref{Qz}), (\ref{Kperp}), (\ref{Kz}), differ from
standard LF relative three-momenta of constituents
in the CRF.\footnote{In LF dynamics, CRF differs
from the bound state rest frame because
conservation of $P^+$ implies that $P^3$ is not
conserved by interaction terms in LF
Hamiltonians.} For example, $\vec k_{standard}$ in
Eq. (\ref{standardk}) differs in length and
direction from $\vec k$. But there is no
difference between these momentum variables when
they are small in comparison to masses. Thus, the
new variables are identical to standard variables
in the NR dynamics at $\lambda \gtrsim \lambda_c$.
In addition, they fully describe a relativistic
relative motion of constituents, always providing
an exact representation of their free invariant
mass, but in a different way than the standard
momentum variables do. Standard variables
$k^z_{standard}$, $K^z_{standard}$, and
$Q^z_{standard}$,\footnote{E.g., see Ref.
\cite{GlazekNamyslowski}.} depend on the
transverse relative momenta. The new variables
$k^z$, $K^z$, and $Q^z$, are defined using the
constituent masses and fractions $x$ of their
total $P^+$, independently of the transverse
relative momenta. At the same time, the transverse
relative momenta are scaled by mass ratios and
$\sqrt{x(1-x)}$. The square-root factor is
precisely the one that appears in Brodsky-Teramond
holography \cite{FAdS3}.

To illustrate the possibility of correspondence
between the reinterpreted gluon-condensate induced
harmonic potential in LF QCD and a soft-wall potential 
in Brodsky-Teramond AdS/QCD holography, we use example of a
meson form factor. The discussion is far from
complete and raises many questions regarding
interpretation of simple equations. However, it is
needed for showing similarities and differences
among different approaches.

Let the struck quark and the spectator quark carry
nearly whole hadron momentum. The glue component
$G$ provides additional smearing in the form
factor expression, Eq. (\ref{ff1}), unless it is
totally absorbed in the constituent
quarks.\footnote{See Appendix
\ref{OverlappingSwarmsModel}.} Let us assume here
that the latter option holds and all one needs to
consider is two constituent quarks. 

In terms of the new variables,
\beq
k^\perp \es \kappa^\perp/\sqrt{4x(1-x)} \, , \\
k^z \es (x-1/2)\,2m/\sqrt{4x(1-x)} \, , 
\eeq
the LF eigenvalue equation for a meson of mass $M$
reads 
\beq
(4m^2 + 4\vec k^{\,2} + \kappa^4 \vec r^{\,2} ) \,
\psi \es M^2 \, \psi \, ,
\eeq
where $\kappa^2 = m\omega_M$ and $\vec r =
i\partial/\partial \vec k $. Using a factorized
wave function, $ \psi(k^\perp, k^z) = N \phi(k^\perp) 
\, f_\eta(x)$, where $f_\eta$ is an eigenfunction 
of the oscillator in $z$-direction,  
\beq
f_\eta(x)
\es
H_\eta(k_z)\, e^{-
{k_z^2 \over \kappa^2}}  
\rs
H_\eta\left[m(x-1/2)\over \kappa \sqrt{x(1-x)}\right]\, 
e^{- {m^2 (x-1/2)^2 \over \kappa^2 x(1-x)}} \, ,
\eeq
one arrives at the eigenvalue equation for
$\phi(k^\perp)$,
\beq
\label{BT1}
\left[4m^2 + 4k^{\perp\,2} + \kappa^4 r^{\perp\,2}
+(4\eta + 2)\kappa^2 \right] \,
\phi \es M^2 \, \phi \, ,
\eeq
which is the eigenvalue problem to compare with 
the Brodsky-Teramond holographic eigenvalue problem, 
e.g., Eq. (10) in \cite{FAdS3}. We denote quantities 
used by Brodsky and Teramond with subscript $BT$, 
except for their $\zeta$,
\beq
k^\perp \es k^\perp_{BT}/2 \, , \\
r^\perp \es 2 \zeta^\perp \, .
\eeq
Eq. (\ref{BT1}) can be written using $\kappa_{BT} = \sqrt{2} \, \kappa$ 
as 
\beq
\left(k^{\perp\,2}_{BT} 
+ \kappa_{BT}^4 \zeta^{\perp\,2} \right) \,
\phi \es M_{BT}^2 \, \phi \, , \\
M_{BT}^2 \es M^2 - 4m^2 - (2\eta + 1)\kappa^2_{BT} \, ,
\eeq
where $k^{\perp\,2}_{BT} = - \left(\partial/ \partial
\zeta^\perp \right)^2$. For angular momentum around 
$z$-axis equal $l_z$, using $\zeta = |\zeta^\perp|$, 
one has
\beq
\left[ - { \partial^2 \over \partial \zeta^2 } 
- 
{1 \over \zeta} {\partial \over \partial \zeta}
+
{l_z^2 \over \zeta^2}
+ 
\kappa_{BT}^4 \zeta^2 \right] \,
\phi \es M^2_{BT} \, \phi \,  .
\eeq
Defining $ \phi = \phi_{BT}/\sqrt{\zeta}$, 
we arrive at 
\beq
\label{BTeigenvalue}
\left[ - { \partial^2 \over \partial \zeta^2 } 
- 
{1 - 4l_z^2 \over 4\zeta^2}
+ 
\kappa_{BT}^4 \zeta^2 
+ (2\eta + 1)\kappa^2_{BT}
\right] \, \phi_{BT} 
\es 
(M^2 - 4m^2)\, \phi_{BT} \, , \nn
\eeq
which by comparison with Eq. (11) in 
\cite{FAdS3} for massless quarks renders 
\beq
U(\zeta)
\es
\kappa_{BT}^4 \zeta^2 
+ 
(2\eta + 1)\kappa^2_{BT}
\, ,
\eeq
as a counterpart of the potential $U(\zeta)$ in 
Ref. \cite{FAdS3}, with
\beq
\label{kappaphi}
\kappa_{BT}^2 \es 2 m \omega_M \rs {2 \pi \over 3}
\, \varphi_{glue} \, .
\eeq
The role of quark masses on the right-hand side of
Eq. (\ref{BTeigenvalue}) requires explanation. A
sound explanation is currently not available in the 
sense that it is not clear why the quark mass is set
to 0 in Eq. (10) and (11) in Ref. \cite{FAdS3}. But 
it is plausible for constituent quarks that the 
condensate mass-advantage constant $M_\varphi$ in 
Eq. (\ref{hadronspectrum}) can be considered a result 
of incorporation of $G$ in the constituent quarks 
entirely. $M_\varphi$ may be so large that it reduces 
the contribution of the quark masses to the hadron 
mass and thus reduces the term $-4m^2$ in the 
eigenvalue.

On the other hand, the SW model eigenvalue Eq. (12) 
in Ref. \cite{SW} reads
\beq
\label{SWeq}
- \psi'' + \left[ z^2 + {\lambda_z^2 -1/4 \over
z^2}\right] \psi \es E \psi \, , \\
E \rs 4n_z + 2\lambda_z + 2 \, . &&
\eeq
By dividing Eq. (\ref{BTeigenvalue}) by $\kappa_{BT}^2$, 
and using variable $z = \kappa_{BT}\zeta$, one
obtains
\beq
-  \phi''_{BT} 
+ 
\left[ W(z) + {l_z^2 -1/4\over z^2} \right] \, \phi_{BT} 
\es  E \, \phi_{BT} \, , \\
E \rs { M^2 - 4m^2 - (2\eta + 1)\kappa_{BT}^2 
\over \kappa_{BT}^2 } \, . && 
\eeq
Comparing this result with the SW model result, Eq.
(\ref{SWeq}), and identifying $\lambda_z$ with $l_z$,
one obtains
\beq
W(z) \es z^2  \, .
\eeq
The SW model quadratic potential appears to have
a coefficient given by the gluon condensate inside
hadrons, Eq. (\ref{kappaphi}).

If $\eta = 0$, which means no excitation along the
LF, only $-\kappa_{BT}^2$ enters the eigenvalues.
This is a sizable term (see below) and its
compensation is not guaranteed by $M_\varphi$. 

Baryon form factors are described in terms of an
active constituent and spectators. The analysis
proceeds in a similar way to the case of a meson
built from two constituents of different masses.
This simplification is available because the free
LF invariant mass can be written in terms of the
active constituent and the rest of a hadron as if
the rest of a hadron was a single particle with 
its mass equal to the invariant mass of the rest of 
a hadron. The closest analogy to consider is that
a baryon form factor and eigenvalue problem can 
be represented in terms of a quark and a di-quark.

It should be stressed, however, that it is far too
early for drawing a firm conclusion regarding
connection between gluon condensation in hadrons
and SW models on the basis of a pure RGPEP
calculation. The reason can be seen using power
counting \cite{Wilsonetal}. Besides uncertainties
concerning $x^-$ and inverse of $i\partial^+$,
spin-independent quark-anti-quark interaction term
in a LF Hamiltonian density with a quadratic
potential, which is absent in canonical QCD, may
have a coefficient proportional to
$\Lambda_{QCD}^2$, to compensate the dimension of
$x^{\perp \, 2}$. There also exists a possibility
\cite{NonlocalH} that a dimensionless square of a
product $r_\mu P^\mu$ is used instead of $x^{\perp
\, 2}$, where $r$ is the relative position of
arguments of two field operators on the LF,
$r^+=0$, and $P$ is the momentum carried by a
product of two quark fields, i.e., two
constituents interacting by the potential. This
possibility implies that the harmonic potential
does not necessarily requires a factor
$\Lambda_{QCD}^2$ because a gradient cancels
dimension of a distance. Another $\Lambda_{QCD}^2$
may be needed to cancel the dimension of
additional integration measure $d^2x^\perp$.
Still, the potential term must vanish when
$\lambda \rightarrow \infty$. If it vanishes as
$1/\lambda^2$, another $\Lambda^2_{QCD}$ may be
expected in the coefficient. There always exists a
possibility to use quark masses as dimensionful
quantities. Together, one may need powers of
$\Lambda_{QCD}$ as high as 6. Such terms may be
hard to establish quickly in any renormalization
group procedure.

In any case, the above results certainly allow one
to entertain the possibility that the
small-$\lambda$ RGPEP picture of hadrons, at
strong coupling, can actually be the one that
corresponds to the AdS picture, including the SW
model \cite{SW}. On the other hand, even if the SW
model does correspond to the Brodsky-Teramond LF
holographic picture with a harmonic oscillator
potential between an active constituent and
spectators (in transverse directions), and even if
these pictures can be related to the gluon
condensation in hadrons, how is RGPEP to explain
emergence of the 5th dimension in the AdS/QCD
analogy on the QCD side?

It has been pointed out that the 5th dimension in
AdS may correspond to a renormalization group
scale in gauge theory \cite{Maldacena1,Polyakov1}.
More recent calculations, such as
\cite{Akhmedov1,Akhmedov2}, also point in this
direction. We observe in this context that RGPEP
provides an opportunity to explicitly introduce a
5th dimension in QCD with a simple physical
interpretation. Namely, Eqs.
(\ref{quantumfieldpsilambda}) and
(\ref{quantumfieldAlambda}) for effective fields
can be rewritten using the observation that
$\lambda$ corresponds to the inverse size of
effective particles. The effective particles
exhibit this size in their interactions; the
effective interactions contain vertex form factors
whose momentum width is determined by the RGPEP
parameter $\lambda$. It is convenient to think
about the size of the effective particles using
variable $s=1/\lambda$.\footnote{ The particle 
size parameter $s$ enters the non-perturbative 
RGPEP equations through replacement of the 
derivative $d/d\lambda^{-4}$ on the left-hand 
side of Eq. (\ref{npRGPEP}) by $d/ds^4$.}

So, instead of labeling the field and particle
operators with $\lambda$, we can label them with
$s$ that corresponds to the size of effective
particles. The effective fields on the LF can be
rewritten according to a rule
\beq
\label{psixs}
\psi_s(x) 
\es  \int \hspace{-15pt}\sum 
[k] \, b_s(k) \, e^{ikx}
\rs
\psi(x,s) \, .
\eeq
In this rule, the 5th argument of the quantum 
field is the RGPEP size of effective particles. 
This size plays the role of a renormalization 
group parameter in RGPEP. 

While the AdS dual picture involves its own
renormalization issues \cite{BoerVerlinde1,
SkenderisLecture2002, Skenderis2004,
KiritsisPart2}, and RGPEP methodology may
eventually find applications in these issues, too,
the 5th dimension of particle size makes QCD {\it
a priori} a 5-dimensional theory, with scale
invariance broken by $\Lambda_{QCD}$. The
structure of such a theory may or may not be
similar to an AdS field theory. On the other hand,
RGPEP is certainly available for studying the
5-dimensional QCD.

The issue that emerges immediately is that in
order to evaluate observables in QCD one in
principle can use just one value of $s$ or
$\lambda$. There is no need for integration over
the 5th coordinate. So why would one integrate
over the 5th dimension in the dual picture? 

The Brodsky-Terramond LF holography for hadronic
form factors suggests that the integration over
the 5-th dimension in AdS field theory corresponds
to integration over relative transverse momentum
of an active hadronic constituent with respect to
spectators. But RGPEP suggests that the LF wave
functions have the form shown in Eq.
(\ref{psihadron}), which is Gaussian in invariant
mass of the hadronic constituents. Integration
over the transverse relative momenta in form
factors amounts to integration over invariant
mass, with a proper rescaling by $\sqrt{x(1-x)}$.
The invariant mass appears in ratio to a square of
the width parameter $\lambda$ or, equivalently, in
product with the square of the effective particle
size parameter $s = 1/\lambda$. The challenge is
to explain how the integration over $\cM$ can be
turned into integration over $\lambda$ or
$s=1/\lambda$.

Although we do not provide here the ultimate
response to this challenge, we offer a qualitative
argument that suggests where one can look for the
solution in future studies. Consider a fixed scale
of an observable, such as $Q$ in a form factor.
For a given $\lambda = 1/s$, one can evaluate the
observable in QCD by integrating wave functions
over $\cM$. But if the wave functions are
functions of $\cM/\lambda = s\cM$, they are
constant on hyperbolas defined by the condition
that $s\cM$ is fixed. One can consider a plane of
variables $\cM$ and $s$ and plot a wave function
in a perpendicular direction, forming a 3-dimensional
plot of the wave function. Select a value of $s
= 1/Q$ and draw a profile of the 3-dimensional
plot for the selected value of $s$ as a function
of $\cM$. Draw another profile as a function of
$s$ for $\cM = Q$. These two profiles are
identical for all wave functions that only depend
on the variables $\cM$ and $s$ through their
product $t$. In such circumstances, integration
over relative motion of quarks of fixed size $Q$
is equivalent to integration over all sizes $s$ of
quarks that have fixed invariant mass $Q$. In QCD,
this pure scaling picture for lightest quarks
must be broken by the scale set through
$\Lambda_{QCD}$.

The small-$\lambda$, or large-$s$ picture of
hadrons in RGPEP appears in accordance with
expectations that there exists a low-energy regime
with a stable strong coupling constant
\cite{BrodskyCoupling1}.\footnote{RGPEP may be
used to seek a precise definition of such coupling
constant using quark-gluon, three-gluon, and
inter-quark potential terms in the corresponding
Hamiltonians, cf. \cite{AFandBS,gluons}.} In the
stabilization region, the gluon condensation
parameter $\varphi_{glue}$ inside hadrons is
considered constant as a function of $\lambda$.
The hadronic wave functions are Gaussian in the
invariant mass with the width provided by quark
masses and $\varphi_{glue}$. The overall result is
that at small $\lambda$ the effective parameters
are frozen at values comparable with
$\Lambda_{QCD}$. On the other hand, when $\lambda$
increases, and $s$ decreases, the Hamiltonian
changes and it is expected to eventually simplify
to the QCD canonical form with counterterms at
$\lambda \rightarrow \infty$, i.e., for point-like
quarks and gluons, $s \rightarrow 0$. Therefore,
the wave functions of hadronic eigensolutions
change with $\lambda$, or $s$. One may expect that
when the wave function width in momentum space
increases and the sizes of effective quarks and 
gluons decrease, the invariant mass dependence 
turns from Gaussian to a power-law behavior at 
large virtuality, perhaps as indicated by 
perturbation theory for exclusive processes 
in a collinear approximation \cite{LepageBrodsky}.

Finally, regarding the Brodsky-Teramond holography
and SW models, we wish to mention that one considers
different values of the SW parameter $\kappa^4$ in
front of $\zeta^2$ in the AdS equations for masses
of mesons (M) and baryons (B) \cite{SWkappa}.
Namely, $\kappa_M \sim 0.54-0.59$ GeV and
$\kappa_B \sim 0.49$ GeV, with 
\beq
{\kappa_M \over \kappa_B } & \sim & 1.15 \pm 0.5 \, .
\eeq
In the reinterpreted gluon condensate image for hadrons, 
assuming a universal value of mass for $u$ and $d$ 
constituent quarks, one has
\beq
{\kappa_M \over \kappa_B }\es \left({8 \over 5}\right)^{1/4} 
\sim 1.125 \, .
\eeq
This agrees quite well with the phenomenological
result.

\section{Conclusion}
\label{c}

Reinterpretation of the gluon condensate in RGPEP
has several implications that matter in theory of
hadrons. Instead of the entire vacuum, the gluons
condense only inside hadrons. If this option were
actually realized in QCD, there would be no need
anymore to construct the quantum vacuum state in
Minkowski space that satisfies fundamental
requirements of invariance with respect to
Poincar\'e transformations, a construction that so
far eluded theoreticians. At the same time, the LF
Hamiltonian formulation of QCD would become free
from any obligation to produce a non-trivial
vacuum that is commonly assumed to provide
physically important expectation values. When
reinterpreted, the same expectation values come
from distribution of matter that is limited to the
interior of hadrons. Simultaneously, it becomes
not clear how one should interpret expectation
values that are measured in lattice formulation of
gauge theories; they might also correspond to the
hadron interior rather than a state of infinite
volume. In any case, the reinterpretation is
available within the RGPEP that provides tools for
a close inspection what actually happens. Thus,
the picture of hadrons that is sketched here in
the context of reinterpretation of gluon
condensate can be viewed as providing an idea how
the CQM may be an approximation to QCD and a
method for verifying this idea.

The constituent picture based on gauge symmetry
dictates Gaussian LF wave functions that are
exponentials of free invariant masses of
constituents. The associated oscillator
frequencies are in agreement with quark models
phenomenology when the gluon condensate inside
hadrons is set to the value found in phenomenology
using QCD sum rules. The corresponding eigenvalue
problems appear naturally in terms of variables in
which AdS/QCD models are developed, especially the
SW model. The ratio of baryon and meson oscillator
frequencies in SW models agrees with the ratio
implied by the reinterpreted gluon condensate in
RGPEP. 

This article provides no comparable evidence for
reinterpretation of the quark condensate. But one
may hope that RGPEP can generate constituent quark
masses when $\lambda$ is lowered to values
comparable with $\Lambda_{QCD}$ and shed this way
some light on the mechanism of absence of chiral
symmetry in hadronic spectrum and the nature of
$\pi$-mesons as nearly Goldstone particles.

The author's opinion that RGPEP provides a method
for verifying the reinterpretation of the gluon
condensate and solving for hadronic structure in
QCD is mainly based on the fact that RGPEP is now
available in both perturbative and
non-perturbative versions, both being boost
invariant. The step beyond perturbation theory
maintaining boost invariance is seen as a chance
for simultaneously taking advantage of knowledge
about hadrons in the constituent picture and in
the parton picture, while RGPEP appears capable of
generating required scale dependence dynamically.
Seen this way, RGPEP provides an operator calculus
for deriving quark and gluon wave functions of
hadrons in the LF Fock space and unify
calculations of the hadron mass spectrum and
partonic distributions inside hadrons. At the same
time, new variables identified in the
reinterpretation of the gluon condensate, relative
momentum three-vectors $\vec k$ in mesons and
$\vec K$ and $\vec Q$ in baryons, provide a frame
of reference in which LF solutions can be
classified in terms of well-known angular momentum
classification of constituent wave functions. The
same transverse variables occur in LF AdS/QCD
holography. Longitudinal variables are new and
require further testing. 

A summary of RGPEP program for verifying the
reinterpretation of gluon condensate is following.
Start with canonical QCD. Set up regulated
$H_{QCD}$ on the LF. Apply RGPEP to lower the
scale parameter $\lambda$ toward $\Lambda_{QCD}$.
Find counterterms in $H_{QCD}$ and evaluate
effective $H_\lambda$. Solve eigenvalue problem of
$H_\lambda$ and see if the gluon-condensate
induced oscillator picture is reproduced by gluon
components in hadronic states. So far, the RGPEP
scheme is known to work only in some crudely
approximate calculations in QCD for heavy
quarkonia. Studies of hadrons built in QCD from
light quarks require breaking an entirely new
ground.

\begin{appendix}

\section{ Universality in RGPEP }
\label{RGPEPuniversalityappendix}

This Appendix explains the RGPEP scheme in which
CQMs can be sought as limited representatives of the
universality class which QCD is thought to belong
to as a quantum field theory with a Hamiltonian
acting in a Hilbert space, cf.
\cite{DiracDeadWood}. The scheme is composed of
two interrelated parts.

In the first part of RGPEP, one constructs a
sequence of rotations $U_\lambda$ for creation and
annihilation operators in order to discover
counterterms in $H_\infty$ and obtain the
effective Hamiltonian $H_\lambda$. The second part
involves solving the eigenvalue problem of
$H_\lambda$. Solutions to the eigenvalue problem
of $H_\lambda$ are needed to fix finite parts of
counterterms in $H_\infty$
\cite{GlazekWilson1,GlazekWilson2}.

The first part of the RGPEP calculation involves
dealing with ultraviolet divergences. Instead of
standard renormalization group procedure based on
Gaussian elimination in linear problems
\cite{Wilson1,Wilson2}, RGPEP is based on a
unitary transformation of field or particle
operators. In terms of matrix elements of a
Hamiltonian, this means that one rotates the basis
states like in the similarity renormalization
group approach (SRG) \cite{GlazekWilson1,GlazekWilson2}.

Formulation of RGPEP in terms of operators for
effective particles organizes otherwise
un-intelligible variety of Hamiltonian matrix
elements in a physically motivated way; one traces
coefficients of specific operators instead of only
calculating matrix elements to which many
operators may contribute. It also provides the
concept of effective quantum field; see Eqs.
(\ref{quantumfieldpsilambda}),
(\ref{quantumfieldAlambda}), and below. The
operator structure of RGPEP results in unitarity
features in the calculated interactions that are
absent in the general formulation of SRG, since
SRG applies also to Hamiltonians that cannot be
written in terms of creation and annihilation
operators. For example, $U_\lambda(b_\infty) =
U_\lambda(b_\lambda)$. The operator structure is
also helpful in demonstrating a connection between
the momentum space formulation of the theory in
terms of quanta (effective particles) and position
space formulation in terms of quantum fields
(effective fields). The interaction terms in
effective Hamiltonian densities are non-local.
RGPEP provides a method for calculating the
effective non-local interactions that correspond
to renormalized canonical theories
\cite{NonlocalH}.

In contrast to the first part of the RGPEP scheme
that is akin to the SRG scheme, the second part
resembles the standard Wilsonian procedure because
it involves Gaussian elimination, in the
mathematical sense as part of solving a linear
problem. However, in sharp distinction from the
standard procedure, the second step of RGPEP does
not involve ultraviolet divergences. These are
eliminated in the first step. 

The RGPEP differential equation for $H_\lambda$
in its generic form reads
\beq 
\label{differentialeq}
{d \over d\lambda} \, \cH_\lambda
\es 
\cF_\lambda \left[ {\cal H}_\lambda \right] \, ,
\eeq 
where ${\cal H}_\lambda$ denotes $H_\lambda(b_\infty)$ 
and $\cF_\lambda$ is a suitable functional.\footnote{E.g., 
see \cite{RGPEPform}. Appendix \ref{NPRGPEP} provides 
a non-perturbative definition of $\cF_\lambda[\cH_\lambda]$. 
One needs to adjust the dimension of parameter $\lambda$ 
to dimensions of the Hamiltonian $\cH_\lambda$ and 
functional $\cF_\lambda \left[ {\cal H}_\lambda \right]$. 
The parameter $\lambda$ is defined in terms of a momentum 
scale that plays the role of width in momentum-space vertex 
form factors, denoted also by $\lambda$, or by the inverse of 
the width. The inverse of $\lambda$ corresponds to the 
size of effective particles that they exhibit in the 
interactions. The inverse is denoted by letter $s$ that 
refers to the word {\it size}.} The solution 
is thus also of the generic form 
\beq 
\label{allsteps} 
{\cal H}_\lambda \es
{\cal H}_\infty + \int_\infty^\lambda d \lambda' \, \cF_{\lambda'}
\left[ {\cal H}_{\lambda'} \right] \, . 
\eeq 
$H_\lambda(b_\lambda)$ is obtained by
replacing $b_\infty$ in ${\cal H}_\lambda$ by
$b_\lambda$. 

Irrespective of the theory one considers, for
$\lambda$ much greater than the regularization
parameter $\Delta^2/\epsilon$ (see Section
\ref{effectiveparticles}) times the largest
allowed number of creation or annihilation
operators in any product of them in the initial
interaction Hamiltonian, $\cF_\lambda[\cH_\lambda]$ is
exponentially close to zero. The reason is that
the regulated interactions are exponentially close
to zero when the invariant mass of interacting
particles is grater than allowed by the
regularization. 

So, one can always introduce some $\Lambda \gg
\Delta/\sqrt{\epsilon}$ and arrange a series of
cutoffs $\lambda_0 = \Lambda$, $\lambda_1 =
\Lambda/2$, $\lambda_2 = \Lambda/4$, etc., until
one reaches $\lambda = \lambda_n = \Lambda/2^n$
for some large $n$.\footnote{Instead of the factor
2, one can choose $e$ or any other similar number.
The exponential spacing of $\lambda$s in the
sequence of cutoffs is introduced for the purpose
of handling logarithmic singularities. In the case
of simple models in which one can easily introduce
an exponential grid in momentum space, such as 
the NR Schr\"odinger equation for one particle in
a $\delta$-function potential, it is most
convenient to use a grid in which the chosen
factor for cutoff reduction, similar to 2 or $e$
above, is also equal to an integer power of the
constant used in the grid, i.e., if $p_k = p_0
a^k$ in the grid, then $\Lambda_n/\Lambda_{n+1} =
a^l$ with an integer $l$ such that $a^l$ equals 2,
$e$, or some other convenient factor chosen for the
cutoff reduction. This relates the value of
$a$ chosen for the exponential grid in momentum
variables with the RGPEP cutoff reduction factor in
terms of an integer $l$.} This is done in analogy
to Refs. \cite{Wilson1, Wilson2} for the purpose
of sequencing the RGPEP procedure of evaluating
$H_\lambda$ into manageable steps that are free
from huge terms resulting from ultraviolet
divergences of the local theory. Namely, step
number $k$ in the sequence that builds up the
integral in Eq. (\ref{allsteps}), contains only
integration over scales between $\lambda_{k+1}$ to
$\lambda_k$. When one evaluates effects of
interactions that allow changes of invariant
masses only in the range between $\lambda_{k+1}$
and $\lambda_k$, the ratio $\lambda_0/\lambda_n
\rightarrow \infty$ does not contribute. Moreover,
one can carry out each and every one of the finite
steps using well-known techniques, such as
perturbation theory or numerical methods. 

Although the sequence one obtains is analogous to
the sequence considered in Wilsonian
renormalization group procedure \cite{Wilson2}, it
differs considerably because no states are
eliminated from the domain of the Hamiltonian. The
RGPEP steps are automatically free from the
dependence on eigenvalues that appear in Gaussian
elimination in matrix notation for eigenvalue
problems and there is no need for an additional
(and in principle determined only up to a nearly
arbitrary rotation of basis) procedure required
for maintaining formal hermiticity of effective
Hamiltonians.\footnote{ In Ref. \cite{Wilson2},
the operation chosen to maintain Hermiticity is
denoted by $R$, Eq. (4.1).}

The ratio of subsequent parameters, $\lambda_{k+1}
/ \lambda_k$ does not depend on $k$ and one can
seek regularity in how successive integrations
transform $H_{\lambda_k}$ to $H_{\lambda_{k+1}}$.
Every element in the resulting sequence of
Hamiltonians can be written in terms of
dimensionless momentum variables $y^\perp =
p^\perp / \lambda_k $ instead of momenta
$p^\perp$. Creation and annihilation operators
also require rescaling in order to keep their
commutation relations in terms of variables
$y^\perp$ independent of the step number $k$. In
principle, such rescaling corresponds to
scale-dependent renormalization constants for
fields in standard perturbative formulation of
quantum field theory. The coefficients
$c_{\lambda_k}$ of products of creation and
annihilation operators $b_{\lambda_k}$ in
$H_{\lambda_k}$ can be studied as functions of
$y^\perp$ and $x$. These functions evolve in $k$.
Their evolution can be classified in terms of
characteristic dominant behavior associated with
universality classes, presumably associated with
fixed points, limit cycles, and even chaotic
behavior. Such analysis is also expected to help
in identifying dominant effects due to small-$x$
cutoff parameter $\delta$ in gauge theories
quantized on the LF (see Section
\ref{effectiveparticles}). No farther comments are
offered here regarding the evaluation of
$H_\lambda$, except for stressing that the
establishment of finite parts of ultraviolet
counterterms eventually requires a reference to
the second part of the RGPEP scheme. 

The second part involves solving for the spectrum
of $H_{\lambda_n}$. This is facilitated using
techniques similar to the ones used in Ref.
\cite{Universality} with the following qualitative
distinction. The Hamiltonian $H_{\lambda_n} = H_{0
\lambda_n} + H_{I \lambda_n}$ is so narrow on
energy (invariant mass) scale (defined by
eigenvalues of $H_{0 \lambda_n}$), that the wave
functions of its eigenstates are typically
exponentially suppressed outside the range of
energies (invariant masses) that are similar in
size to the observable energies (invariant masses)
one is interested in \cite{lcet}. Therefore, one
is no longer in need to solve the severe
ultraviolet renormalization problem when carrying
out calculation that removes residual
(exponentially small) cutoff effects using a
procedure such as in \cite{Universality}. This
suggests, for example, that the non-linear
operators that produce corrections to scaling in
Wegner's sense \cite{WegnerOperators} only lead to
small corrections to the scaling properties (if
any are obtained) that result already from the
first step of the RGPEP scheme.

The expectation of lack of significant sensitivity
of the second part of RGPEP scheme to the actual
values of smallest eigenvalues can be forecast by
analogy with (and it is confirmed in) an elementary
model of a harmonic oscillator with an additional
potential proportional to the fourth power of
distance (a Higgs-like field dynamics in zero
dimensions) \cite{WojcikLicencjat}. Excited states
of the pure oscillator (excited by multiples of
$\hbar \omega$) serve as a model of the Fock space
basis with different numbers of effective
particles, each of mass $m = \hbar \omega$. The
Hamiltonian with the quartic term models
$H_{\lambda_n}$ after the entire first part of
RGPEP scheme is completed. Namely, the interaction
is narrow on the energy scale because it is able
to change the number of particles only by 0, 2, or
4, which means that the size of energy changes due
to interactions is limited from above by $4\hbar
\omega$, or $4m$. The result of numerical
computation of eigenstates in that model
\cite{WojcikLicencjat} is that for obtaining
smallest eigenvalues accurately it is sufficient
to use Gaussian elimination assuming that the
smallest eigenvalues are simply zero. One can come
down this way to cutoffs on energy of basis states
not much greater than about 10 times $\hbar
\omega$, or $10m$. This holds even for quite large
values of the coupling constant that characterizes
the size of the quartic coupling. 

Since the entire RGPEP procedure is in fact a
sequence of successive approximations with an
increasing space of variables when regularization
is lifted, both parts of it, the rotation and
solution, need to be iterated, in the sense of
Wilsonian triangle of renormalization, until a
stable result is established. The iteration
involves finite parts of counterterms. They are
fixed by comparison with data in the solution
part. But the finite parts influence the rotation
in the first part. This is why the two parts of
the procedure are interrelated and neither the
part using unitary rotation of particle operators
nor the solution part with elimination of states
can be distinguished as determining the path of
successive approximations for the Hamiltonian on
the triangle. As a result, the evolution of a
Hamiltonian on the triangle is neither purely
unitary nor a plain solving of an {\it a priori}
specified eigenvalue problem. In other words, the
ultimate self-consistent solution to the triangle
of renormalization defines the theory one is
interested in solving to explain observables.

RGPEP requires extensive studies in order to
verify if it can produce a universal low-energy
theory that explains success of the CQMs starting
from the Lagrangian for QCD. In addition, all of
these models could also be subjected to RGPEP or
SRG as long as they can be written in a
Hamiltonian form.\footnote{This is not immediately 
obvious in the case of approaches based on the 
Dyson-Schwinger and Bethe-Salpeter equations.} 
Therefore, there arises a question if all CQMs that 
do reproduce a considerable number of observables for
smallest-mass states in hadronic spectrum actually
converge on a single, universal model with some
specific shape of quark potential. Then, the
question would be if a universal model, if it is
obtained, matches the effective theory obtained
from QCD using RGPEP.

It should be stressed, however, that the procedure
outlined here is not limited to RGPEP for quarks
and gluons in LF QCD. It can be applied to studies
of universality in all Hamiltonians that can be
expressed in terms of creation and annihilation
operators acting in some Fock space and, even more
generally, to all Hamiltonians that require
renormalization and can be subject to the SRG
procedure \cite{GlazekWilson1,GlazekWilson2}. In
particular, one can apply the same procedure when
using beautiful Wegner's flow equation
\cite{Wegner, WegnerReview,Kehrein,Bach}, and the
latter can also be suitably altered in order to
improve weak-coupling expansion that may apply in
the case of asymptotically free theories
\cite{AlteredWegnerPRD,AlteredWegnerAPP}.

\section{ Calculations of $W$ }
\label{AppendixW}

Assuming that $\Lambda_{QCD}$ formally tends to 0
in comparison to quark masses, so that $g_\lambda$
can be treated as extremely small when $\lambda$
is comparable with the masses, $W$ could be
calculated in RGPEP using perturbative formulae
given in Section \ref{W1inQCD} and expanding LF
quantum fields into bare creation and annihilation
operators. In this case, the dominant term besides
1 is $g_\lambda W_1$. However, for a realistic
value of $\Lambda_{QCD}$, which is much larger
than $u$ and $d$ quark mass parameters in the SM,
and for $\lambda \sim \lambda_c$ that is expected
to correspond to the CQMs, perturbative expression
cannot be considered reliable on the basis of
smallness of $g_\lambda$. Nevertheless,
perturbative calculations described here
illustrate the structure of $W$ that carries over
to values of $\lambda$ comparable with $\lambda_c$
for as long as ${\cal H}_\infty$ as an initial
condition in RGPEP equations is replaced by the
corresponding ${\cal H}_\lambda$ and $\lambda$ is
sufficiently near in magnitude to $\lambda_c$ so
that $W$ is not violently different from 1. The
key argument is that the strength of the
interaction is not fully determined by $g_\lambda$
alone. Namely, when the RGPEP form factor $f_\lambda$ 
is narrow as function of momentum variables, the net
strength of the interaction is suppressed and the
interaction can be weak even if $g_\lambda$ is
sizable. Thus, the calculation given below
illustrates not only how the lowest-order perturbative
$W$ looks like in the case of small $g_\lambda$
but also how the structure of $W$ is related to
the structure of ${\cal H}_\lambda$ that is weak
because $\lambda$ is small, even though ${\cal
H}_\lambda$ is expected to significantly differ
from ${\cal H}_\infty$ when $\lambda$ approaches
$\lambda_c \sim \Lambda_{QCD}$ for light quarks.

For the canonical quark field, we have on the LF
\beq
\label{quantumfieldpsi}
\psi \es \sum_{\sigma c f} \int [k] 
  \left[ \chi_c u_{fk\sigma} b_{k\sigma c f} e^{-ikx} + 
  \chi_c v_{fk\sigma} d^\dagger_{k\sigma c f} e^{ikx}
  \right]_{x^+=0} \, .
\eeq
The momentum and spinor notation involves
(e.g., see Ref. \cite{GlazekMlynik})
\beq
{[}k{]} \es \theta(k^+) dk^+ d^2 k^\perp/(16\pi^3 k^+) \, , \\
u_{fk\sigma} \es B(k,m_f) u_{f\sigma} \, , \\
v_{fk\sigma} \es B(k,m_f) v_{f\sigma} \, , \\
v_{f\sigma}  \es C u^*_{f\sigma} 
\, , \\
u_{f\sigma}  \es
\sqrt{2m_f} \,\, \zeta_f \, \chi_\sigma \, ,
\eeq
where $\zeta_f$ is a normalized vector in flavor space, 
$\chi_\sigma$ is a spinor that corresponds to one of 
two spin states of a fermion at rest, both normalized 
to 1 and orthogonal to each other, $C$ is the charge 
conjugation matrix,
\beq
B(k,m)       \es {1\over \sqrt{k^+ m}}
  [ \Lambda_+ k^+ + \Lambda_- (m + k^\perp \alpha^\perp)] 
\eeq
is the LF boost matrix, and $\Lambda_\pm = {1\over 2}\gamma_0 \gamma^\pm$. 
For the canonical gluon field, we have
\beq
\label{quantumfieldA}
A^\mu \es \sum_{\sigma c} \int [k] \left[ t^c \varepsilon^\mu_{k\sigma} 
  a_{k\sigma c} e^{-ikx} + t^c \varepsilon^{\mu *}_{k\sigma} 
  a^\dagger_{k\sigma c} e^{ikx}\right]_{x^+=0} \, , \\
\varepsilon^\mu_{k\sigma} \es (\varepsilon^+_{k\sigma}=0,
  \varepsilon^-_{k\sigma} \rs 2k^\perp \varepsilon^\perp_\sigma/k^+,
  \varepsilon^\perp_\sigma) \, .
\eeq
The same expansions can be used for effective
quantum fields in which creation and annihilation
operators correspond to some scale $\lambda$
\cite{NonlocalH}\footnote{In order to relate the above
notation to the one introduced in Ref. \cite{NonlocalH},
one needs to treat $d^\dagger$ with positive $k^+$
as $b$ with negative $k^+$, and $a^\dagger$ with
positive $k^+$ as $a$ with negative $k^+$, taking
into account the well-known constraint relations
for fermion field $\psi_- = \Lambda_- \psi$ and
gauge boson fields $A^-$ \cite{BjorkenKogutSoper,Yan1,Yan2,
spinors2,Wilsonetal}.}, see Eqs. (\ref{quantumfieldpsilambda})
and (\ref{quantumfieldAlambda}).

To evaluate $W_1$ using Eq. (\ref{W1example}), one first 
needs to evaluate ${\cal H}_{I\infty 1}$ and ${\cal H}_0 
= {\cal H}_{0 \infty}$. According to Eq. (\ref{calHinfty}),
\beq
{\cal H}_{0\infty} \es \int dx^- d^2 x^\perp \,
h_{0\infty} \, , \\
{\cal H}_{I\infty 1} \es \int dx^- d^2 x^\perp \,
h_{I \infty 1} \, .
\eeq
Using Eq. (\ref{hinfty0}) and the field expansions, one obtains 
the normal-ordered expression for ${\cal H}_0$,
\beq
{\cal H}_0 \es
\sum_{\sigma c f} \int [k] \, {k^{\perp \, 2} +
m^2_f \over k^+} \,
  \left[b^\dagger_{k\sigma c f}b_{k\sigma c f} + 
  d^\dagger_{k\sigma c f}d_{k\sigma c f} \right]
\nn
\ps
\sum_{\sigma c} \int [k] \, {k^{\perp \, 2} \over k^+} \,  
  a^\dagger_{k\sigma c}a_{k\sigma c} \, .
\eeq
Similarly, using Eq. (\ref{hinfty1}), one obtains
${\cal H}_{I \infty 1}$,
\beq
\label{resultH1}
{\cal H}_{I \infty 1} 
\es 
\sum_{123}\int[123] \, 2(2\pi)^3 \delta^3( P_c - P_a)\, 
r_{\Delta\delta}(1,2,3) \,
\left[\,\bar u_2 \not\!\varepsilon_1^* u_3 \,
t^1_{23} \, b^\dagger_2 a^\dagger_1 b_3 \right. 
\nn
\ms
\left.
\bar v_3 \not\!\varepsilon_1^* v_2 \, \cdot t^1_{32} \cdot
\, d^\dagger_2 a^\dagger_1 d_3 
+
\bar u_1 \not\!\varepsilon_3 v_2 \, \cdot t^3_{12} \cdot
\, b^\dagger_1 d^\dagger_2 a_3 
\right.
\nn
\ps
\left.
Y_{123}\, a^\dagger_1 a^\dagger_2 a_3 + h.c. \right]
\eeq
where $P_c$ denotes the total momentum of created
particles, $P_a$ denotes the total momentum of annihilated
particles, $r_{\Delta \delta}(1,2,3)$ denotes the
regularization factors described in Section 
\ref{effectiveparticles}, $t^a_{ij} =\chi^\dagger_{ic} 
t^a \chi_{jc}$, 
\beq
Y_{123} \es i f^{c_1 c_2 c_3} \left[ 
  \varepsilon_1^*\varepsilon_2^* \cdot \varepsilon_3\kappa 
  - 
  \varepsilon_1^*\varepsilon_3 \cdot \varepsilon_2^*\kappa {1\over x_{2/3}} 
  - 
  \varepsilon_2^*\varepsilon_3 \cdot \varepsilon_1^*\kappa {1\over x_{1/3}} 
  \right] ,
\eeq
with $\varepsilon \equiv \varepsilon^\perp$ and 
$\kappa = k_1^\perp - k_1^+ k^\perp_3/k_3^+$, and
$h.c.$ denotes Hermitian conjugation of the 4
explicitly written terms in the square bracket
in Eq. (\ref{resultH1}).

The result for ${\cal H}_{I \infty 1}$ needed in Eq. 
(\ref{W1example}), can be written in an abbreviated 
form 
\beq
{\cal H}_{I \infty 1} 
\es 
\sum_{123}\int[123] \, 2(2\pi)^3 \delta^3( P_c - P_a)\, 
r_{\Delta\delta}(1,2,3) \,
\left\{1,2,3 \right\} \, ,
\eeq
where $\left\{1,2,3 \right\} $ denotes the $4 + h.c.$ terms
with $b_\infty$, $d_\infty$, and $a_\infty$ replaced by
$b_\lambda$, $d_\lambda$, and $a_\lambda$, respectively. 
Using this notation, the exact perturbative result for 
$W_1$ in RGPEP reads
\beq
\label{W1exact}
W_1 
\es 
\sum_{123}\int[123] \, 2(2\pi)^3 \delta^3( P_c - P_a)\, 
r_{\Delta\delta}(1,2,3) 
\nn
\ts
{ f_\lambda - f_{\lambda_c} \over P_a^- - P_c^- } \,
\left\{1,2,3 \right\} \, ,
\eeq
where $P_a^-$ and $P_c^-$ are eigenvalues of ${\cal H}_0$ 
and $f$ denotes the RGPEP form factor,
\beq
f_\lambda \es e^{ -(P_c^2 - P_a^2)^2/\lambda^4} \, , 
\eeq
in which the squares of invariant masses of
created particles, ${\cal M}^2_c = P_c^2$, and
annihilated particles, ${\cal M}^2_a = P_a^2$, 
are evaluated using eigenvalues of ${\cal H}_0$ 
associated with the particle momentum components
$k_i^+$ and $k_i^\perp$ for $i = 1, 2, 3$ in 
every term \cite{NonlocalH}.

The structure of $\left\{1,2,3 \right\}$ implies
that $W_1$ can replace a quark by two particles:
a quark and a gluon. It can replace a gluon by 
a quark-anti-quark pair, or by two gluons. 
Reversed changes are also possible.

According to Eq. (\ref{W2}), evaluation of $W_2$ 
involves action of ${\cal H}_{I\infty 1}$ twice 
and ${\cal H}_{I \infty2}$ once. This means that 
$W_2$ can replace one virtual particle by 3, or
vice versa, change the number of virtual particles
by 1, or do not change their number at all but 
only alter their individual momenta and other 
quantum numbers as dictated by ${\cal H}_{I\infty}$.

When one evaluates $W$ for $\lambda$ and
$\lambda_c$ using expansion in powers of
$g_\lambda$, creation and annihilation operators
in Eq. (\ref{W1exact}) being those corresponding
to $\lambda$, the coefficient $(f_\lambda -
f_{\lambda_c}) /(P_a^- - P_c^-)$ only appears with
first power of $g_\lambda$. Higher powers involve
different coefficients and the entire sum of the
series in powers of $g_\lambda$ is not known.
Moreover, a complete answer may contain dependence
on $\Lambda_{QCD}$ that perturbation theory cannot
identify. Therefore, the question is what
structure of $W$ one can expect for $\lambda \sim
\lambda_c \sim \Lambda_{QCD}$. In particular,
seeking a connection between CQMs and QCD, one
needs to estimate the structure of $W$ that is
responsible for changing the number of effective
particles, since CQMs do not include such
interactions.

When $\lambda$ is near $\lambda_c$, one knows that
$W$ is near 1 irrespective of the size of both
$\lambda$s and $g_\lambda$, because $W$ is 1 when
$\lambda = \lambda_c$. However, the deviation of
$W$ from 1 is not the same for $\lambda \sim
\lambda_c \sim \Lambda_{QCD}$ as in the
perturbative case discussed above for two
$\lambda$s that are much greater than
$\Lambda_{QCD}$. The significant change in $W$ in
comparison to the perturbative result for large
$\lambda$s comes from the difference between
${\cal H}_\infty$ and full solution for ${\cal
H}_\lambda$. Nevertheless, if one assumes certain
particle-number changing interaction in ${\cal
H}_\lambda$ for $\lambda$ near $\lambda_c$, $W$
will put this structure of ${\cal H}_\lambda$ to
action in a similar way to the one described
above. The reason is not the smallness of
$g_\lambda$ but the presence of the form factors
$f_\lambda$. They weaken the interaction by
limiting the range of momenta that interacting
particles may have. Thus, one can look at a
candidate for ${\cal H}_\lambda$ for some value of
$\lambda$ near $\lambda_c$ as an initial condition
for integration over a relatively small range of
$\lambda$ near $\lambda_c$ and treat the whole
interaction that changes the number of particles
as weak even if $g_\lambda$ appears large.

The weakness of particle-number changing
interactions with small $\lambda$ is reinforced by
the particle masses as described in Section
\ref{effectiveparticles}. Both quark, $m$, and
gluon, $m_g$, mass parameters in the region of
$\lambda \sim \lambda_c$ may be sizable. Gluon
mass terms for effective gluons are not excluded
by local gauge invariance of Lagrangian for QCD
because the interaction terms in LF QCD
Hamiltonians with finite $\lambda$ are not local
and their non-locality is not of canonical type.
Since regularization of LF QCD Hamiltonian
requires gluon mass counterterms and the latter
have finite parts that are expected to depend on
$\Lambda_{QCD}$, perturbation theory for
scattering in femtouniverse \cite{femtouniverse}
cannot be used to argue that effective gluons
cannot have sizable mass terms for $\lambda \sim
\lambda_c$. The spectrum of masses of lightest
hadrons, which is quite sparse in comparison to
the QED near-threshold atomic spectra with
massless photons, suggests instead that there is a
considerable mass gap involved in excitation of
gluon degrees of freedom. Since QCD is assumed to
describe these data, it seems reasonable to expect
that RGPEP equations in QCD lead to massive
effective gluons, instead of massless ones. 

As an illustration of what one might expect of $W$
in QCD in the range of $\lambda \sim \lambda_c
\sim \Lambda_{QCD}$ when one reasons by analogy
with the perturbative solution at large $\lambda$,
consider the structure of local three-gluon
interaction term that originates from the
gauge-invariant Lagrangian density $- Tr F^{\mu
\nu} F_{\mu \nu} /2$ in canonical theory. The
interaction causes splitting of a bare gluon into
two bare gluons. Its structure is given in Eq.
(\ref{resultH1}), as the term with $Y_{123}$. In
distinction from this local canonical term (in
which $\lambda = \infty$), the corresponding term
in an effective Hamiltonian at small $\lambda \sim
\Lambda_{QCD}$, is non-local \cite{NonlocalH}.
Suppose the complete non-local three-gluon vertex
includes the same structure $Y_{123}$ and a vertex
factor $V_\lambda(1,2,3)$ times the form factor
$f_\lambda$. Such structure is known to emerge
through order $g_\lambda^3$ \cite{gluons}. The
corresponding 1-to-2-gluon term in $W$ of lowest
order in the effective interaction reads 
\beq 
\label{WY1ra}
W_{Y1 \, 21} \es \sum_{123}\int[123] \, 2(2\pi)^3
\delta^3( P_c - P_a)\, { f_\lambda - f_{\lambda_c}
\over P_a^- - P_c^- } \nt \label{WY1} Y_{123} \,
V_\lambda(1,2,3) \, \, a^\dagger_{\lambda 1}
a^\dagger_{\lambda 2} a_{\lambda 3} \, , 
\eeq
where $V_\lambda(1,2,3)$ plays the role of initial
condition, instead of $V_\infty(1,2,3)$ that
provided the initial condition in ${\cal
H}_\infty$. Note that now the regularization
factor $r_{\Delta\delta}(1,2,3)$ is absent. This
follows since the form factors $f_\lambda$ and
$f_{\lambda_c}$ entirely eliminate the ultraviolet
divergences and they eliminate the small-$x$ 
divergences when one adopts the assumption that 
effective gluons are heavy \cite{Wilsonetal}. 
In these circumstances, the regularization 
factors are immaterial.

Suppose $\lambda_c$ is smaller than the gluon mass
$m_g$ at $\lambda_c$. In this case, for $\lambda \sim 
\lambda_c$, one can approximate $P_a^-$ by $m_g^2
/p_3^+$ and $P_c^-$ by $4 m_g^2/p_3^+$ in the 
difference in denominator while the form factors can 
be approximated using (see \cite{NonlocalH} for
details)
\beq
f_\lambda
\es
e^{ - \left[ {\cal M}_{12}^2 - m_g^2
\right]^2 / \lambda^4 } \, ,
\eeq
with ${\cal M}^2_{12} = 4(m_g^2 + \vec k\,^2)$,  
rewritten as 
\beq
f_\lambda \es e^{-(3m_g^2/\lambda^2)^2} \,
e^{ - { \vec k\,^2 \, + \,\, 3m_g^2/2 \over (\lambda/2)^2 } \, {
\vec k\,^2 \over (\lambda/2)^2 } } \, .
\eeq
For $\lambda$ smaller than $m_g$, the relative momentum 
$|\vec k\,|$ of gluons 1 and 2 is smaller than 
$\lambda/2 < m_g/2$ and the form factor can 
be very well approximated by Gaussian
\beq
f_\lambda \es e^{-9m_g^4/\lambda^4} \, 
e^{ - 24 m_g^2 \vec k\,^2 /\lambda^4 }  \, .
\eeq
The exponential in front causes the form factor to
vanish when $\lambda$ is much smaller that $m_g$,
as discussed in Section \ref{effectiveparticles}.
This exponential is a part of the mechanism by
which the presence of RGPEP form factors
suppresses interactions that change the number of
effective particles. The other part of the
mechanism is the Gaussian factor whose width
decreases when $\lambda$ decreases. This factor
causes the range of interaction in momentum space
to decrease and thus to diminish the net effect of
it. For the form factor suppression mechanism to
work, the coupling constant $g_\lambda$ must not
blow up to huge values and compensate for the
smallness of the form factor. However, such
compensation should not be expected in QCD if the
theory is to explain the phenomenological success
of the CQMs.

For $\lambda_c$ much smaller than $m_g$, one could 
literally set $f_{\lambda_c}$ to 0 in Eq. (\ref{WY1}). 
The resulting estimate for the structure of $W_{Y1 \, 21}$ 
reads
\beq
\label{WY1a}
W_{Y1 \,  21} 
\es 
e^{-9m_g^4/\lambda^4} \,
\sum_{123}\int[123] \, 2(2\pi)^3 \delta^3( P_{12} - p_3)\, 
{ e^{ - 24 m_g^2 \vec k\,^2 /\lambda^4 } \over 3m_g^2/p_3^+ } \,
\nt
\label{WY2}
Y_{123} \, V_\lambda(1,2,3) 
\, a^\dagger_{\lambda 1} a^\dagger_{\lambda 2} a_{\lambda 3} \, .
\eeq
This term in $W_1$ replaces a gluon with a pair of
gluons. The relative motion of the two emerging
gluons corresponds to a Gaussian wave function of 
the form $e^{ - c \vec k\,^2 /\lambda^2 }$
with $c = 24 m_g^2 /\lambda^2 $. Assuming that 
$\lambda$ is comparable with $m_g \sim 1$ GeV,
the width of the Gaussian wave function is about
$\lambda/5 \sim 200 $ MeV, which is not
unreasonable. Nevertheless, one has to remember
that the full vertex contains also the function 
$V_\lambda(1,2,3)$ which causes that the
non-locality of $W_{Y1 \, 21}$ is not determined solely
by the Gaussian factor in Eq. (\ref{WY1a}).

Another illustration is provided by consideration
of the first term in $\{1,2,3\}$, i.e., the term
that replaces a quark of scale $\lambda$ by a
quark and a gluon of the same scale. The only new
elements in the analysis of this term, in
comparison to the analysis described above in the
case of gluons, are different values of masses and
spinor factors instead of $Y$. However, the
conclusion is similar: $W$ changes a quark into a
quark and a gluon. 

The larger the difference between $\lambda$ and
$\lambda_c$, the more important particle-changing
interactions in $H_\lambda$. The full solution for
$W$ may replace a state of two or three
constituent quarks by a superposition of states
with many different numbers of gluons (and
additional quark-anti-quark pairs created from
additional gluons) at $\lambda$, as dictated by
the interaction terms in $H_\lambda$. 

The perturbative calculation of $W$, and calculations
for small difference between $\lambda$ and $\lambda_c$ 
using assumed simple picture at $\lambda_c$ and guesses
concerning interactions at $\lambda \gtrsim \lambda_c$, 
can be replaced by non-perturbative solutions of RGPEP 
equations, see Appendix \ref{NPRGPEP}. 

\newpage

\section{ RGPEP beyond perturbation theory }
\label{NPRGPEP}

For the principles of renormalization group
procedure for Hamiltonians \cite{GlazekWilson1} to
apply in a boost-invariant LF dynamics, the
right-hand side in Eq. (\ref{differentialeq}) must
guarantee that matrix elements of $H_\lambda$
vanish when changes of the invariant masses exceed
$\lambda$ and become comparable with the invariant
masses themselves. This feature will be called
narrowness in invariant mass of interacting
effective particles of width $\lambda$, or just
narrowness. One way of securing narrowness of width
$\lambda$ in perturbation theory is to work with
vertex form factors in RGPEP as described in
Appendix \ref{AppendixW}. While this option is
welcome in perturbative calculations
\cite{AlteredWegnerAPP}, it involves a derivative
of the Hamiltonian in the definition of functional
$\cF_\lambda$ and the derivative is not easy to
calculate outside perturbation theory. The
perturbative approach of Ref. \cite{GlazekWilson2}
defines differential equations in which the
functional $\cF_\lambda$ guarantees narrowness of
width $\lambda$ without depending on derivatives
of the Hamiltonian on the right-hand side, but the
generator of the transformation is again defined
recursively in terms of a commutator of the
generator with the Hamiltonian. The recursion
formally secures an expansion to all orders in the
coupling constants, but it is difficult to solve
beyond perturbation theory. 

Non-perturbative formulation of RGPEP for operators in 
the Fock space can be defined in the form of a double commutator 
\cite{doublecommutator1,doublecommutator2,doublecommutator3}
equation that reads
\beq 
\label{npRGPEP}
{d \over d\lambda^{-4}} \, \cH_\lambda \es
\left[ [ \cH_{0\lambda}, \cH_{P\lambda} ], 
\cH_\lambda \right] \, .
\eeq 
The Hamiltonian $H_\lambda$ is obtained when 
$b_\infty$ in $\cH_\lambda$ is replaced by 
$b_\lambda$. The required transformation $U_\lambda$ 
is thus meant to be always incorporated with 
accuracy dictated by the accuracy of solving 
Eq. (\ref{npRGPEP}). Strictly speaking, nothing 
is known yet about the accuracy one can actually 
achieve this way beyond perturbative RGPEP in QCD.

The structure of Eq. (\ref{npRGPEP}) resembles the
beautiful equation for Hamiltonian matrices that
Wegner proposed for solving theoretical problems
in condensed matter physics
\cite{Wegner,WegnerReview,Kehrein}.\footnote{ To
avoid a misunderstanding, it should be clarified
that in his original work Wegner did not relate
his equation to renormalization group procedures
in condensed matter physics or quantum field
theory.} In Wegner's proposal, $\cH_{0\lambda}$ is
a diagonal part of the Hamiltonian matrix,
$\cH_{I\lambda}$ is the off-diagonal part, and
$\cH_{P\lambda}$ is the matrix $\cH_{I\lambda}$
itself. Attempts have been made to apply the
Wegner equation to Hamiltonian matrices obtained
from quantum field theory for the purpose of
eliminating matrix elements that involve a change
in the number of virtual particles. Such matrix
elements are found in the initial conditions
set using interaction terms from canonical
Hamiltonians in quantum field theory. The
Tamm-Dancoff truncation to a few Fock components 
was used to define a sufficiently small Hamiltonian 
matrix for using Wegner's equation \cite{GubankovaWegner,
GubankovaPauliWegner, GubankovaJiCotanch1,
GubankovaJiCotanch2}.

Wegner's equation with matrix $\cH_{P\lambda} =
\cH_{I\lambda}$ implies that the matrix
$\cH_\lambda$ with small $\lambda$ is narrow with
a width order $\lambda$ on energy scale (it would
be $P^-$ scale in LF dynamics). However, Wegner's
equation for Hamiltonian matrices does not agree
with kinematical LF symmetries, including the Lorentz
boosts that are essential for explaining a
connection between the constituent picture and the
parton model picture of hadrons. Also, one desires
an equation for a Hamiltonian operator that in
principle may act in the entire Fock space rather
than only in a severely truncated space in a 
Tamm-Dancoff approach. These two issues will be 
addressed below by defining the operator $\cH_{P\lambda}$ 
and showing that the resulting equation leads to 
narrowness of $\cH_\lambda$ that respects all 
kinematical LF symmetries and is narrow in the 
entire Fock space.

\subsection{ Boost invariance }
\label{boosts}

The origin of the difficulty with LF boost
symmetry in Wegner's equation is that the
left-hand side in his equation is supposed to be a
derivative of a Hamiltonian with respect to a
parameter, while the right-hand side is tri-linear
in a Hamiltonian, an object clearly depending on a
frame of reference in a different way than a
Hamiltonian does. LF power counting
\cite{Wilsonetal} for Hamiltonian terms suggests
that a straightforward application of Wegner's
equation in the SRG scheme may create a host of
complex counterterms required for restoring
kinematical LF symmetries. An attempt to cure the
situation was undertaken \cite{AllenPerry,
KylinAllenPerry} by assuming a double-commutator
equation for matrix elements of the invariant mass
squared rather than a Hamiltonian. It was assumed
that the LF boost invariance could be maintained
because powers of the invariant mass are invariant
with respect to boosts if the mass itself is.
However, invariant masses of Fock states depend on
spectators, resulting in effective interactions
that depend on spectators. This means that the
resulting effective interactions require extra
care to recover the cluster decomposition
principle \cite{cluster} and it is not clear that
they can help in literally solving quantum field
theory where this principle is expected to be
valid in effective theories, including effects of
a complete renormalization procedure. 

The issue of LF symmetries in RGPEP was considered
before \cite{RGPEP}. Here, the operator ${\cal
H}_{P\lambda}$ is defined using the Hamiltonian
${\cal H}_\lambda$. Assuming that ${\cal
H}_\lambda$ conserves momentum and is of the form 
\beq
\label{Hstructure} 
{\cal H}_\lambda \es
\sum_{n=2}^\infty \, \int \hspace{-15pt}\sum \,
[1...n] \, h_\lambda(1,...,n) \, \, q^\dagger_1
\cdot \cdot \cdot q_n \, , 
\eeq 
where $q$ denotes annihilation operators $b_\infty$, 
the operator $\cH_{P\lambda}$ is defined to be of 
the form 
\beq
\label{HPstructure} 
{\cal H}_{P\lambda} \es
\sum_{n=2}^\infty \, \int \hspace{-15pt}\sum \,
[1...n] \, h_\lambda(1,...,n) \,\left( {1 \over
2}\sum_{k=1}^n p_k^+ \right)^2 \, q^\dagger_1
\cdot \cdot \cdot q_n \, , 
\eeq
and the symbol $\int \hspace{-10pt}\Sigma$ denotes
integration over momenta and summation over
discrete quantum numbers that label creation and
annihilation operators while $p_k^+$ denotes the
$+$-component of momentum that labels the creation
or annihilation operator number $k$. In words,
${\cal H}_{P\lambda}$ differs from ${\cal
H}_\lambda$ by multiplication of its terms by the
square of $+$-component of total momentum carried
by the particles that enter a term, which is the same 
as the $+$-component of total momentum carried
by the particles that leave the term. In the
resulting Eq. (\ref{npRGPEP}), both sides behave
in the same way with respect to operations of
kinematical LF symmetries, and the RGPEP width
parameter $\lambda$ is invariant with respect to
boosts of kinematical LF symmetry.

The operator ${\cal H}_{0\lambda}$ in Eq.
(\ref{npRGPEP}) is now set equal to a free-particle
Hamiltonian $\cH_0$, i.e., an operator built from
products of one creation operator and one
annihilation operator per particle species. The
notation $\cH_{0\lambda} = \cH_0$ also indicates 
that $\cH_{0\lambda}$ in Eq. (\ref{npRGPEP}) is
chosen here to be independent of $\lambda$.\footnote{This 
assumption simplifies the analysis that follows.} It 
is a diagonal operator in the Fock space spanned by
states created by operators $b_\infty$ from the LF
vacuum state.  Diagonal matrix elements of $\cH_0$ 
are not equal to the diagonal matrix elements of the 
full Hamiltonian. Typical Hamiltonians include 
interactions that contribute to diagonal matrix 
elements.\footnote{It is possible to include 
interactions in $\cH_{0\lambda}$ in Eq. (\ref{npRGPEP})
in the form of mass effects (self-interactions) and 
beyond (potentials), but these options are ignored 
here for simplicity.}

\subsection{ Non-perturbative narrowness in the Fock space }
\label{NonperturbativeNarrowness}

Eq. (\ref{npRGPEP}) is designed for operators that
can act in the entire Fock space (action on every
state is well-defined). Eq. (\ref{npRGPEP}) with a
constant $\cH_0$ renders $\lambda$-dependent
interaction terms that die out when the change of
the invariant mass squared (evaluated using the
kinematical momenta and eigenvalues of $\cH_0$) of
particles involved in an interaction exceeds
$\lambda$ (spectators do not influence Hamiltonian
interaction terms). The narrowness feature is
obtained through a universal mechanism for double
commutator evolution equations. The factors of
$P^{+\,2}$ in $\cH_{P\lambda}$ do not change the
general mechanism.

We start from the observation that Eq.
(\ref{npRGPEP}) can be solved using a method of
successive approximations. An approximation is
defined by how many terms are kept in the
Hamiltonian. In principle, a Hamiltonian with
finite $\lambda$ contains infinitely many terms.
The terms that are kept in an approximation can be
defined by the condition that they contain no more
than a prescribed number $N$ of creation and
annihilation operators in a product. The number
$N$ labels the approximation. Terms that contain
more operators in a product are ignored in the
approximation. Successive approximations are
labeled by increasing numbers $N$.

Suppose one wants $\cH$ to contain at most $N$
creators and at most $N$
annihilators.\footnote{One can limit the number of
particle operators for different species of
particles differently. In this case $N$ becomes a
vector with natural components corresponding to
species.} Such Hamiltonian will be denoted by 
$\cH_{\lambda N}$. Using notation corresponding to Eq.
(\ref{Hstructure}), this means that
\beq
\label{Hstructure1} 
{\cal H}_{\lambda N} \es
\sum_{n_c = 1}^N \sum_{n_a = 1}^N \, 
\int \hspace{-15pt}\sum \,
[1...n_c+n_a] \, h_{\lambda N n_c n_a}(1,...,n_c+n_a) \, \, t_{n_c n_a} \, ,
\eeq 
where 
\beq
\label{tac}
t_{n_c n_a} \es \prod_{k=1}^{n_c} q^\dagger_k \prod_{l= n_c+1}^{n_c+n_a} q_l \, ,
\eeq
$n_c$ is the number of creation operators, 
$n_a$ is the number of annihilation operators,
and subscripts denote also all relevant quantum 
numbers of the operators they label.

Suppose one wants to know the terms in $\cH_{\lambda N}$ 
that contribute to the dynamics of a physical state 
that certainly contains a specified Fock component 
$| \psi \rangle $ of the form 
\beq
\label{psizero}
|\psi \rangle \es 
\int \hspace{-15pt}\sum \,
[1...n] \, \psi_P(1,...,n) \, \,  
\prod_{k=1}^{n} q^\dagger_k \, |0 \rangle \, ,
\eeq
where $P$ denotes the fixed total kinematical momentum 
of the state and otherwise the wave function 
$\psi_P(1,...,n)$ is not known. For example, in the
case of mesons one may select the state $|\psi \rangle$
that contains a quark and an anti-quark, and in the case 
of a baryon a state $|\psi \rangle$ that contains 
three quarks. But one could also consider a state
that contains 1182 quarks and many gluons and 
quark-anti-quark pairs in order to describe a 
collision of two nuclei of gold neglecting all 
interactions but the strong.

Having specified the set of operators that are
included in $\cH_{\lambda N}$, with $n_c \le N$, $n_a
\le N$, and unknown coefficients $h_{\lambda N n_c
n_a}$, one can enumerate forms of all possible
states that can be generated from $|\psi \rangle$
by action of $\cH_{\lambda N}$ on it once. All these
states are eigenstates of the total kinematical
momentum operator with one and the same eigenvalue
$P$ that characterizes $|\psi \rangle$. Knowing
the set of states that are obtained from $| \psi
\rangle$ by acting on it with $\cH_{\lambda N}$ once,
one can construct the set of all states that can
be obtained by acting with $\cH_{\lambda N}$ on $|
\psi \rangle$ twice, and so on. 

We introduce the set of states, denoted by $R$,
that can be generated from $|\psi\rangle$ by
acting with $\cH_{\lambda N}$ on it $\tau$ times
and are normalized in a limit of infinite volume.
Then we introduce a minimal subspace in the Fock
space (a space of smallest possible basis) that is
sufficient to build the set $R$. This subspace in
the Fock space is denoted by $R_{\tau \psi N}$.
For example, if one starts with a state of 3
quarks, action just one time by the $\cH_{\lambda
N}$ that is equal to regulated canonical
Hamiltonian of LF QCD, $\tau = 1$ and $N=3$, on
$|3q\rangle$ produces states with 3 quarks, 3
quarks and a gluon, 3 quarks and a
quark-anti-quark pair, and 3 quarks and 2 gluons.
The space $R_{1\, 3q \, 3}$ is built from the Fock
space basis states that are needed to construct
states with these particles. Action twice, $\tau =
2$, generates additional gluons and
quark-anti-quark pairs. The corresponding space
$R_{2\, 3q \, 3}$ is built by adding the Fock
space basis states that are needed to construct the
additional states. When one assumes that
$\cH_{\lambda N}$ has more terms than the
canonical LF Hamiltonian for QCD
\cite{Wilsonetal}, the space $R_{\tau \psi N}$ is
greater than in the canonical case. States created
from $|\psi \rangle$ with $k$ particles by $\tau$
actions of $\cH_{\lambda N}$ may contain up to $k
+ \tau(N-2)$ particles.

The set $R$ of states one obtains for any $\tau$
is a priori infinite, because one can have an
arbitrary relative motion of particles in a state
and the range of relative momentum is infinite.
The corresponding subspace in the Fock space 
is also unlimited. This happens despite that a 
regulated Hamiltonian cannot change a relative 
momentum by an arbitrarily large amount. The reason 
is that the unknown wave function $\psi_P(1,...,n)$ 
in Eq. (\ref{psizero}) does not limit the magnitude 
of relative momenta. For example, a square-integrable
function, such as Gaussian, quickly falls off as a
function of relative momentum of two particles but
the range of relative momentum is infinite. 

In order to introduce a space of states with a
finite range of relative momenta, we impose a
cutoff $\Delta$ on the invariant mass. Namely, we
define the space of states that are in $R_{\tau
\psi N}$ and whose invariant mass $\cM < \Delta$.
$\cM$ is calculated using kinematical momenta and
eigenvalues of $\cH_0$. The resulting space with
this cutoff is denoted by $R_{\Delta \tau \psi N}$, 
or shortly $R$. We introduce the projection operator 
on the space $R$ and denote this projector also by 
$R$. Hamiltonian $H_{\lambda N}$ projected on the 
space $R$ is denoted by 
\beq 
\cH_{\lambda N R} \es R \, \cH_{\lambda N} R \, . 
\eeq

The successive approximation to $\cH_\lambda$
order $N$ in application to states of the type
$|\psi \rangle$ with accuracy to $\tau$ actions
of $\cH_\lambda$ on $|\psi \rangle$ with invariant 
mass cutoff $\Delta$ is obtained by solving
equation
\beq 
\label{narrowR1}
\cH_{\lambda N R}' \es
\left[ \, [\cH_0, R \,\, \cH_{\lambda N P}R \,], \, R \,\, \cH_{\lambda
N} R \, \right] \, ,
\eeq 
with initial conditions specified by the canonical
Hamiltonian of LF QCD with counterterms. 

Structure of the counterterms is found by
demanding that all matrix elements of the
effective Hamiltonian $\cH_{\lambda N R}$ for a
large ratio $\Delta/\lambda$ among basis states
with invariant masses much smaller than $\Delta$
do not depend on the regularization applied in the
canonical Hamiltonian in the limit of this
regularization being removed \cite{GlazekWilson1}.

The degree of dependence of the counterterms on
$N$ and $\tau$ that are used in defining an
approximation order $N$ to $\cH_\lambda$ using
$\tau$ actions on various states $|\psi \rangle$
requires studies. Specific choices of states
$|\psi \rangle$ may simplify identification of
structure of counterterms in the canonical LF
Hamiltonian for QCD. However, fixing all finite
parts of the counterterms will require systematic
studies of symmetries that are expected to relate
finite parts of different counterterms, and input
from phenomenology in order to fix mass parameters
as required, in addition to the value of
$\Lambda_{QCD}$. We proceed to showing narrowness
of $\cH_{\lambda N R}$ that satisfies Eq.
(\ref{narrowR1}). 

For notational brevity, Eq. (\ref{narrowR2}) is written as 
\beq 
\label{narrowR2}
H' \es
\left[ [H_0, H_{IP}], H \right] \, ,
\eeq 
where
\beq
H   \es R \, \, \cH_{\lambda N } R \, , \\ 
H_0 \es R \, \, \cH_0            R \, , \\
H_I \es         H - H_0            \rs R \, \, \cH_{I\lambda N } R \, , \\ 
H_{IP} \es R \, \, \cH_{I\lambda NP} R \, .
\eeq

Evolution in $\lambda$ according to Eq.
(\ref{narrowR1}), or (\ref{narrowR2}), 
preserves traces of powers of $H$. Trace is defined by 
summing diagonal matrix elements in the set $R$. The 
diagonal matrix elements are proportional to $\delta^3(0)$ 
in momentum space which has interpretation of volume and 
can be divided out. The argument 0 corresponds to the 
conservation of total momentum. 

Using the condition that the trace of $H^2$ does
not depend on $\lambda$, one has
\beq
\label{TrHsquarex}
0 \es 
Tr \left( H_0^2 + 2 H_0 H_I + H_I^2 \right)' \, . 
\eeq
Since $H_0$ is fixed,
\beq
\left( Tr H_I^2 \right)'
\es - 2 \, Tr \, H_0 H_I'  \, . 
\eeq
Thus, using the basis in space $R$ that is 
built from normalized eigenstates $|m\rangle$ 
of $H_0$, with eigenvalues $E_m$, so that 
$H_0 |m\rangle = E_m |m\rangle$ and 
$H_{mn} = \langle m | H | n \rangle$, etc., 
we have
\beq
\label{start2x}
\left( \sum_{mn} |H_{Imn}|^2 \right)'
\es
- 2 \sum_m E_m H'_{Imm} \, .
\eeq
But Eq. (\ref{narrowR2}) implies that 
\beq
\label{start3x}
H'_{Imn} \es 
\left[ [H_0, H_{IP}], H_0 \right]_{mn}
+
\left[ [H_0, H_{IP}], H_I \right]_{mn}  \\
\es 
-(E_m - E_n)^2 P_{mn}^{+2} H_{Imn}
\np
\sum_k \left[ (E_m - E_k)P_{mk}^{+2} 
           + (E_n - E_k)P_{kn}^{+2} \right] 
           H_{Imk} H_{Ikn}  \, .
\eeq
The momentum $P^+_{ij}$ is the $+$ component of
the total momentum of the particles that undergo
interaction in the relevant matrix element
$\langle i | H_I | j \rangle$. Momenta of the 
spectators of the interaction do not contribute 
to $P^+_{ij}$. The result needed on the right-hand 
side of Eq. (\ref{start2x}) is 
\beq
\label{start4x}
H'_{Imm} 
\es 
\sum_k (E_m - E_k) \, \left( P_{mk}^{+2} + P_{km}^{+2} \right) \, 
           H_{Imk} H_{Ikm}  \, . 
\eeq
Therefore, the sum on the right-hand side of Eq.
(\ref{start2x}) is
\beq
\sum_m E_m H'_{Imm}
\es
\sum_{km} (E_k - E_m)^2 |H_{Ikm}|^2  \left( P_{mk}^{+2} + P_{km}^{+2} \right)/2 \, ,
\eeq
and 
\beq
\label{start5x}
\left( \sum_{mn} |H_{Imn}|^2 \right)'
\es
- \sum_{km} (E_k - E_m)^2 |H_{Ikm}|^2 (P_{mk}^{+2}+P_{km}^{+2}) \le 0 \, .
\nn
\eeq
This result leads to the following conclusion:
{\it The sum of moduli squared of all matrix
elements of the interaction Hamiltonian decreases
with $\lambda$ until all off-diagonal matrix
elements of the interaction Hamiltonian between
states with different free invariant masses vanish.} 
Thus also: {\it Matrix elements of the interaction
Hamiltonian on the diagonal and between states of
equal invariant masses and between states of small
invariant masses as measured by $H_0$ may stay constant 
or even increase when $\lambda$ decreases but only at
the expense of still faster decrease of the
off-diagonal and large invariant mass matrix elements 
and only until all off diagonal matrix elements
between non-degenerate states are eliminated.}

Another way of writing relation (\ref{start5x}) is
\beq
\label{start6x}
\left( \sum_{mn} |\cH_{Imn}|^2 \right)'
\es
- 2 \sum_{km} (\cM^2_{km} - \cM^2_{mk})^2
|\cH_{Ikm}|^2 \le 0 \, ,
\eeq
where $\cM_{ab}$ denotes an invariant mass of
particles in state $a$ that interact in an 
action of $\cH_I$ once on particles in state 
$b$. This form concludes our demonstration that 
RGPEP provides effective Hamiltonians which are 
narrow in the invariant mass of interacting 
particles beyond perturbative calculus. The 
invariant masses result from cancellation of 
spectator contributions to $P^-$ and multiplication 
only by $P^+$ of particles that are involved in 
an interaction, i.e., not including spectators.

Initial comparison between the non-perturbative RGPEP
calculus described in this Appendix and perturbative 
RGPEP calculus described earlier, can be done by 
solving Eq. (\ref{npRGPEP}) using expansion in 
powers of a coupling constant, such as $g_\lambda$ 
in the case of QCD, and observing how results of
one way of calculating compare with the other for 
terms order $g_\lambda$ and $g_\lambda^2$. Using 
notation introduced earlier (e.g., see 
\cite{gluons}, Sec. III.C, or \cite{GlazekScalar}, 
Sec. II.B), one rewrites Eq. (\ref{npRGPEP}) as
\beq 
\label{EQstep14}
\cH_{ac}' \es
- ac^2 \, [\cH_I]_{ac} 
+ \sum_b (p_{ab} \, ab + p_{cb} \, cb) \,\left[ \cH_I \, \cH_I \right]_{ac} 
\, ,
\eeq 
where letters $a$, $b$, and $c$, denote configurations of
particles. Using expansion
\beq
\cH \es \cH_0 + g \cH_1 + g^2 \cH_2 + ... \, ,
\eeq
one obtains for the first two terms equations
\beq 
\cH_{1ac}' \es
- ac^2 \, \cH_{1ac} \, , \\
\cH_{2ac}' \es
- ac^2 \, \cH_{2ac} 
+ \sum_b (p_{ab} \, ab + p_{cb} \, cb) \,\left[ \cH_1 \, \cH_1 \right]_{ac} 
\, .
\eeq 
Using the particle size parameter $s = 1/\lambda$, 
one obtains the solutions  
\beq
\cH_{1ac} \es e^{-ac^2 s^4} \, \cH_{1ac}(0) \, , \\
\cH_{2ac} \es 
e^{-ac^2 s^4} \, \cH_{2ac}(0)
+
\sum_b  \left[ \cH_1(0) \, \cH_1(0) \right]_{ac} 
\nt
e^{-ac^2 s^4} \, { p_{ba} \, ba + p_{bc} \, bc  \over ba^2 + bc^2 - ac^2 }
\, 
\left[ e^{-(ab^2 + bc^2 - ac^2)s^4} - 1 \right]
\, .
\eeq
The first-order result is Gaussian in the change
of invariant masses squared, identical to the first-order
result in perturbatively defined RGPEP. The second-order 
result matches the perturbatively defined RGPEP,
e.g., see Eq. (2.22) in Ref. \cite{GlazekScalar},
when $ac$ can be neglected in comparison to 
$ab$ or $bc$. When $ac$ refers to a dominant Fock
component and $ba$ and $bc$ refer to a component
with additional constituents, this condition may
be satisfied very well if the additional
constituents introduce a significant contribution 
to the invariant mass of the subsystem that counts.
Such situation is encountered in the dynamics of 
heavy quarkonia \cite{GlazekMlynik}, where the 
additional constituent is an effective gluon
while kinetic energies of quarks cannot change
by much because quarks move slowly with respect to 
each other and their masses are large in comparison 
to $\Lambda_{QCD}$. This means that the non-perturbative 
formulation of RGPEP can be directly applied to heavy 
quarkonia with arbitrary motion as a bound state in 
QCD, and implies that the non-perturbative formulation 
can be used to derive corrections to the harmonic 
oscillator force in Ref. \cite{GlazekMlynik} beyond 
perturbation theory to describe the spectrum of 
charmonium and bottomonium states that results from 
gluon dynamics neglecting light quarks.

The procedure described here differs from
truncation in powers of a bare coupling constant
$g$, truncation in the number of Fock sectors, and
combinations of both these truncations. Although
we use the Fock space subspace $R$ for solving 
Eq. (\ref{narrowR1}), and it is useful to use an 
expansion for $\cH_\lambda$ in powers of the coupling 
constant $g_\lambda$ \cite{Wilsonetal}, the solutions 
provide in principle not just the matrix elements of 
Hamiltonian terms in a limited space of states but 
the matrix elements of Hamiltonian terms from which 
one extracts the coefficients $h_\lambda$ of particle 
operators in the Hamiltonian terms that a priori 
act in the entire Fock space. Neither the Fock space 
nor perturbation theory limitations prevent us from
seeking a non-perturbative solution for the
coefficients $h_\lambda$. 

Some comments are in order here. One can alter the
generator in Eq. (\ref{narrowR1}) by introducing
convergence-improving factors that are required
for perturbative approximations
\cite{AlteredWegnerPRD,AlteredWegnerAPP}. For
bound states of low-mass hadrons alone, the
relative motion of constituents can be described
using other basis functions than plane waves, such
as the oscillator basis. Scattering states of
hadrons may require a more elaborate basis in
coupled channels. Reasoning described in this
Appendix justifies methods used in nuclear physics
with only kinetic energy in SRG generator
\cite{OSU1, OSU2, OSU3, OSU4, OSU5}, including the
impact of bound states \cite{impact}. The impact
can only lead to increase of interaction matrix
elements that are small since the combined
strength of all interaction matrix elements must
only decrease until it reaches a minimum. This
result extends the range of applicability of SRG
with a kinetic energy or a similar term in the
generator. It suggests applications of simpler
flow equations than Wegner's in relevant areas of
physics.

\section{ Visualization of RGPEP scale dependence of 
          hadronic structure }
\label{OverlappingSwarmsModel}

\begin{figure}
\begin{center}
\includegraphics[scale=.5]{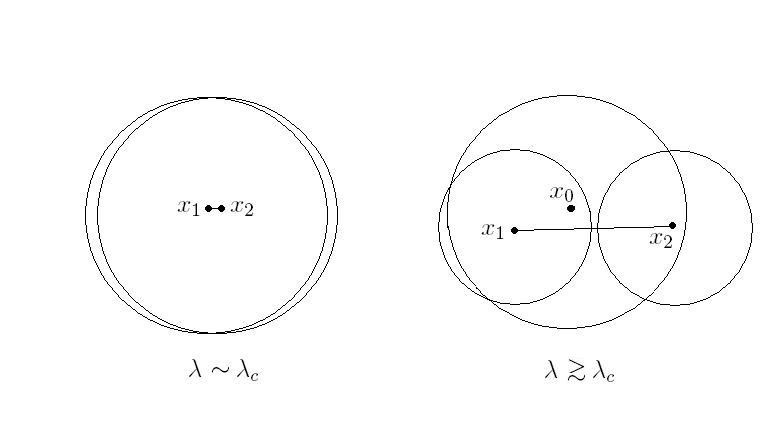}
\end{center}
\caption{The RGPEP image of a meson seen at $\lambda$ near $\lambda_c$.}
\label{fig:MQ}
\end{figure}

\begin{figure}
\begin{center}
\includegraphics[scale=.5]{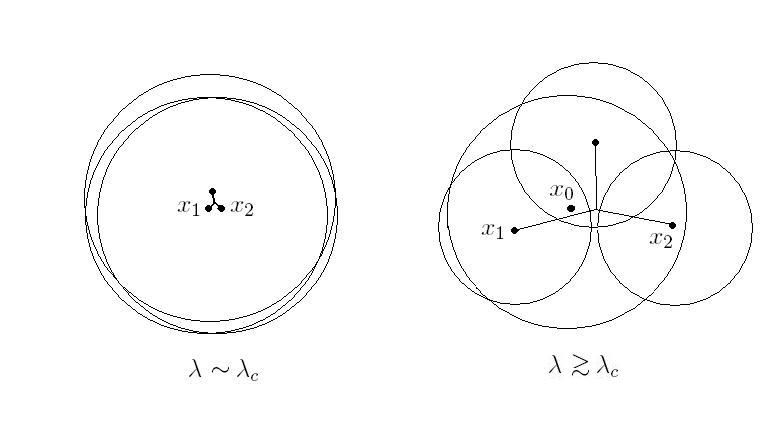}
\end{center}
\caption{The RGPEP image of a nucleon seen at $\lambda$ near $\lambda_c$.}
\label{fig:NQ}
\end{figure}

Quantum fields for effective particles of size $s
= 1/\lambda$ in RGPEP, see Eqs.
(\ref{quantumfieldpsilambda}),
(\ref{quantumfieldAlambda}), and (\ref{psixs}),
are the degrees of freedom from which one can
build effective Hamiltonian densities. An
effective Hamiltonian at scale $\lambda$
describes a hadron only in terms of quarks of
size $s = 1/\lambda$. Therefore, when the scale
parameter $\lambda$ approaches $\lambda_c \sim
\Lambda_{QCD}$, the size of quarks, as measured by
the strong interaction range in interaction
vertices \cite{NonlocalH}, becomes comparable with
$1/\Lambda_{QCD}$. This in turn means that the
effective quarks become as large as the whole
hadron. Such quarks must nearly overlap and this
is how they form a white mixture. There is some
imbalance of color on the boundary, perhaps able
to couple to $\pi$-mesons that form a cloud around
the quarks. Visualization of these circumstances
is provided in Figs. \ref{fig:MQ} and
\ref{fig:NQ}. When $\lambda$ is near but greater
than $\lambda_c$, the constituent quarks become
smaller and part of the color-active medium inside
a hadron is no longer fully overlapped by quarks.
Therefore, when $\lambda \gtrsim \lambda_c$, there
must be additional matter density involved in
filling the volume of a hadron.

One can use an analogy with swarms of bees. For
example, a baryon can be seen at scale $\lambda_c$
as built from three swarms. One swarm is made of
red bees, one of green, and one of blue. One of
the three swarms corresponds to one constituent
quark. There is also a fourth swarm, built from
bi-colored glue bees. But the four swarms overlap
and in nearly every part of the baryon volume one
has one red, one green and one blue bee, or glue
bees. The regions where the three quark swarms are
overlapping are locally white. 

When one changes the scale $\lambda=\lambda_c$ to
a nearby $\lambda \gtrsim \lambda_c$ as described
in Section \ref{3ways}, the size $s$ of each of
the three swarms becomes smaller than $s_c =
1/\lambda_c$, but the medium around the quarks of
size $s$ remains filled with the glue and quark 
bees that can still balance to white, too. When 
one decreases $\lambda$ below $\lambda_c$, the mass 
of the glue component may decrease toward 0 or
stabilize while the constituent quarks fully
saturate the volume of a hadron. 

In distinction from the parton model
\cite{partonmodel} and related scaling pictures
\cite{Wilsonpartonmodel,KogutSusskindscaling}, the
new element of this visualization is that the
constituent swarms are much larger in size than
the distances between their centers. It seems to
the author that this feature of RGPEP deserves a
visualization because none of the visualizations
of hadrons known to the author has this feature.
Namely, common visualizations represent a baryon
as made of three constituent quarks that fly like
apples in a bag, sometimes accompanied by some
strands of glue in some kind of a cavity of
unspecified nature. Images associated with an
increase of momentum scale typically include more
tiny objects in two ways: either as finer pieces
scattered around or as interaction processes among
quarks and gluons (such as on the cover of Ref.
\cite{PeskinSchroeder}). It is hard to visualize
the origin of constituent quarks and CQM
potentials in these ways. 

In the context of RGPEP, the analogy with swarms
of colored bees provides the following
visualization for the mechanism of formation of
potentials. We start with a meson built from a
quark and an anti-quark. A quark has opposite color
charge density to anti-quark. When the effective
particles are completely overlapping like two
densities of opposite charges, their state is
locally neutral. This picture oversimplifies the
non-Abelian picture to the Abelian one but
still offers some intuition. Imagine now
that the centers of the initially overlapping
swarms move a little bit apart. It is well
known that the Coulomb force between two
spherically symmetric distributions of opposite
charges grows linearly with the distance between
their centers for as long as this distance is
small in comparison to the individual sizes of
the distributions. Therefore, the potential energy
of two nearly overlapping spheres is described by a
harmonic oscillator potential. It should be
stressed that the distance between the centers of
the swarms is much smaller than the size of
each swarm. In the case of baryons, one needs
to consider three centers of three overlapping
swarms being relatively close to each other.

Of course, this visualization has many drawbacks:
it is classical, Abelian, non-relativistic, and
does not provide any insight concerning the
difference between quarks and gluons. But it does
model the transformation $W$ that increases the
number of quarks and gluons of decreasing size $s$
when $\lambda$ increases. These smaller quarks and
gluons form the medium that provides the
expectation value of the gluon field that is
interpreted in this article as a gluon condensate.

\end{appendix}



\end{document}